\documentclass[fleqn,12pt]{wlscirep}
\usepackage[utf8]{inputenc}
\usepackage[T1]{fontenc}
\usepackage{enumitem}
\usepackage{marginnote}

\newcommand{\notereviewone}[1]{\reversemarginpar\marginnote{\textcolor{LimeGreen}{R1:#1}}\normalmarginpar}
\newcommand{\notereviewtwo}[1]{\marginnote{\textcolor{ForestGreen}{R2:#1}}}
\renewcommand{\marginnote}[1]{}

\usepackage{float}
\usepackage[format=hang]{caption}
\usepackage[format=hang]{subcaption}
\graphicspath{{../graphics/}}
\usepackage{url}
\usepackage{xspace}
\newcommand{\suppi}{SI\xspace}
\newtheorem{definition}{Definition}
\usepackage{array}
\newcommand{\PreserveBackslash}[1]{\let\temp=\\#1\let\\=\temp}
\newcolumntype{C}[1]{>{\PreserveBackslash\centering}p{#1}}
\newcolumntype{R}[1]{>{\PreserveBackslash\raggedleft}p{#1}}
\newcolumntype{L}[1]{>{\PreserveBackslash\raggedright}p{#1}}
\title{Complex Societies and the Growth of the Law}

\author[1,2,5,*]{Daniel Martin Katz}
\author[3]{Corinna Coupette}
\author[4]{Janis Beckedorf}
\author[2,5]{Dirk Hartung}
\affil[1]{Illinois Tech -- Chicago Kent College of Law}
\affil[2]{CodeX -- The Stanford Center for Legal Informatics}
\affil[3]{Max Planck Institute for Informatics}
\affil[4]{Faculty of Law, Ruprecht-Karls-Universit\"at Heidelberg}
\affil[5]{Bucerius Law School}

\affil[*]{dkatz3@kentlaw.iit.edu}


\begin{abstract}
While a large number of informal factors influence how people interact, modern societies rely upon law as a primary mechanism to formally control human behaviour. 
How legal rules impact societal development depends on the interplay between two types of actors: 
the people who create the rules and the people to which the rules potentially apply.
We hypothesise that an increasingly diverse and interconnected society might create increasingly diverse and interconnected rules,
and assert that legal networks provide a useful lens through which to observe the interaction between law and society.  
To evaluate these propositions, we present a novel and generalizable model of statutory materials as multidimensional, time-evolving document networks. 
Applying this model to the federal legislation of the United States and Germany, we find impressive expansion in the size and complexity of laws over the past two and a half decades. 
We investigate the sources of this development using methods from network science and natural language processing. 
To allow for cross-country comparisons over time, we algorithmically reorganise the legislative materials of the United States and Germany into cluster families that reflect legal topics. 
This reorganisation reveals that the main driver behind the growth of the law in both jurisdictions is the expansion of the welfare state, backed by an expansion of the tax state.
\end{abstract}
\begin{document}

\flushbottom
\maketitle

\thispagestyle{empty}

\section{Introduction}
\label{section:introduction}

Modern societies rely upon law as the primary mechanism to control their development and manage their conflicts. 
Through carefully designed rights and responsibilities, institutions and procedures,
law can enable humans to engage in increasingly complex social and economic activities. 
Therefore, law plays an important role in understanding how societies change. 
To explore the interplay between law and society, 
we need to study how both co-evolve over time. 
This requires a firm quantitative grasp of the changes occurring in both domains. 
But while quantifying societal change has been the subject of tremendous research efforts in fields such as sociology, economics, or social physics for many years \cite{bowers1937,bogue1952,tuma1984,palla2007,castellano2009,ebrahim2019},
much less work has been done to quantify legal change. 
In fact, legal scholars have traditionally regarded the law as hardly quantifiable, 
and although there is no dearth of empirical legal studies \cite{heise2011,ho2013,epstein2014}, 
it is only recently that researchers have begun to apply data science methods to law \cite{whalen2016,coupette2019,livermore2019,frankenreiter2020}.
To date, there have been relatively few quantitative works that explicitly address legal change \cite{cross2007,buchanan2014,rockmore2017,ruhl2017,rutherford2018,fjelstul2019}, 
and almost no scholarship exists that analyses the time-evolving outputs of the legislative and executive branches of national governments at scale. 
Unlocking these data sources for the interdisciplinary scientific community will be crucial for understanding how law and society interact.

Our work takes a step towards this goal.
As a starting point, we hypothesise that an increasingly diverse and interconnected society might create increasingly diverse and interconnected rules. 
Lawmakers create, modify, and delete legal rules to achieve particular behavioural outcomes, often in an effort to respond to perceived changes in societal needs.
While earlier large-scale quantitative work focused on analysing an individual snapshot of laws enacted by national parliaments \cite{bommarito2010a,katz2014},
collections of snapshots offer a window into the dynamic interaction between law and society.
Such collections represent complete, time-evolving populations of statutes at the national level. 
Hence, no sampling is needed for their analysis, 
and all changes we observe are direct consequences of legislative activity.
This feature makes collections of nation-level statutes particularly suitable for investigating temporal dynamics.

To preserve the intended multidimensionality of legal document collections and explore how they change over time,
legislative corpora should be modelled as dynamic document networks
\cite{bommarito2010a,katz2014,boulet2011,koniaris2014,sweeney2014,winkels2014,koniaris2018}.
In this paper, we develop an informed data model for such corpora,
capturing the richness of legislative data for exploration by social physics. 
We leverage our data model to analyse the evolution of federal statutes in the United States and Germany. 
Here, we find extensive growth in legal complexity as a function of volume, interconnectivity, and hierarchical structure of the legislation in both countries%
---evidence that the highly industrialised countries we study seek to manage behaviour by building increasingly complex bodies of legal rules.
Searching for the sources of the growth we observe, 
we draw on graph clustering techniques
to locate those legal topics that contribute most to the complexity increase and trace their development over time.
Our work highlights the potential of legal network data for studying the interaction between law and society when viewed through the lens of Complex Adaptive Systems (CAS)
\cite{post2000,porter2005,fowler2007,schaper2013,ruhl2017,lee2019}, 
and it opens novel research avenues to the interdisciplinary scientific community.

\section{Results}
\label{section:results}

\subsection{Dynamic network model of legislation}\label{subsection:model}
\label{subsection:datamodel}
We model 25 years of statutory materials from two advanced industrial countries, the United States and Germany, as time-evolving document networks. 
To build our original datasets, 
for the United States, we collect annual snapshots of the United States Code (US Code) from 1994 to 2018 from the Office of the Law Revision Counsel of the U.S. House of Representatives.  
For Germany, we create a parallel set of yearly snapshots for all federal statutory laws in effect at the beginning of the year in question based on documents from Germany's primary legal data provider, \emph{juris GmbH}. 
For details on our data sources, see Section~1 of the Supplementary Information (SI). 

Each individual law or section (that has not been repealed) contains at least some text, 
and it may contain nested subsections as well as ingoing and outgoing references.  
For each country and yearly snapshot, we construct a network of all federal statutes. 
The entities in this network are the structural elements of the statutes we collect, some of which contain text (i.e., the stipulation of a legal rule). 
These entities are interconnected by inclusion relationships (e.g., a section containing several paragraphs) and cross-references (i.e., the text of an element referencing another element), 
and they can be sequentially ordered by their labels.
Notably, only one level of the inclusion hierarchy in legislative corpora is uniquely sequentially labelled (this is the Section level in the United States and the § or Article level in Germany). 
We refer to the structural elements in this layer as \emph{seqitems}.
For excerpts from United States law and German law that illustrate their inherent structure, see Section~1.3 of the \suppi. 

\begin{figure}[H]
	\notereviewone{3}
	\centering
	\includegraphics[width=\textwidth]{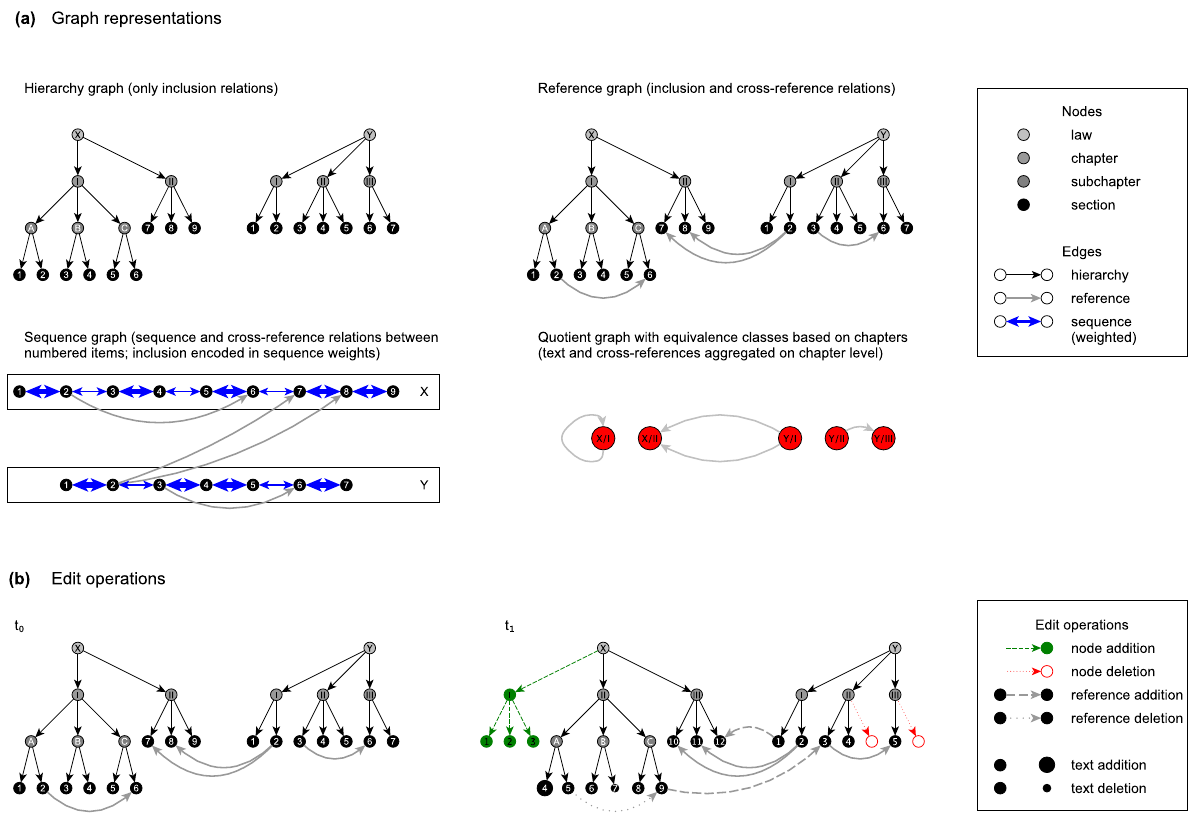}
	\caption{Dynamic network data model for legislative document collections. All figures created by the authors.}
	\label{fig:conceptual_sequence_graph}
\end{figure}

In the legislative process, 
the structure of legislative texts is controlled by the administrative officials drafting the rules. 
Therefore, \emph{hierarchy}, \emph{reference}, and \emph{sequence} within a corpus of legislative documents contain information about the content of the corpus 
that is less noisy and easier to parse than its language. 
To unlock this information for large-scale comparative and dynamic analysis, 
we model a body of legislation at a certain point in time as a document collection following the Document Object Model (DOM) standard \cite{domspec} (for our domain-specific XML Schema Definition [XSD], see Section~2.4 of the~\suppi). 
With each document collection, we associate four graphs 
as depicted in Figure~\ref{fig:conceptual_sequence_graph}~(a),
\notereviewone{2}
whose formal definitions are given in Section~\ref{subsection:methods:modelling}.
Our simplest graph is the \emph{hierarchy graph}, 
which models the inclusion relationships between the structural elements of legal texts. 
It is a subgraph of the \emph{reference graph}, which models inclusion and cross-reference relationships. 
From a network science perspective, the reference graph is perhaps the most intuitive representation of a legislative document collection, and all of our other graphs can be derived from it. 
The \emph{sequence graph} contains only the \emph{seqitems} from the reference graph, 
which are connected by cross-reference edges and bidirectional \emph{sequence edges} 
(Section~\ref{subsection:methods:modelling} introduces a parametrized definition of this graph for greater analytical flexibility). 
The cross-reference edges are unweighted, 
while the sequence edges have weights proportional to the distance between their endpoints in the undirected version of the hierarchy graph 
(e.g., a sequence edge between Section $i-1$ and Section $i$ in Chapter $A$ weighs more than a sequence edge between Section $i$ in Chapter $A$ and Section $i+1$ in Chapter $B$). 
\notereviewone{11}
The sequence graph expresses how legal practitioners work with a legal text (i.e., they approach a topic through one particular rule,
scan its vicinity as long as it is also hierarchically close,
possibly follow a cross-reference, 
then scan the hierarchically close vicinity of a referenced rule).
Finally, we define \emph{quotient graphs} based on attributes attached to the elements of our reference graphs. 
In these graphs, all elements with the same attribute value(s) (e.g., all \emph{seqitems} belonging to the same Chapter) are collapsed into one node, 
and edges are rerouted accordingly. 

The graphs sketched above allow us to compare legislative document collections both \emph{horizontally} (i.e., across nations) and \emph{vertically} (i.e., across time).
In particular, hierarchy graphs and reference graphs let us track basic statistics over time (cf. Section~\ref{subsection:methods:growth}), 
which change when lawmakers add, update, or delete legal rules as depicted in Figure~\ref{fig:conceptual_sequence_graph}~(b).
Sequence graphs help us align basic elements of legal texts across years (cf. Section~\ref{subsubsection:methods:temporal}).
Along with quotient graphs, they also facilitate the reorganisation of legislative materials via graph clustering (cf. Section~\ref{subsubsection:methods:clustering}), 
where they allow us to focus on different topics or levels of granularity depending on the research question to be investigated.
To the best of our knowledge, there exists no comparably flexible explicit model for legislative document collections in the document network analysis literature. 
Since we do not use all features of the model in our analysis, 
exploiting the power of our data model to a greater extent is a natural direction for future work (see Section~\ref{section:discussion} for details).

\subsection{Substantial growth in volume, connectivity, and hierarchical structure}
\label{subsection:growth}

\notereviewone{12}\notereviewtwo{5}
The data model introduced in Section~\ref{subsection:model} enables us to track the development of our legislative corpora over time. 
As Table~\ref{tab:descriptive-statistics} shows, the absolute size of these corpora has grown substantially in the past two and a half decades, 
whether measured by the number of tokens (whitespace-delimited chunks of text that roughly correspond to words), 
the number of structural elements, or the number of cross-references contained therein. 
Judging merely by the number of tokens, 
in both jurisdictions,
the law in 2018 is more than $1.5$ times as large as the law in 1994. 
Given the fact that the legal systems of both countries were already fully developed twenty-five years ago, 
the sheer magnitude of this growth is striking. 

\begin{figure}[H]
	\centering
	\includegraphics[width=\textwidth]{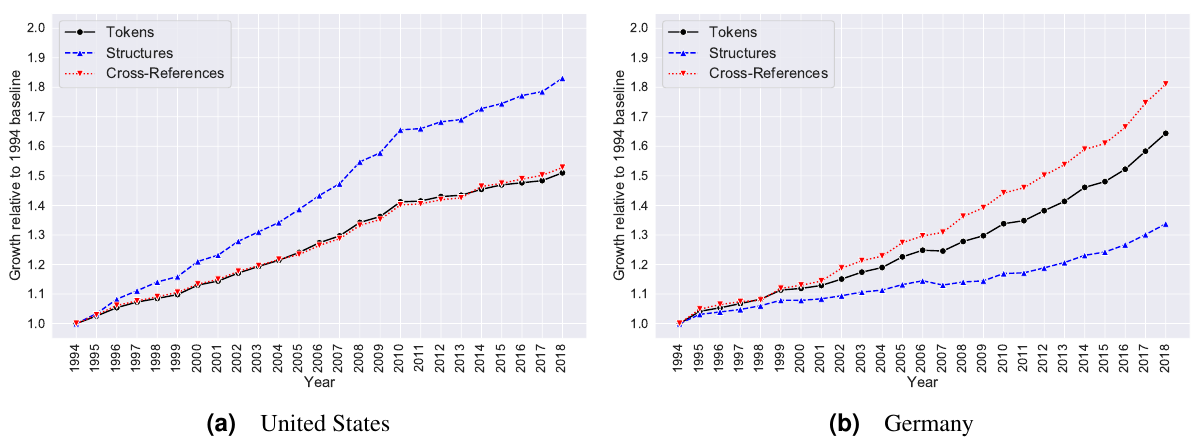}
	\caption{Federal legislation in the United States and Germany: growth statistics  (1994--2018).}
	\label{fig:legislative-growth}
\end{figure}

Inspecting the statistics in Table~\ref{tab:descriptive-statistics} along with the relative growth over time illustrated in Figure~\ref{fig:legislative-growth} 
further reveals two distinct growth patterns:
In the United States, the number of tokens and the number of cross-references grow at the same rate, 
which is considerably lower than the growth rate for the number of structural elements. 
In contrast, the German corpus exhibits its highest growth rate for the number of cross-references, 
and growth in the number of tokens is noticeably faster than growth in the number of structural elements.
Thus, the volume increase in the federal statutory legislation of the United States is accompanied primarily by an increase in the number of entities, 
whereas the volume increase in the federal statutory legislation of Germany is accompanied primarily by an increase in the number of relationships in the legislative network. 
This suggests that cross-references and hierarchical elements function as substitutes when it comes to integrating new content into an existing legal corpus.

\notereviewone{6}
However, increasing the number of hierarchical elements or the number of cross-references also tends to correlate with a decrease in navigability as it may be indicative of content fragmentation: 
Anyone trying to understand a legal rule will more often be forced to combine information from multiple places in the law to obtain a complete picture of its content. 
This difficulty is only exacerbated by the dominant legal information systems, 
which often force users to click through hierarchies of legal elements and seldom allow them to display a custom selection of structural units in a single browser tab for joint appreciation.
Therefore, our statistics for both countries support the intuition that their legislative apparatuses are growing also in complexity---%
although the complexity increase is driven by different design choices in both jurisdictions. 
While the difference in legislative drafting styles is of natural interest for comparative legal scholarship, 
the common growth trend we observe begs a broader question: 
What is its source?

\begin{figure}[H]
	\centering
	\includegraphics[width=\textwidth]{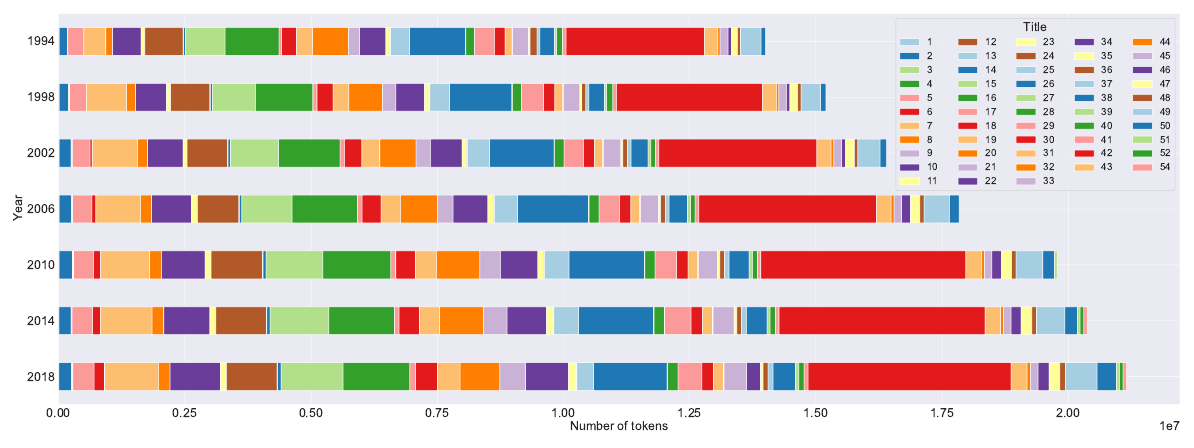}
	\caption{Federal legislation in the United States by Title (1994--2018), measured in tokens.}
	\label{fig:us-tokens-per-title}
\end{figure}

This question has no meaningful answer within the current formal organisation of the legislative materials. 
In fact, the US Code as the primary organisational system for legislation in the United States has barely changed in the time period under study. 
The US Code comprised 50~Titles in 1994;
since then, three Titles have been added (51, 52, and 54), 
two formerly empty Titles have been reassigned (6, 34), 
and two Titles have experienced small name changes (36, 47). 
Apart from that, US federal legislation has been codified in the same Titles since 1994,
with the total number of Chapters existing across all Titles rising from 2000 to 2723 (for an average growth of 30 Chapters per year). 
Figure~\ref{fig:us-tokens-per-title} localizes the growth over four-year intervals within the existing, 
content-based organisation of the US code. 
Based on raw token counts (excluding notes and appendices), 
the biggest growth has occurred in Title~42 (The Public Health and Welfare), 
Title~7 (Agriculture), 
and Title~15 (Commerce and Trade). 
The relative growth in the number of tokens has been highest in 
Title~4 (Flag and Seal, Seat of Government, and the States), 
Title~46 (Shipping), 
and Title~7 (Agriculture).

This gives an interesting first impression on the macro level,
but the Title headings are so general, 
and the content placed in the individual Titles is so diverse (e.g., the current Title~42 contains provisions on~Social Security [Chapter~7], Energy Policy [Chapter 134], and Aeronautics and Space Activities [Chapter 155]), 
that it tells us little about the triggers and the nature of the growth we observe. 
The situation further deteriorates if we want to compare the German developments with those in the United States: 
Germany does not codify its federal legislation in a single official collection but publishes only individual acts and classifies them into subject areas for navigation
(details can be found in Section~1.2 of the~\suppi). 
The number of consolidated acts with more than $500$ characters (roughly a paragraph, effectively excluding laws with purely formal content) grew from around $1550$ to over $1800$ in the period from 1994 to 2005, 
then was intentionally shrunk to around $1550$ until 2011,
and has resumed slow growth since 2011, reaching around $1600$ in 2018---%
so we do not even see a monotone growth pattern in this data.
To uncover the sources of the growth of the law, 
and compare our findings between the United States and Germany,
we thus need to reorganise the legislative materials of both nations.

\subsection{Clustering for comparative and dynamic analysis}
\label{subsection:reorganisation}

\begin{figure}[H]
	\vspace*{-40pt}
	\notereviewone{7}
	\centering
	\includegraphics[width=\textwidth]{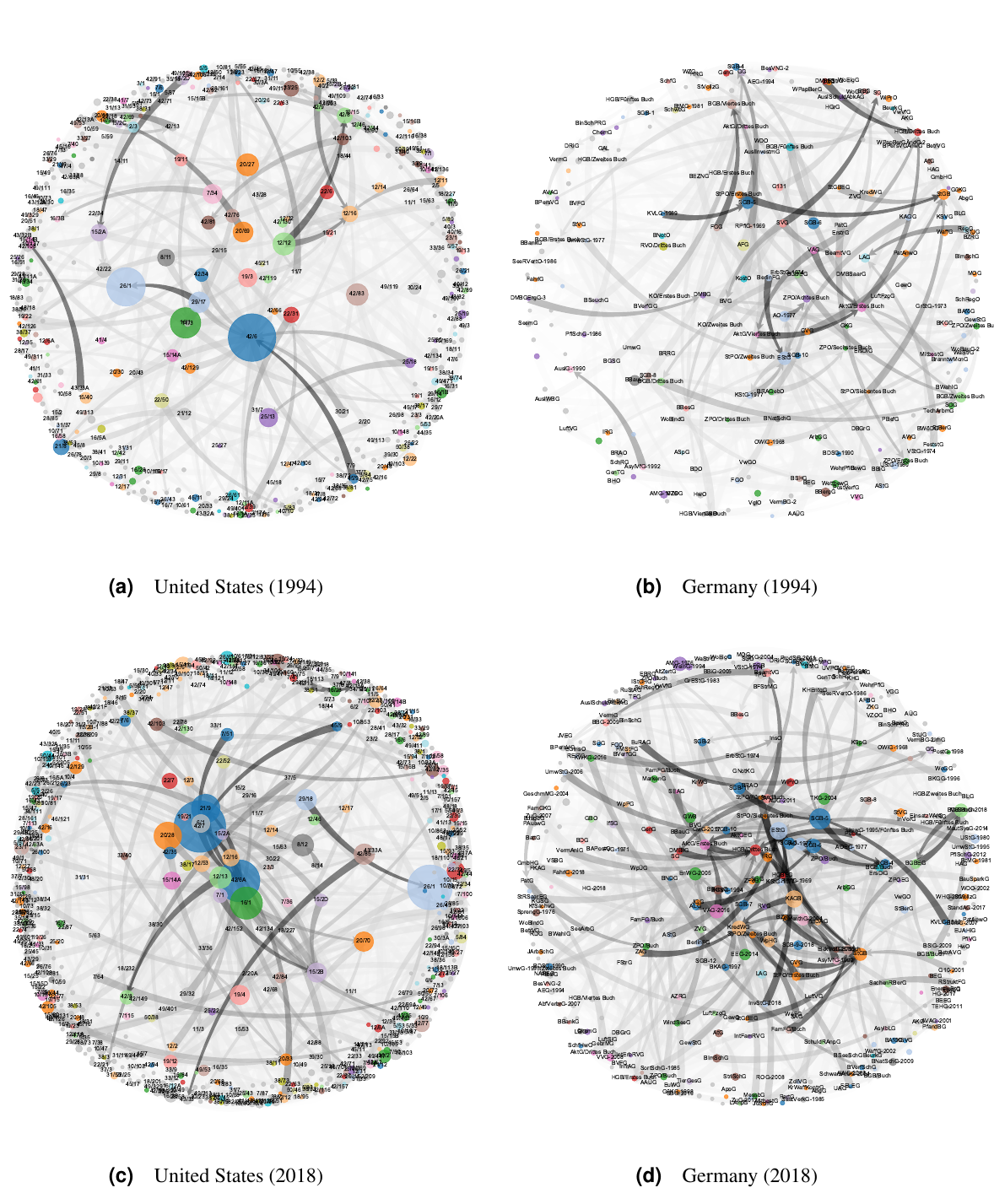}
	\caption{Federal legislation in the United States and Germany: quotient graphs by Title/Chapter (United States) and Law Name/Book (Germany) (1994 and 2018),
		with arrows running between nodes indicating that text contained in one node cites text contained in another node.
		Node sizes indicate token counts (larger $=$ more tokens), where only nodes with at least $5000$ tokens (corresponding to roughly ten pages) are shown.
		For each nation separately, nodes share the same colour if they are placed in the same cluster family,
		and nodes not in one of the $20$ largest cluster families are coloured in grey.
		Only the labels of nodes that get cited by or cite other nodes at least $20$ times (i.e., nodes with a combined in- and out-degree of $20$) are drawn.
	}
	\label{fig:us-de-chapter-quotient}
\end{figure}

A first, straightforward way to reorganise the US Code is to aggregate it not at the Title level but rather at the Chapter level. 
This is especially convenient because the number of Chapters in the US Code is comparable to the number of individual laws in Germany,
which we only break up into smaller units if they contain several Books (a common feature of large German codifications such as the German Civil Code [BGB] and the German Commercial Code [HGB]).
\notereviewone{7}
The node-link diagrams of the quotient graphs corresponding to this reorganisation for the United States and Germany in 1994 and 2018 are shown in Figure~\ref{fig:us-de-chapter-quotient}.
In these graphs, nodes share the same colour if they belong to the same \emph{cluster family}.
Broadly speaking, cluster families are sets of clusters (a cluster is a set of nodes), 
mostly from different snapshots, 
which contain many identical, similar, or related rules (cf. Definition~\ref{def:cluster-family} in Section~\ref{subsubsection:methods:temporal})%
---and as such, they approximate legal topics. 
We identify cluster families using node and cluster alignments (cf. Section~\ref{subsubsection:methods:temporal}). 
Cluster families will help us assess which legal topics are driving the growth we report in Section~\ref{subsection:growth}. 
The cluster family colouring scheme will be used in all remaining graphics; 
a full legend mapping colours to legal topics can be found in Section 5.1 of the~\suppi. 
In Figure~\ref{fig:us-de-chapter-quotient}, 
nodes of the same colour can generally be thought of as belonging together (i.e., \emph{same colour} $\Leftrightarrow$ \emph{(roughly) same legal topic}), 
and node colours can be compared across years but not across nations (e.g., the legal topic of red nodes in the graphs for the United States may differ from the legal topic of red nodes in the graphs for Germany). 

The node-link diagrams allow us to identify interesting connections between individual parts of the law at the Chapter level---%
e.g., Book Three of the German Commercial Code (HGB/Drittes Buch), which regulates books of accounts, is much more central as a reference target in 2018 than it was in 1994, 
and the central role of Title 42, Chapter 6 of the US Code in 1994 (The Children's Bureau) has been taken by Title 42, Chapter 6A (Public Health Service) and Chapter 7 (Social Security) in 2018. 
But since there are well over $1000$ nodes in both jurisdictions, 
the quotient graphs are difficult to analyse in their entirety,
and related content remains scattered over different nodes.
To coherently group related content, 
we need more sophisticated reorganisation method. 
Therefore, we cluster our annual Chapter quotient graphs for each country based on their cross-references. 

\notereviewone{11}
In (non-overlapping) graph clustering, 
the goal is to divide the elements of the graph (typically the nodes, here: Chapters in the United States and Books or individual laws in Germany) 
into groups such that elements in the same group are relatively densely connected, 
whereas elements in different groups are relatively sparsely connected.
We use the \emph{Infomap} algorithm based on the \emph{map equation} \cite{rosvall2008,rosvall2009} to find our clusters for three reasons. 
First, in using random walks (i.e, sequences of random steps using the edges of the graph) as a basic ingredient, 
\emph{Infomap} mimics how lawyers navigate legal texts. 
The legal navigation process is similar to how scholars navigate papers or web surfers navigate the World Wide Web (WWW), 
with the additional quirk that sequence edges play a large role in steering the search 
(think of reading the next paper in the special issue of a journal or clicking through a series of blog posts). 
Second, by leveraging the connection between finding clusters in a graph and minimizing the description length of a random walk on the graph, \emph{Infomap} has a solid information-theoretic foundation.
And third, the algorithm scales to large graphs. 

When running \emph{Infomap}, we use the default configuration, with one exception:
We pass $100$ as a parameter for the preferred number of clusters, 
which roughly corresponds to the number of top-level categories with which legal databases structure their content.  
This parameter choice allows us to determine the legal topics driving the growth we observe at a sufficiently high granularity while maintaining an overview of the entire corpus, 
and it protects against sudden jumps in the cluster granularity between years due to small differences in the description lengths of competing solutions, which are more likely to occur when no preferred cluster size is given. 
\notereviewone{5}\notereviewtwo{8}
As detailed in the sensitivity analysis included in Section~4.1 of the~\suppi, 
the precise number of input clusters has little impact on the overall results, 
as long as the numbers of clusters are comparable across years (e.g., tracing changes between a clustering in which most of the text is contained in $5$ clusters and a clustering with $50$ clusters is an invidious task). 
To increase the stability of our results, we obtain our final clustering for each country and year as the consensus clustering of $1000$ \emph{Infomap} runs, 
where the consensus clusters are the connected components of a graph whose nodes are the quotient graph nodes, 
and whose edges indicate which nodes co-occurred in the same cluster in $95~\%$ of all runs.
\notereviewtwo{9} 
As shown in Section~4.2 of the~\suppi, 
there is little variance both across those runs and across multiple consensus clusterings using $1000$ runs to find the consensus, 
indicating that our results are robust against the randomness inherent in the \emph{Infomap} algorithm.  

Based on our consensus clusterings, 
we can compute alignments between the clusters we find in subsequent snapshots for each of our countries. 
These \emph{cluster} alignments allow us to track the temporal evolution of individual clusters.
They are based on \emph{node} alignments of a fine-grained variant of sequence graphs, 
which leverage that most rules do not change most of the time---%
i.e., we can match many \emph{seqitems} between adjacent snapshots based on their (nearly) identical texts or (nearly) identical keys. 
For details on our node alignment heuristic and the cluster alignment procedure that builds upon it, see Section~\ref{subsubsection:methods:temporal}.

\begin{figure}[H]
	\centering
	\vspace*{-16pt}
	\notereviewone{8}
	\includegraphics[width=0.9\textwidth]{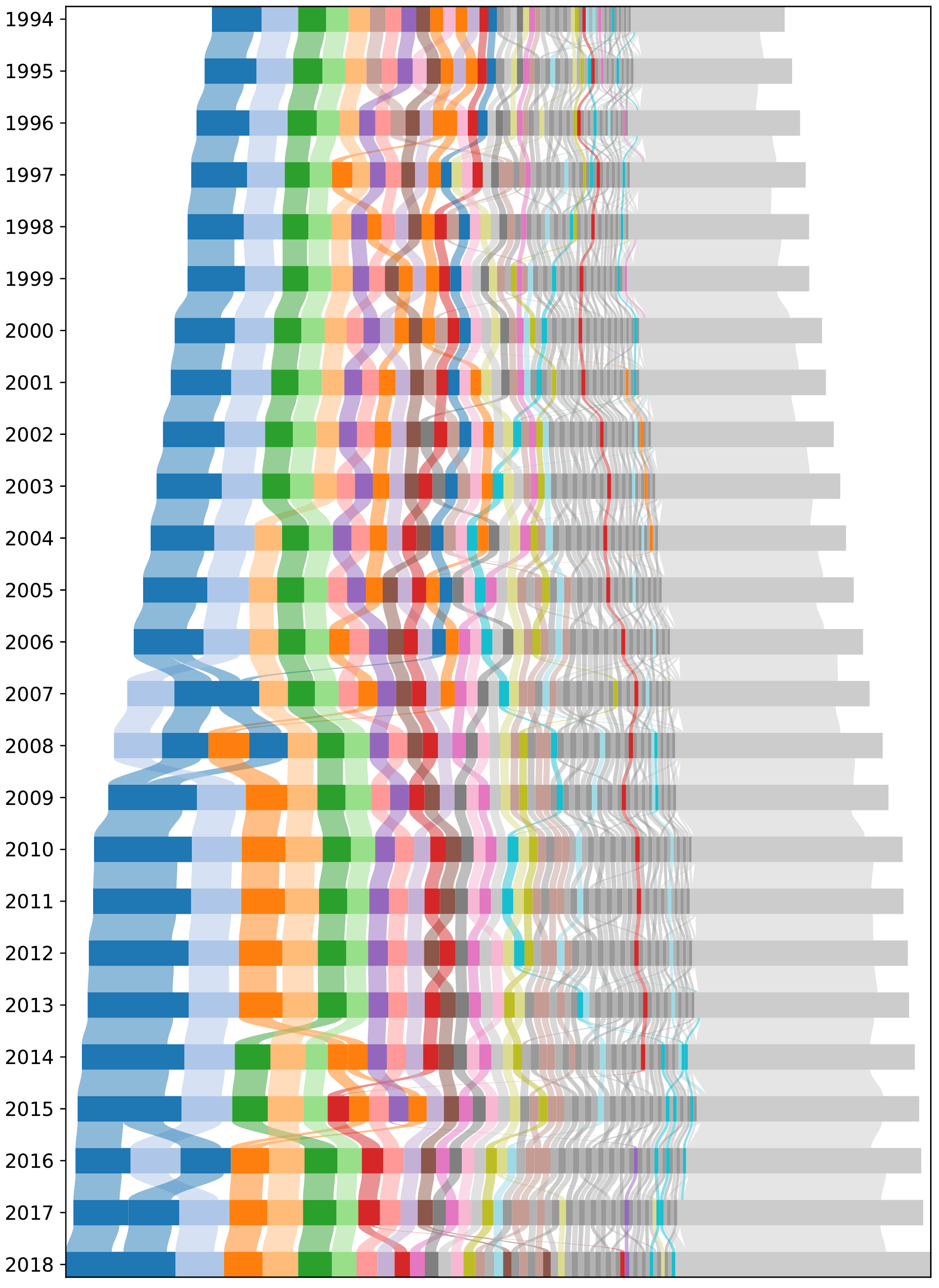}
	\caption{%
		Federal legislation in the United States by cluster (1994--2018). 
		Each block in each year represents a cluster. 
		Clusters are ordered from left to right by decreasing size (measured in tokens) and coloured by the cluster family to which they belong, 
		where clusters not in one of the $20$ largest cluster families are coloured in alternating greys. 
		Small clusters are summarised in one miscellaneous cluster, which is always the rightmost cluster and coloured in light grey.
		A full legend mapping colours to legal topics can be found in Section 5.1 of the~\suppi.
	}
	\label{fig:sankey}
\end{figure}

\notereviewone{8}
The fine-grained year-to-year cluster alignment facilitates a meso-level analysis of the growth reported in Section~\ref{subsection:growth}.
Figure~\ref{fig:sankey} provides a comprehensive overview of the aligned clusters for the entire United States corpus (an analogue figure for the German corpus can be found in Section~3.3 of the~\suppi): 
The corpus in a certain year is modelled by a horizontal bar, 
which is composed of blocks representing clusters with width proportional to the number of tokens they contain.
\notereviewone{4}
The year-to-year movement of tokens between clusters---%
i.e. the volume of text associated with one cluster in one year and another cluster in the next year, identified using the alignment between the items below the \emph{seqitem} level (\emph{subseqitems}) of the clusters---%
is indicated by splines connecting the blocks of adjacent years, 
where we only plot token movements corresponding to at least $15~\%$ of the tokens from both the ingoing and the outgoing cluster to filter out noise and isolate largely self-contained strands of the law as cluster families (see also Section~\ref{subsubsection:methods:temporal}).
The width of the plotted splines is again proportional to the number of tokens moving.
Within each horizontal bar, 
the blocks representing the clusters are sorted in descending order by their size, 
i.e., the clusters with the largest numbers of tokens are always pushed to the left.
To reduce visual clutter, we summarize very small clusters in one  \emph{miscellaneous} cluster. 
This cluster is always the rightmost cluster, depicted in light grey;
\notereviewtwo{6} 
more information on its contents can be found in Section~5.2 of the~\suppi. 
The blocks and splines belonging to the $20$ largest cluster families are uniquely coloured, 
whereas smaller cluster families are alternately coloured in alternating greys.
The absolute growth of the United States corpus is reflected in the increasing width of the bars over time, 
whereas changes in cluster compositions and relative cluster sizes are visible as diagonal year-to-year movements.  

\begin{figure}[H]
	\centering
	\includegraphics[width=\textwidth]{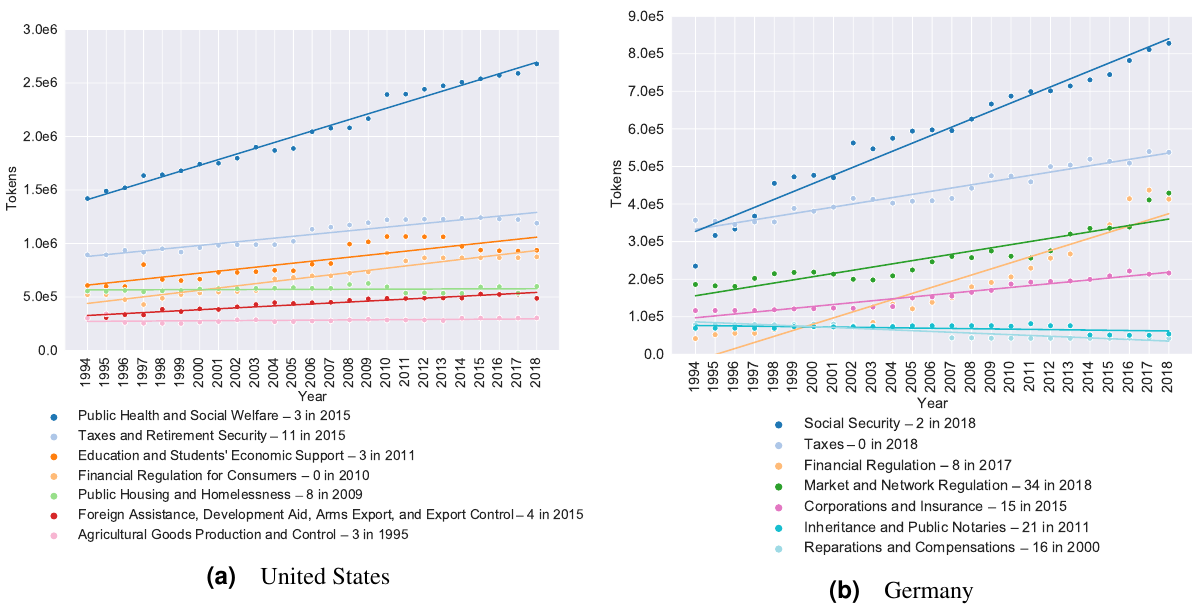}
	\caption{Federal legislation in the United States and Germany: growth statistics by cluster family for selected cluster families (1994--2018).
		The legends are sorted by the $y$-values of the regression lines in $2018$.
		The colours are comparable across countries, i.e., \emph{same colour} $\Leftrightarrow$ \emph{(roughly) same topic}.
	}
	\label{fig:legislative-growth-cluster}
\end{figure}

Inspecting the numbers behind Figure~\ref{fig:sankey}, 
we find that our clusters grow linearly with respect to their size, 
i.e., bigger clusters gain more tokens than smaller clusters, 
but that the growth rates differ depending on the legal topic represented by the cluster. 
To understand which legal topics are driving the overall growth, 
we determine the growth rate of our cluster families via an ordinary least squares (OLS) regression.
We select the 20 largest cluster families for both countries,
where the size of a cluster family is the size of its largest cluster (measured in tokens), called its \emph{leading cluster}.
For each of these cluster families, we inspect its content composition, 
and label it with the dominant legal topic.
\notereviewone{9}
More information on our labelling process, including a list of all labels, can be found in Section~5.1 of the~\suppi.
Together, the labelled cluster families account for roughly $50~\%$ of the total growth in the United States and roughly $80~\%$ of the total growth in Germany.
Figure~\ref{fig:legislative-growth-cluster} displays a selection of the most and least growing cluster families in the United States and Germany,
while detailed results can be found in Section~3.4 of the~\suppi. 
The colouring scheme for the United States is identical to that used in Figure~\ref{fig:sankey}, 
while the colours for Germany are chosen to match those for the United States for similar topics and avoid colour clashes otherwise.

\notereviewone{13}
Notably, in both jurisdictions, 
growth rates are highest for the cluster families concerning social welfare and financial regulation, 
and cluster families dealing with taxes, environmental protection, and immigration also display strong growth in both countries.
In addition to these similarities, we also find some differences in the growth patterns of both countries. 
As one might expect, 
the United States has cluster families concerned with Native Americans (shrinking) and student loans (growing),
while no analogous families exist in Germany.
Likewise, Germany has a cluster family concerned with war restitution (shrinking) that has no counterpart in the United States.
The unmatched growth of the criminal law and corporate and insurance law cluster families in Germany, which may be counterintuitive at first sight, 
is probably a result of differences in legislative competencies (criminal law and corporate law including insurance are largely state law in the United States, while they are federal law in Germany).
In addition, insurance regulatory law on the federal level in the United States is primarily enforced through federal regulations, 
which are not part of our dataset as they are kept in a separate collection (the Code of Federal Regulations).
That the United States has a growing cluster family concerned, inter alia, with foreign assistance and export control will not surprise those working in international development or international politics, 
and the fast-growing cluster dealing with renewable energy, power grid regulation, and related administrative procedures in Germany will not surprise those following the nation's political discourse (although in both cases, the unexpectedness could be impacted by hindsight bias).
Overall, the differences we observe seem to be in line with differences in the prominence of certain policy debates in both countries,
reflecting social, political, and cultural divergences.  
As such, they invite in-depth analysis by subject matter experts.

Finally, the year-to-year cluster alignment underlying Figure~\ref{fig:sankey} allows us to observe different types of growth.
For example, some clusters or cluster families witness \emph{intrinsic growth}, 
i.e., growth by addition of tokens without large gains of tokens from other areas; 
the cluster family containing veteran's benefits is a case in point, 
as is a cluster family on small business support and civil and military public procurement.
Such cluster families, which have been rather self-contained in the past $25$ years, address issues of sustained or increasing societal importance.
Other clusters or cluster families, however, witness \emph{extrinsic growth}, 
i.e., growth by gaining tokens from clusters in other families.
One example is a United States cluster concerned with the environmental protection of national parks and rivers, 
which grew substantially when rules about national forests as well as prospecting permits and leases joined it from clusters concerned with forestry and mining, 
indicating a shift in perspective from land use as resource exploitation to land use as resource conservation. 
To capture such differences in change processes, 
an elaborate cluster change event taxonomy is needed. 
Such a taxonomy could build on the work by Palla et~al. \cite{palla2007},
and developing it provides an interesting opportunity for further research.

\section{Discussion}
\label{section:discussion}

This paper investigates the growth of federal legislation in two industrial countries over a period of 25 years. 
As such, it is limited in \emph{geographic} scope (United States and Germany), 
\emph{temporal} scope (1994--2018), and \emph{institutional} scope (legislative bodies on the federal level). 
\notereviewtwo{2}
This makes it hard to assess to which extent the growth we observe is \emph{particular} to our data or rather \emph{universal}. 
The trend we identify applies to the recent history of federal legislation in at least two countries, the United States and Germany, 
and our findings in one country provide context for our findings in the other. 
Thus, we can establish that the growth we find is not a singular phenomenon, but we can only guess how it relates to the trends we might find in the legal document networks of other countries, time frames, or institutions.

The document networks that are most closely related to legislative networks are networks of regulations (produced by executive agencies) and networks of judicial decisions (produced by courts)---%
and for all of them, growth statistics that are directly comparable to ours are lacking. 
Some growth statistics are known for patent citation networks \cite{strandburg:2006, strandburg:2009,torrance2017,uspto2020},
where, e.g., the number of patents granted annually by the United States Patent and Trademark Office has roughly tripled in the 25 years from 1994 to 2018 \cite{uspto2020}. 
Since the generating processes of patent citation networks are very different from those of legislative networks 
(patent applicants need to cite prior art in their filings, patent examiners can add further citations, and too much prior art might risk patentability) 
and the units of analysis are not the same (structural elements in legislative networks vs. individual patents in patent citation networks), 
however, this result has little bearing on our findings. 
For similar reasons, comparing our findings with results on non-legal document networks, 
such as the World Wide Web or scholarly networks, is potentially misleading.
To put our findings into perspective, 
extending the scope of our data to other legal document networks is therefore an important direction for future work.
\notereviewtwo{3}
For example, investigations in the following directions are supported by our legal network data model:
\begin{enumerate}
	\item Analysing legislative activity on levels above and below the federal level
	and comparing the results with our findings will advance the search for invariants that characterize the development of legislative systems. 
	It can also help us understand the division of labour within the legislative pyramid (e.g., the federal, the state, and the local level).
	Does state law grow even faster than federal law? 
	If so, are the growth mechanisms similar or different? 
	How do the answers to these questions depend on the allocation of legislative competencies?
	
	\item Integrating documents from the executive and judicial branches of government with our datasets could help us explore how different parts of the legal system interact. 
	How does the evolution of a legislative network compare to that of a network of administrative regulations, a network of executive orders, or a network of judicial decisions? 
	In what areas of law is the development driven by the executive or the judiciary, rather than the legislative?
	What does this tell us about the distribution of power between the different branches of government?
	
	\item Combining our legislative network data 
	with data collected in other fields of quantitative social science 
	might improve our understanding of the interaction between legal rules and other rule sets that impact the behaviour of individuals and societies.
	When, where, and how do legislative changes impact how people behave on the ground? 
	When, where, and how do changes in how people behave prompt legislative changes? 
	In other words: What \emph{causal} relationships can we establish between legal change and societal change? 
	These questions are inherently multidisciplinary, 
	and to separate causes from confounders, 
	legal network data would need to be combined with data reflecting public sentiment (e.g., social media data or public news data) and data reflecting individual or collective choices (e.g., 
	financial network data, company reports, or economic panel data on households, firms, and non-governmental organisations). 
	Similarly, a multi-pronged strategy could be pursued to investigate the relationships between legal change and \emph{technological} change. 
	Here, combining legal network data with patent citation network, patent litigation, and R\&D investment data appears to be particularly promising.
\end{enumerate}

Methodologically, 
our approach emphasizes the structural features of legislative texts.
In particular, for the results we report in this work, 
the content of the legal texts has been only of indirect interest, 
e.g., as reflected in raw token counts or in reference structures that characterize legal topics.
As demonstrated in Section~\ref{subsection:reorganisation}, however,
qualitative analyses of the legal rules contained in our document networks can yield further insights, 
and this opens opportunities for normative legal research in areas such as comparative law and legal theory \cite{armour2009,spamann2009,cabrelli2015}.
In these legal disciplines, 
the United States and Germany are usually classified as following different legal traditions, also referred to as \emph{legal families},
and the categorization, though commonly accepted, has not been corroborated by empirical studies \cite{zweigertkoetz1998,husa2016,siems2016,glenn2019}.

Last but not least, the findings reported in this paper are based on a set of choices for methods and parameters.
For example, we examine growth by analysing year-to-year net gain of tokens, 
as this difference can be determined reliably.
The amount of legislative activity, however, is likely much higher (e.g., deletions and additions cancel out from the net gain perspective), 
and developing tools that allow for a fine-grained accounting of legislative changes constitutes an interesting research direction.
While we explored our model space extensively (as detailed in Section~4 of the~\suppi),
the parametrisation of the clustering required numerous decisions based on our experience and familiarity with the subject matter.
Other parametrisations are possible, 
and they might be needed in other analytical settings.
In particular, future work could examine selected parts of our data in greater detail, 
zoom in on a particular legal topic, 
and therefore choose very different parameters to operate at a higher level of resolution.

\section{Methods}
\label{section:methods}

\subsection{Modelling legislative document collections}\label{subsection:methods:modelling}

To formalize the intuition that is given in Section~\ref{subsection:model} and illustrated in Figure~\ref{fig:conceptual_sequence_graph}, 
we use the following definitions.
Let $D$ be a document understood by the Document Object Model (DOM) standard, 
with elements $\mathcal{E}_D$ of types $\{$\texttt{document}, \texttt{item}, \texttt{seqitem}, \texttt{subseqitem}, \texttt{text}$\}$ and root $r_D$ of type \texttt{document}.
We interpret $D$ as a directed rooted tree $T_D$ in the graph theoretical sense,
where the nodes of $T_D$ are the elements of $D$ that are \emph{not} of type \texttt{text}, 
and an arc between two nodes indicates that the source contains the target---i.e., $T_D$ contains all structural elements of $D$ with their containment relations.
With each node in $T_D$, we associate a unique identifier and three attributes: 
The \emph{type} of a node is its type in $D$, 
the \emph{level} of a node is its distance $d$ from the root, with $d(r_D,r_D) = 0$, and the \emph{text} of a node is the text of all its children (which can be used to derive additional statistics as necessary). 
Nodes of type \texttt{seqitem} (short for \emph{sequence item}) typically have \emph{cite~keys}, i.e., sequentially ordered unique identifiers by which they are commonly referenced. 
All nodes may also have \emph{headings} (representing the headings in the original document), and \texttt{documents} may have abbreviations by which they are commonly referenced.
The custom XSD expressing this document model can be found in Section~2.4 of the \suppi.

Now let $\mathcal{D}^i_t$ be collection $i$ of documents at time $t$ with their tree representations $\mathcal{T}^i_t$.
We define the following graphs for $\mathcal{D}^i_t$:

\begin{definition}[Hierarchy Graph $H^i_t$]
	The \emph{hierarchy graph} of collection $\mathcal{D}^i_t$, denoted $H^i_t$, is a \emph{digraph}
	\begin{align*}
	H^i_t = (V^i_{t,H}, E^i_{t,H})~,
	\end{align*} 
	where 
	\begin{align*}
	V^i_{t,H} = \underset{T\in\mathcal{T}^i_t}{\bigcup} V(T) \cup \{~\bar{r}_i~\}
	\end{align*}
	with a structural element $\bar{r}_i$ on level $-1$ representing the identity of the collection, 
	and
	\begin{align*}
	E^i_{t,H} = \underset{T\in\mathcal{T}^i_t}{\bigcup} E(T) \cup \{~(\bar{r}_i,r_D)\mid~D\in\mathcal{D}^i_t~\}~.
	\end{align*}	
\end{definition}

That is, the hierarchy graph is the union of all document trees' structural elements equipped with their containment relation, joined by a meta root node identifying the collection.

\begin{definition}[Reference Graph $R^i_t$]
	The \emph{reference graph} of collection $\mathcal{D}^i_t$, denoted $R^i_t$, 
	is a \emph{directed multigraph}
	\begin{align*}
	R^i_t = (V^i_{t,H}, E^i_{t,R})~,
	\end{align*} 
	where 
	\begin{align*}
	E^i_{t,R} = E^i_{t,H}\cup C^i_{t}~,
	\end{align*}	
	with $C^i_{t}$ a multiset given by
	\begin{align*}
	C^i_{t} =\{~(v,w)^m \mid~\text{text of~}v~\text{makes m references}\phantom{~.}\\\text{to}~w~\text{in  }\mathcal{D}^i_t~\wedge~type(v)=type(w)=\texttt{seqitem}~\}~. 
	\end{align*}
\end{definition}

That is, the reference graph is the hierarchy graph, 
augmented by reference relations between its nodes.

\begin{definition}[Sequence Graph $S^i_t(\rho, w, \alpha)$]\label{def:sequencegraph}
	The \emph{sequence graph} of collection $\mathcal{D}^i_t$ with parameters $\rho$, $w$, and $\alpha$, denoted $S^i_t(\rho, w, \alpha)$, is a \emph{directed multigraph}
	\begin{align*}
	S^i_t(\rho, w, \alpha) = (V^i_{t,S}(\rho), E^i_{t,S}(\rho, w, \alpha))~.
	\end{align*} 
	Here, $V^i_{t,S}(\rho)$ initially contains all nodes of type \texttt{seqitem}, 
	and nodes that are neighbours in the sequence are merged if and only if 
	they meet the \emph{merge condition} $\rho$.
	$E^i_{t,S}(\rho, w, \alpha)$ contains the arcs of $R^i_t$, projected onto the node set of $S^i_t$, with containment relations now represented as a pair of sequence arcs between nodes with adjacent \emph{cite keys}. 
	The sequence arcs in $E^i_{t,S}(\rho, w, \alpha)$ are weighted according to a weight function $w$ (specifying the weight decay of sequence arcs as a function of the distance between the source node and the target node in the undirected graph underlying $H^i_t$), 
	and the reference arcs are weighted according to a weight ratio $\alpha$ (specifying the weight of reference arcs in relation to sequence arcs of maximum weight). 
\end{definition}

As mentioned in Section~\ref{subsection:model},
the sequence graph representation of a legislative document collection is inspired by how practitioners work with legislative texts.
Furthermore, the parameters of the sequence graph allow us to incorporate knowledge about legal users into our model 
(e.g., by weighting reference arcs less heavily than the highest-weight sequence arcs, we can express the intuition that looking up a reference is less likely than simply reading on).
To compute the node alignments mentioned in Section~\ref{subsection:reorganisation}, 
we use a more granular variant of the sequence graph:

\begin{definition}[Subsequence Graph $\bar{S}^i_t(\rho, w, \alpha)$]\label{def:subsequencegraph}
	The \emph{subsequence graph} of collection $\mathcal{D}^i_t$ with parameters $\rho$, $w$, and $\alpha$, denoted $\bar{S}^i_t(\rho, w, \alpha)$, 
	is defined as the sequence graph $S^i_t(\rho, w, \alpha)$, with \emph{seqitems} being replaced by \emph{subseqitems} (i.e., structural elements one level below the \texttt{seqitem} level) if they exist.
\end{definition}

\notereviewone{2}
Finally, we use a multigraph version of the standard graph theoretical notion of a quotient graph (see also Section~\ref{subsubsection:methods:temporal}):

\begin{definition}[Quotient Graph $Q(G,R)$]
	Given a graph $G$ and an equivalence relation $R$ on its node set $V$ (i.e., a reflexive, symmetric, and transitive binary relation),
	a quotient graph is the graph $Q(G,R)$ with
	\begin{align*}
	V_{Q(G,R)} = V / R = \{~[u]_R \mid u\in V~\}
	\end{align*}
	and 
	\begin{align*}
	E_{Q(G,R)} = \{~([u]_R, [v]_R)^m \mid~|\{~(x,y) \in E_G \mid x\in [u]_R \wedge y\in [v]_R~\}| = m > 0~\}~, 
	\end{align*}
	where $[u]_R := \{~x\in V\mid (u,x)\in R~\}$ and $[v]_R := \{~y\in V\mid (v,y)\in R~\}$ are equivalence classes of $V$ under $R$.
\end{definition}

As shown in Section~\ref{subsection:reorganisation} for aggregating legal texts at the Chapter level,
the equivalence relations of our quotient graphs are generally given by the attributes associated with the structural elements contained in our reference graphs. 
Another example of quotient graphs, based on the cluster identifiers produced by our graph clustering as node attributes, can be found in Section~3.2 of the~\suppi. 

\subsection{Assessing legislative growth}\label{subsection:methods:growth}

To assess legislative growth in Section~\ref{subsection:growth}, 
we track three statistics for the United States and Germany from 1994 to 2018: 
the number of tokens, 
the number of hierarchical structures, 
and the number of references contained in the federal statutory legislation of both countries. 
For the token counts, we concatenate the text of all statutory materials for one country and year, 
ignoring the extensive appendices to some Titles or laws,
and split on whitespace characters. 
The hierarchical structure counts reflect the number of nodes in our hierarchy graphs, 
and the reference counts reflect the number of edges in our reference graphs.
Details on our data preprocessing steps can be found in Section~2 of the~\suppi. 

\subsection{Comparing document networks over space and time}

\subsubsection{Clustering document networks}\label{subsubsection:methods:clustering}

To enable our comparative and dynamic analysis in Section~\ref{subsection:reorganisation}, 
we cluster each annual snapshot of the legislative network separately for both countries. 
As mentioned in Section~\ref{subsection:reorganisation},
amongst the plethora of graph clustering methods, 
we choose the \emph{Infomap} algorithm due to its information-theoretical underpinnings, scalability, and interpretability as a legal (re-)search process. 
Details on this algorithm can be found in the original papers \cite{rosvall2008,rosvall2009}.

As the input data to \emph{Infomap}, 
we use the sequence graph representation of an annual snapshot
with a merge condition $\rho$ that collapses into one node 
all rules from the same Chapter (or Title, if the Title has no Chapters) in the United States, 
and all rules from the same Book (or law, if the law has no Books) in Germany. 
This consolidation step densifies the adjacency matrix of the sequence graph and reduces the noise in our data.
As almost all remaining nodes lie at distance $2$ from one another in the hierarchy graph, 
and very few sequence edges would remain,
we base the clustering solely on references.
Legislative network analyses using a different $\rho$ would also require the choice of a weight decay function $w$ 
and a \emph{sequence edge}-to-\emph{reference edge} weight ratio $\alpha$. 
For \emph{Infomap} itself, 
we use the default configuration with a preferred cluster number of $100$ as an additional input parameter. 
As discussed in Section~\ref{subsection:reorganisation},
this parameter choice reflects the level of analytical resolution we seek to operate at, 
and it approximates the number of high-level topics legal database providers utilise to organise their content.
The sensitivity analysis regarding our input parameter can be found in Section~4.1 of the~\suppi. 

As \emph{Infomap} has a stochastic element, 
we use \emph{consensus clustering} \cite{lancichinetti2012} to increase the robustness of our results as follows: 
For each snapshot $t$ in each country $i$, 
we produce $1000$ clusterings with different seeds.
From the results of these clusterings, we produce a \emph{consensus graph}
whose nodes are the nodes of the sequence graph, 
and with an edge connecting two nodes if these nodes are in the same cluster in at least $950 = 95~\%$ of our \emph{Infomap} runs.
For each year and country, the connected components of the consensus graph then constitute our final clusters, 
which represent a careful reorganisation of the law enabling comparative and dynamic analysis.
This leads to more than $100$ final clusters 
because the initial clusters are typically split into a stable core and several smaller satellites, 
each of which becomes an additional separate cluster.

\subsubsection{Tracing temporal dynamics}\label{subsubsection:methods:temporal}

To trace legislative change over time, we need to align the textual contents of our yearly snapshots within each jurisdiction. 
Computing the optimal node alignment between two graphs is generally a hard problem, 
and methods based on tree edit distance do not scale to legislative trees. 
However, we can use sequence graphs with the highest possible granularity 
(using a merge condition $\rho$ that condenses nothing)
along with the text associated with individual nodes, 
and exploit the fact that most rules do not change most of the time 
to construct a practical heuristic that greedily computes a partial node alignment $\phi^i_t$ across two snapshots $S^i_t$ and $S^i_{t+1}$ from corpus $i$. 
Our heuristic operates in at most four sequential passes through these snapshots:

\begin{enumerate}
	\item First pass: 
	If $v$ is a node in $S^i_t$ and we find exactly one node $w$ in $S^i_{t+1}$ with identical text \emph{and} the text is at least 50 characters long, set $\phi^i_t(v) = w$.
	\item Second pass: 
	If $v$ is an unmatched node in $S^i_t$ and we find an unmatched node $w$ in $S^i_{t+1}$ with identical key \emph{and} identical text, 
	set $\phi^i_t(v) = w$.
	\item Third pass: 
	If $v$ is an unmatched node in $S^i_t$ and we find exactly one unmatched node $w$ in $S^i_{t+1}$ such that (i) the text of $v$ contains the text of $w$ (or the text of $w$ contains the text of $v$) and (ii) the text remaining unmatched in $v$ ($w$) is shorter than the matched part, set $\phi^i_t(v) = w$.
	\item Fourth pass: 
	If $v$ is an unmatched node in $S^i_t$ and we find a matched node $v'$ in $S^i_t$ in the five-hop neighbourhood of $v$, 
	search the five-hop neighbourhood of $\phi^i_t(v')$ for the unmatched node $w$ (if any) with the largest Jaro-Winkler string similarity\cite{winkler1990} to $v$; 
	if that similarity is above $0.9$, set $\phi^i_t(v) = w$.
	Repeat recursively with all newly matched nodes until no further matches are found.
\end{enumerate}

With this procedure, we manage to map between $94~\%$ and $100~\%$ of the \emph{subseqitems} between adjacent snapshots in both the United States and Germany, i.e., our partial node alignments are almost complete, 
and the unmatched subseqitems are indicators of larger changes in the code (rather than errors). 
Based on partial \emph{node} alignments $\phi^i_t$ for all relevant $t$, we compute a partial \emph{cluster} alignment across snapshots, 
which we call the \emph{cluster graph} $C^i$:

\begin{definition}[Cluster Graph $C^i$]\label{def:clustergraph}
	Let $C^i_t$ be the consensus clustering obtained for collection $i$ at time $t$. 
	The \emph{cluster graph} of collection $i$ across times $T$, denoted $C^i$, is a weighted digraph 
	\begin{align*}
	C^i = (V^i_C,E^i_C)~,
	\end{align*}
	where
	\begin{align*}
	V^i_C = \underset{t \in T}{\bigcup} \{~c \in C^i_t~\}
	\end{align*}
	and
	\begin{align*}
	E^i_C = \{~(c,c',w) \mid c\in C^i_t~\wedge~c'\in C^i_{t+1}~\wedge~ \Delta(c,c') = w~\}
	\end{align*}
	with
	\begin{align*}
	\Delta(c,c') = \sum_{v~\in~c~\setminus~\{~v~\mid~\phi^i_t(v)~\notin~c'~\}}|\phi^i_t(v)|~,
	\end{align*}
	where $|v|$ denotes the number of tokens in a node $v$ of the sequence graph $S^i_t$ used as input to the clustering at time~$t$.
\end{definition}

That is, the cluster graph $C^i$ contains the clusters resulting from the clusterings of all snapshots as nodes, 
and its weighted edges $(c,c',w)$ indicate how many tokens from a cluster $c'\in C^i_{t+1}$ stem from cluster $c\in C^i_t$.

The cluster graph allows us to identify substantial additions, deletions, and movements of tokens in the United States and Germany over our entire period of study, revealing dynamics at the level of \emph{individual clusters}.
To trace dynamics at the level of \emph{legal topics}, 
we define \emph{cluster families} based on the \emph{family graphs} of our collections:

\begin{definition}[Family Graph $F^i(\gamma)$]\label{def:familygraph}
	Let $C^i$ be the cluster graph of collection $i$ across times $T$.
	The \emph{family graph} of collection $i$ across times $T$, denoted $F^i$, 
	is a weighted digraph 
	\begin{align*}
	F^i(\gamma) = (V^i_C,E^i_F(\gamma))~,
	\end{align*}
	where
	\begin{align*}
	E^i_F(\gamma) = \{~(c,c',w)~\mid~(c,c',w) \in E^i_C~\wedge~\chi(c,c',w) \geq \gamma~\}
	\end{align*}
	with
	\begin{align*}
	\chi(c,c',w)=\min~\Big\{~\frac{w}{|c|}, \frac{w}{|c'|}~\Big\}~,
	\end{align*}
	where $|c|$ denotes the number of tokens in cluster $c$.
\end{definition}

In words, the family graph $F^i(\gamma)$ contains the same nodes as the cluster graph $C^i$ 
but only those edges from $(c,c',w)~\in~E^i_C$ that account for at least a $\gamma$ fraction of the tokens in both $c$ and $c'$. 
We set $\gamma = 0.15$ to filter out noise and isolate parts of the cluster graph that are largely self-contained,
but this threshold can be replaced by any other number between $0$ and $1$ depending for other analyses.

To trace the evolution of legal topics over time, based on the family graph, we define:

\begin{definition}[Cluster Family $V^i_{F,j}$]\label{def:cluster-family}
	Let $F^i(\gamma)$ be the family graph for collection $i$ across times $T$ consisting of cluster families as connected components.
	A \emph{cluster family} $V^i_{F,j}$ is the node set of $F^i(\gamma)$'s $j$\textsuperscript{th} largest connected component (measured in tokens).
\end{definition} 

In addition to the overall size of a cluster family (given by the size of its leading cluster), 
our analysis also uses a temporal notion of cluster family size:

\begin{definition}[Cluster Family Size at Time $t$ $|V^i_{F,j,t}|$]\label{def:cluster-family-size}
	Let $V^i_{F,j}$ be a cluster family $j$ in collection $i$, 
	and let $C^i_t$ be the consensus clustering obtained for time $t$.
	The size of cluster family $j$ at time $t$ is defined as
	\begin{align*}
	|V^i_{F,j,t}| = \sum_{c~\in~(V^i_{F,j}~\cap~C^i_t)}|c|~,
	\end{align*}
	where $|c|$ denotes the number of tokens in a node $c$. 
\end{definition}

With our parametrisation, cluster families are sets of Chapters, Books, or laws that are closely related by cross-references or (almost) textual identity over time.
As such, they approximately correspond to \emph{legal topics}.
Further information on how we label these topics can be found in Section~5.1 of the~\suppi. 

\subsection*{Code availability}

\notereviewtwo{4}
The code used in this study will be made available on GitHub in the following repositories: 
\begin{itemize}[label=--]
	\item Paper: \url{https://github.com/QuantLaw/Complex-Societies-and-Growth}
	\item Data preprocessing: \url{https://github.com/QuantLaw/legal-data-preprocessing}
	\item Clustering: \url{https://github.com/QuantLaw/legal-data-clustering}
\end{itemize}

\subsection*{Data availability}

\notereviewtwo{4}
For the United States, the raw input data used in this study is publicly available from the Annual Historical Archives published by the Office of the Law Revision Counsel of the U.S. House of Representatives,
and is also available from the authors upon reasonable request.

For Germany, the raw input data used in this study was obtained from \emph{juris GmbH} but restrictions apply to the availability of this data, 
which was used under license for the current study, 
and so is not publicly available. 
For details, see Section~1.2 of the SI.

The preprocessed data used in this study (for both the United States and Germany) will be archived with Zenodo.

\bibliography{bibliography}

\section*{Acknowledgements}

The research was supported by a grant from the Interdisciplinary Legal Research Program (ILRP) at Bucerius Law School and benefited from discussions with its director Hans-Bernd Schäfer. 

\section*{Author contributions statement}

All authors conceived of the research project.  
C.C. and J.B. performed the computational analysis in consultation with D.M.K. and D.H.  
All authors drafted the manuscript and it was revised and reviewed by all authors. 
All authors gave final approval for publication and agree to be held accountable for the work performed therein.

\section*{Additional information}

\subsection*{Competing Interests}

D.K. is affiliated with the start-up LexPredict which is now part of Elevate Services.
The other authors declare no competing interests.

\clearpage

\section*{Tables and figures}

\begin{table}[H]
	\centering
	\footnotesize
	\bgroup
	\def\arraystretch{1.5}
	\begin{tabular}{L{0.1\linewidth}R{0.1\linewidth}R{0.1\linewidth}R{0.1\linewidth}R{0.05\linewidth}R{0.1\linewidth}R{0.1\linewidth}R{0.1\linewidth}}
		&\multicolumn{3}{c}{\textbf{United States}}&&\multicolumn{3}{c}{\textbf{Germany}}\\
		&\textbf{1994}&\hfill\textbf{2018}\hfill&$\Delta$&&\hfill\textbf{1994}\hfill&\hfill\textbf{2018}\hfill&$\Delta$\\\hline
		\textbf{Tokens}
		& $14.0$~M
		& $21.2$~M
		& $51~\%$
		&
		& $4.5$~M
		& $7.4$~M
		& $64~\%$
		\\\hline
		\textbf{Structures} 
		&$452.4$~K
		&$828.1$~K
		&$83~\%$
		&
		&$120.6$~K
		&$161.4$~K
		&$34~\%$
		\\\hline
		\textbf{References}
		&$58.0$~K
		&$88.6$~K
		&$53~\%$
		&
		&$76.9$~K
		&$139.1$~K
		&$81~\%$
	\end{tabular}
	\egroup\caption{Federal legislation in the United States and Germany: descriptive statistics ($1994$ and $2018$).}\label{tab:descriptive-statistics}
\end{table}

\clearpage

\newpage

\clearpage

\end{document}


\flushbottom
\maketitle

\thispagestyle{empty}


\section{Data sources}

\subsection{United States}

For the United States, 
we use the United States Code (US Code) as our data source. 
The US Code is a compilation of the general and permanent laws of the United States on the federal level, excluding state legislation.
The Office of the Law Revision Counsel of the U.S. House of Representatives (Office) updates the code continuously and publishes annual versions. 
When Congress passes new legislation, 
this legislation is initially published as a \emph{Slip Law} in the \emph{United States Statutes at Large}. 
If the new legislation is considered general and permanent law, the Office integrates the law into the US Code.
Depending on the Titles of the Code that are modified, 
the work of the Office is approved by Congress, 
whereby the \emph{Slip Law} is repealed and replaced by the US Code as the new primary source of the law. 
In other cases, the US Code is still presumed to be the correct consolidation of the law.

We base our work on the Annual Historical Archives published by the Office, 
which are available on its website:  \url{https://uscode.house.gov/download/annualhistoricalarchives/annualhistoricalarchives.htm}.\newline
The US Code is provided in (X)HTML format as documented at \url{https://uscode.house.gov/download/resources/USLM-User-Guide.pdf}. 
The format is flexible and offers a wide variety of styles to closely represent the printed code. 
The code is split into single files per year and Title.
The Annual Historical Archives date back until 1994, 
and they are published with delay. 
As of 2020, therefore, 2018 is the latest available edition/supplement.
 
While surveying and validating the raw data, we observe and correct the following obvious errors:
\begin{itemize}[itemsep=0em, parsep=0em, topsep=0em,label=--]
\item double-closing \texttt{<div>}-Tags in Title 40 in 2008, and in Title 42 in 1994 and 1995,
\item inconsistent metadata in the appendix of Title 28 in 2017,
\item a duplicate of Title 12 in 1998 that is included in files regarding Title 11 and 12, and
\item inconsistencies regarding the tags \texttt{<statute>} and the comments \texttt{<!-- \/section-head -->}. 
\end{itemize}

\subsection{Germany}

In Germany,
all laws are published in the Federal Law Gazette as amending laws,
which often combine numerous introductions of new as well as changes to and repeals of existing laws.
Individual laws are officially classified into one of nine substantive categories with currently 73 sub-categories containing over 400 subject areas in the ``Fundstellennachweis A'' [engl. Finding Aids A].
These are published on a yearly basis by the Federal Law Gazette upon instruction by the Ministry of Justice.
However, unlike the US Code, there is no official data source that provides all compiled general and permanent laws at the federal level and
their historical versions.
Therefore, we cooperate with the leading German legal data provider, \emph{juris GmbH}, 
to obtain a dataset similar to the annual versions of the US Code. 
Although \emph{juris} is a private company, 
the Federal Republic of Germany is its majority shareholder, 
and all branches of government rely heavily on \emph{juris} to process legislative data. 
According to \emph{juris}, the database contains every German federal law since spring 1990. 
Compared to the US Code, the data is not as structured. 
Instead of providing annual consolidated versions, \emph{juris} provides a new version of a law for all changes that take effect on the same day.
The data we obtained comes in separate files for each law and version.

We may not share the text content of the German dataset together with this paper.
However, a website maintained by the German Federal Ministry of Justice and Consumer Protection in collaboration with \emph{juris} (\url{https://www.gesetze-im-internet.de}) 
provides almost the entire federal legislation as XML files in the most recent version (i.e., without historical versions).
A daily archive of the XML files provided on this website (starting in June 2019) is available at \url{https://github.com/QuantLaw/gesetze-im-internet}.
This dataset allows a partial reproduction of the research with a similar dataset.
We requested the full dataset from \emph{juris GmbH}, 
which required a dedicated contractual framework and non-disclosure agreement to be signed.

\subsection{Samples of Legal Texts}
\label{subsec:textsamples}

\notereviewone{3}
The following fragments illustrate the inherent structure of legal texts. 
Inclusion relationships are marked by indentation, 
labels of \emph{seqitems} are typeset in \textsc{small capitals}, 
cross-references are \underline{underlined},
and text content is set in \emph{Italics}.\\

\noindent\hspace*{0.04\textwidth}\fbox{%
	\parbox{0.9\textwidth}{%
		\textbf{United States Code (2018 Main Edition dated January 14, 2019)}
		
		\hspace*{1em}Title 12—Banks and Banking
		
		\hspace*{2em}Chapter 42—Low-Income Housing Preservation and Resident Homeownership
		
		\hspace*{3em}Subchapter I—Prepayment of Mortgages Insured under National Housing Act
		
		\hspace*{4em}\textsc{§4101. General prepayment limitation}
		
		\hspace*{5em}(a) Prepayment and termination
		
		\noindent\hangindent6em\hspace{6em}\emph{An owner of eligible low-income housing may prepay, and a mortgagee may accept prepayment of, a mortgage on such housing only in accordance with a plan of action approved by the Secretary under this subchapter or in accordance with \underline{section 4114 of this title}. An insurance contract with respect to eligible low-income housing may be terminated pursuant to \underline{section 1715t of this title} only in accordance with a plan of action approved by the Secretary under this subchapter or in accordance with \underline{section 4114 of this title}.}
		
		\hspace*{5em}\dots
		
		\hspace*{4em}\dots
		
		\hspace*{4em}\textsc{§4105. Federal cost limits and limitations on plans of action}
		
		\hspace*{5em}(a) Determination of relationship to Federal cost limits
		
		\hspace*{6em}(1) Initial determination
		
		\noindent\hangindent7em\hspace{7em}\emph{For each eligible low-income housing project appraised under \underline{section 4103(a) of this title}, the Secretary shall determine whether the aggregate preservation rents for the project determined under \underline{paragraph (1) or (2) of section 4104(b) of this title} exceed the amount determined by multiplying 120 percent of the fair market rental (established under \underline{section 1437f(c) of title 42}) for the market area in which the housing is located by the number of dwelling units in the project (according to appropriate unit sizes).}
	}
}

\noindent\hspace*{0.04\textwidth}\fbox{%
	\parbox{0.9\textwidth}{%
		\textbf{German Civil Code (official translated version dated October 1, 2013)}
		
		\hspace{1em}Book 1---General Part
		
		\hspace{2em}Division 1---Persons
		
		\hspace{3em}Title 1---Natural persons, consumers, entrepreneurs
		
		\hspace{4em}\textsc{Section 1---Beginning of legal capacity}
		
		\hspace{5em}\emph{The legal capacity of a human being begins on the completion of birth.}
		
		\hspace{4em}\textsc{Section 2---Beginning of majority}
		
		\hspace{5em}\emph{Majority begins at the age of eighteen.}
		
		\hspace{4em}\dots
		
		\hspace{3em}Title 2---Legal persons
		
		\hspace{4em}Subtitle 1---Associations

		\hspace{5em}Chapter 1---General provisions
		
		\hspace{6em}\dots
		
		\hspace{6em}\textsc{Section 40 ---Flexible provisions}
		
		\noindent\hangindent7em\hspace{7em}\emph{The provisions of \underline{section 26 (2) sentence 1}, \underline{section 27 (1) and (3)}, \underline{sections 28 and 31a (1) sentence 2}, as well as \underline{sections 32, 33 and 38}, do not apply where otherwise provided by the articles of association. It is not possible to derogate from \underline{section 34} through the articles of association, even for the passing of resolutions by the board.}
	}
}

\section{Data preprocessing}

We convert the source data into a structured format that facilitates access for our purposes, 
removes unnecessary details (especially most of the style information for the text),
and unifies the data format across both countries. 
For the purposes of our analysis, 
we focus on three properties that characterise the structure of legislative texts (illustrated in Section~\ref{subsec:textsamples} above):

\begin{enumerate}
	\item \emph{Hierarchy:} They are nested, e.g., into Titles, Sections, Subsections, Paragraphs, and Subparagraphs.
	\item \emph{Reference:} They are interlinked, e.g., one Section can reference (the label of) another Section in its text.
	\item \emph{Sequence:} They are ordered, e.g., Sections have unique labels, and they appear in the text in ascending order of their labels.
\end{enumerate}

To make these properties easily accessible in our data, 
we perform the following preprocessing steps for each dataset:

\subsection{Clean the text}

First, we remove all formatting, annotations, notes, and metadata from the text, 
with the exception of formatting and metadata that we need to extract the hierarchy in the next step. 
For the United States, this results in a more conservative definition of legal text than in \cite{bommarito2010a}, which explains the difference between the reported token counts.
As the German data is not consistently formatted on a more detailed level than text paragraphs (meaning one level below § or Articles), we do not preserve formatting below the paragraph level for this dataset.

\subsection{Extract the hierarchies}

Using the remaining text, formatting, and metadata, 
we extract the hierarchy, i.e., 
we identify boundaries of elements (Chapters, Parts, Sections, etc.) that structure the code, 
determine their parents (Titles, Chapters, Parts, etc.) and, 
if present, their headings (including the alphanumeric ordering identifier), 
and retrieve their children or textual content. 
With this information, 
we generate an XML representation of the code 
in which the text and structural elements are nested inside their respective parents. 

Our data contains explicit information regarding the boundaries and nesting of structural elements above the Section level in its metadata. 
We rely on this metadata down to the Paragraph level in Germany 
and down to the Section level for the United States.
Below the Section level in the United States, we exploit the formatting to derive a nesting of the text. 
Extracting the hierarchy information from the metadata yields all information required to build our hierarchy graphs.

\subsection{Extract the cross-references}

Next, we extract all explicit cross-references in the statutory texts that match a common citation format. 
To simplify the process, we perform the extraction in three steps:

\begin{enumerate}

\item \emph{Find}: We identify parts of the text that contain a potential reference to another text in the same statutory collection. 
We define country specific regular expressions (\emph{regex}) patterns to find the referencing parts. 

Our pattern for the US Code is rather simple, 
as references are mostly formatted consistently and include no headings or names but only numbers (and potentially letters of alphanumeric enumerations). 
The pattern for Germany is more sophisticated. 
The start of a reference is easy to identify as references normally begin with ``§'' or ``Art.''. 
The part of the reference that follows may contain numbers (and letters of alphanumeric enumerations) as well as units (e.g., Satz [engl. sentence], Nummer [engl. number]). 
In the case of a reference to a text in a different law, the reference is followed by the name of the law. 
A list of the law names is generated from the source data, 
but it includes only laws valid at the time the analysed law takes effect. 
Furthermore, 
references to other national regulations, EU legislation, etc., are filtered out,
so that only references within a law and to other laws that are part of the collection remain. 
Detailed documentation regarding the citation format in German laws can be found in the ``Handbuch der Rechtsförmlichkeit'' [engl. Handbook of Legal Formalism] of the Federal Ministry of Justice and Consumer Protection: \url{https://www.bmjv.de/DE/Themen/RechtssetzungBuerokratieabbau/HDR/HDR_node.html}.
Since this guide is not strictly followed by the legislator, 
we used this guide along with the actual data to develop our extraction method.

\item \emph{Parse}: We parse the referencing texts and derive citation keys (\emph{cite keys}) that, for the US Code, consist of a Title and a Section of the referenced text.
In the German case, the keys are composed of the abbreviation of the referenced law and the number of the cited § or Article.
One reference identified in the first step may contain several such citation keys.

\item \emph{Align}: We identify the target structural elements of the parsed references.
To accomplish this, we generate citation keys for each Section, 
§, or Article that can be referenced by a specific version of a law.
In the United States, a Section can be referenced if its structural element is part of the same annual version of the US Code.
In Germany, the structural element must be part of a valid law when the analysed law takes effect.
\end{enumerate}

\subsection{Generate XML files}

We store each preprocessed Title in the United States and each law in Germany in a separate XML file. 
The XML files comply with the following XSD specification:

\lstset{basicstyle=\small}
\begin{lstlisting}
<xs:schema
  attributeFormDefault="unqualified"
  elementFormDefault="qualified"
  xmlns:xs="http://www.w3.org/2001/XMLSchema"
>
  <xs:element
    name="document"
    type="documentType"
  />
  <xs:complexType name="documentType">
    <xs:choice
      maxOccurs="unbounded"
      minOccurs="0"
    >
      <xs:element
        type="itemType"
        name="item"
      />
      <xs:element
        type="seqitemType"
        name="seqitem"
      />
    </xs:choice>
    <xs:attribute
      type="xs:string"
      name="heading"
      use="required"
    />
    <xs:attribute
      type="xs:string"
      name="key"
      use="required"
    />
    <xs:attribute
      type="xs:unsignedInt"
      name="level"
      use="required"
    />
    <xs:attribute
      type="xs:string"
      name="abbr_1"
      use="optional"
    />
    <xs:attribute
      type="xs:string"
      name="abbr_2"
      use="optional"
    />
  </xs:complexType>
  <xs:complexType name="itemType">
    <xs:choice
      maxOccurs="unbounded"
      minOccurs="0"
    >
      <xs:element
        type="itemType"
        name="item"
      />
      <xs:element
        type="seqitemType"
        name="seqitem"
      />
    </xs:choice>
    <xs:attribute
      type="xs:string"
      name="heading"
      use="required"
    />
    <xs:attribute
      type="xs:string"
      name="key"
      use="required"
    />
    <xs:attribute
      type="xs:unsignedInt"
      name="level"
      use="required"
    />
  </xs:complexType>
  <xs:complexType name="seqitemType">
    <xs:sequence>
      <xs:element
        type="subseqitemType"
        name="subseqitem"
        maxOccurs="unbounded"
        minOccurs="0"
      />
      <xs:element
        type="textType"
        name="text"
        maxOccurs="unbounded"
        minOccurs="0"
      />
    </xs:sequence>
    <xs:attribute
      type="xs:string"
      name="citekey"
      use="required"
    />
    <xs:attribute
      type="xs:string"
      name="heading"
      use="required"
    />
    <xs:attribute
      type="xs:string"
      name="key"
      use="required"
    />
    <xs:attribute
      type="xs:unsignedInt"
      name="level"
      use="required"
    />
  </xs:complexType>
  <xs:complexType name="subseqitemType">
    <xs:sequence>
      <xs:element
        type="textType"
        name="text"
      />
    </xs:sequence>
    <xs:attribute
      type="xs:string"
      name="key"
      use="required"
    />
    <xs:attribute
      type="xs:byte"
      name="level"
      use="required"
    />
  </xs:complexType>
  <xs:complexType
    name="textType"
    mixed="true"
  >
    <xs:sequence>
      <xs:element
        type="referenceType"
        name="reference"
        maxOccurs="unbounded"
        minOccurs="0"
      />
    </xs:sequence>
  </xs:complexType>
  <xs:complexType name="referenceType">
    <xs:sequence>
      <xs:element
        type="xs:string"
        name="main"
      />
      <xs:element
        type="xs:string"
        name="suffix"
      />
      <xs:element name="lawname">
        <xs:complexType>
          <xs:simpleContent>
            <xs:extension base="xs:string">
              <xs:attribute
                type="xs:string"
                name="type"
                use="optional"
              />
            </xs:extension>
          </xs:simpleContent>
        </xs:complexType>
      </xs:element>
    </xs:sequence>
    <xs:attribute
      type="xs:string"
      name="parsed"
      use="optional"
    />
    <xs:attribute
      type="xs:string"
      name="parsed_verbose"
      use="optional"
    />
    <xs:attribute
      type="xs:string"
      name="pattern"
      use="optional"
    />
  </xs:complexType>
  <xs:complexType name="lawnameType">
    <xs:simpleContent>
      <xs:extension base="xs:string">
        <xs:attribute
          type="xs:string"
          name="type"
          use="required"
        />
      </xs:extension>
    </xs:simpleContent>
  </xs:complexType>
</xs:schema>
\end{lstlisting}

\subsection{Generate graphs}

We use the XML files along with information regarding the annual version the file belongs to (in the United States) 
or the validity period of a specific version of a law (in Germany)
to generate our hierarchy graphs and our reference graphs. 
We produce these graphs for each annual version of the US Code, 
and for snapshots of the German data taken at the first day of each year from 1994 to 2018.
For the German data, we can generally produce hierarchy graphs and reference graphs for each day in the period under study; 
the chosen dates are designed to match the United States data.
The sequence graphs and the quotient graphs are derived from the reference graphs as simple transformations.
All our graphs are generated using NetworkX (\url{https://networkx.github.io}) and can be exported, e.g.,
as GraphML files (\url{http://graphml.graphdrawing.org/specification.html}). 

\section{Figure generation}

\subsection{Main paper Figure~3} 

\begin{figure}[H]
	\vspace*{-16pt}\centering
	\begin{subfigure}{0.9\linewidth}
		\includegraphics[width=\linewidth]{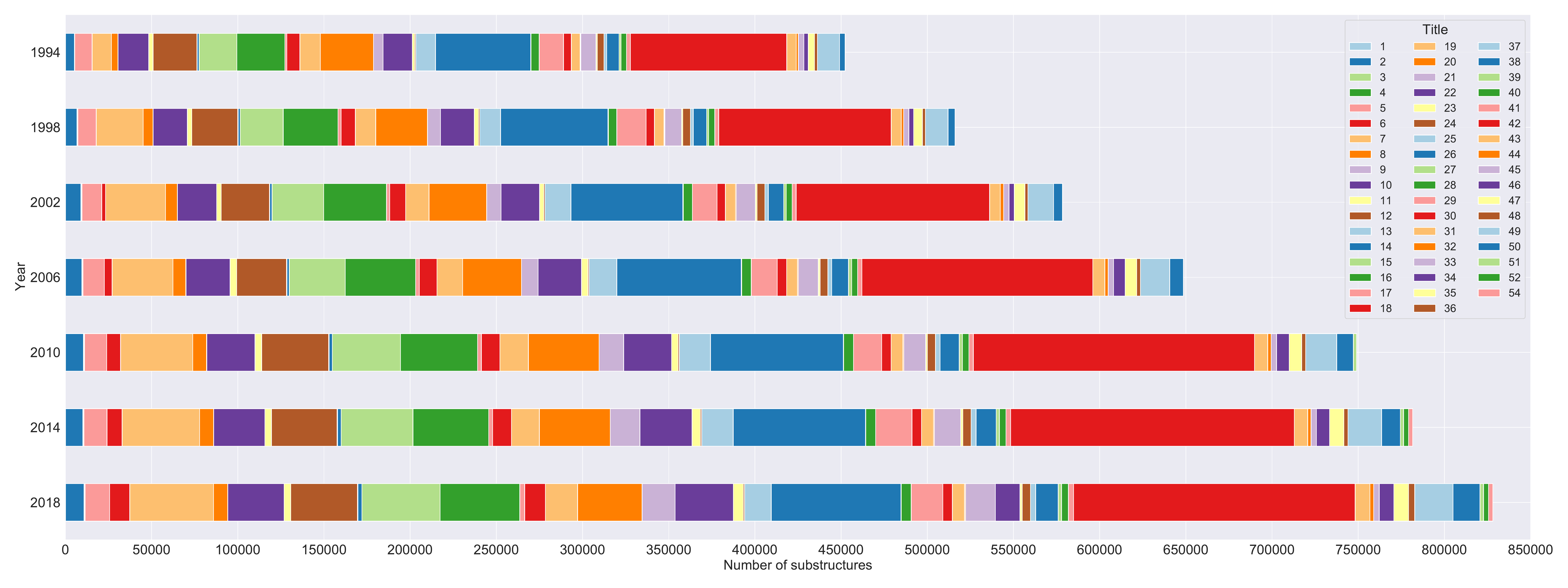}
	\end{subfigure}
	\begin{subfigure}{0.9\linewidth}
		\includegraphics[width=\linewidth]{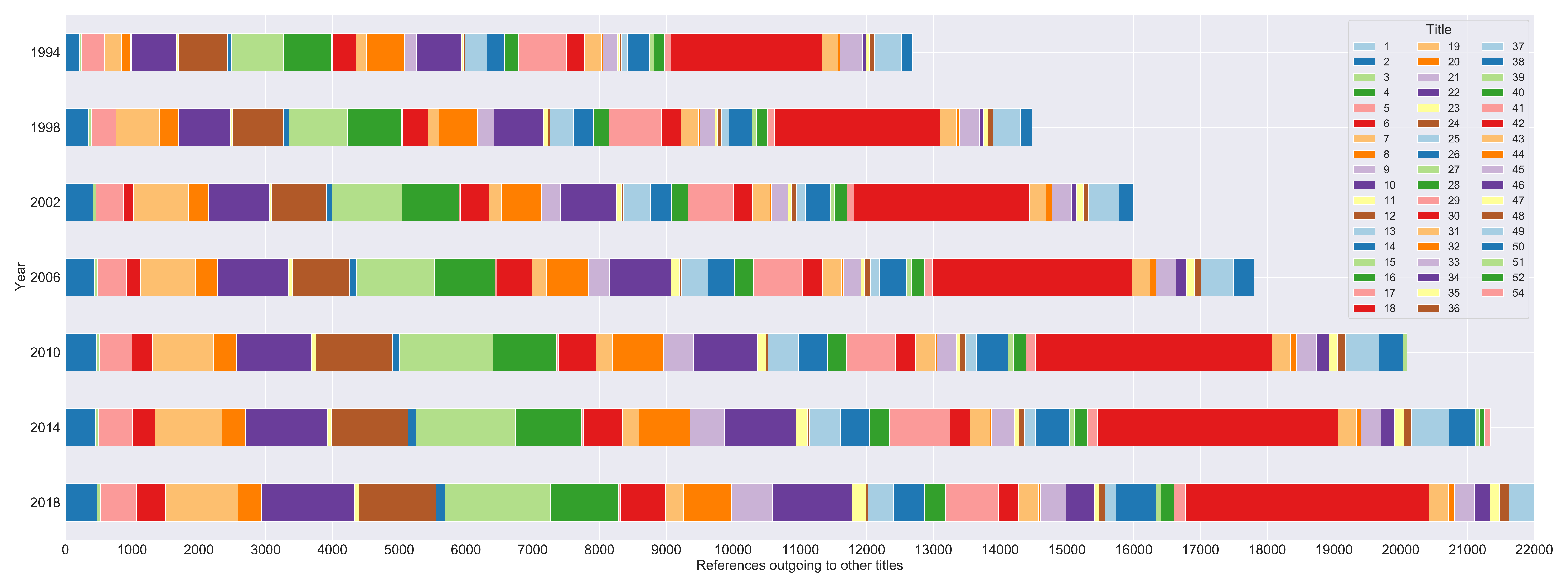}
	\end{subfigure}
	\begin{subfigure}{0.9\linewidth}
		\includegraphics[width=\linewidth]{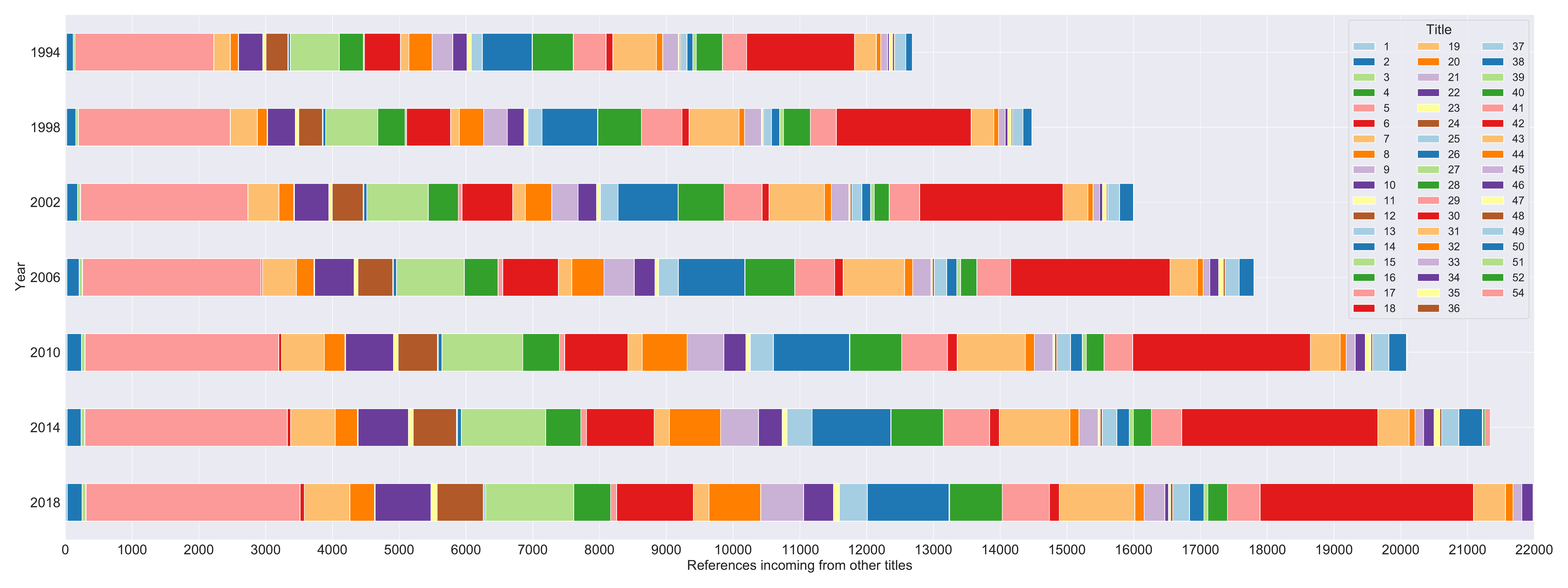}
	\end{subfigure}
	\begin{subfigure}{0.9\linewidth}
		\includegraphics[width=\linewidth]{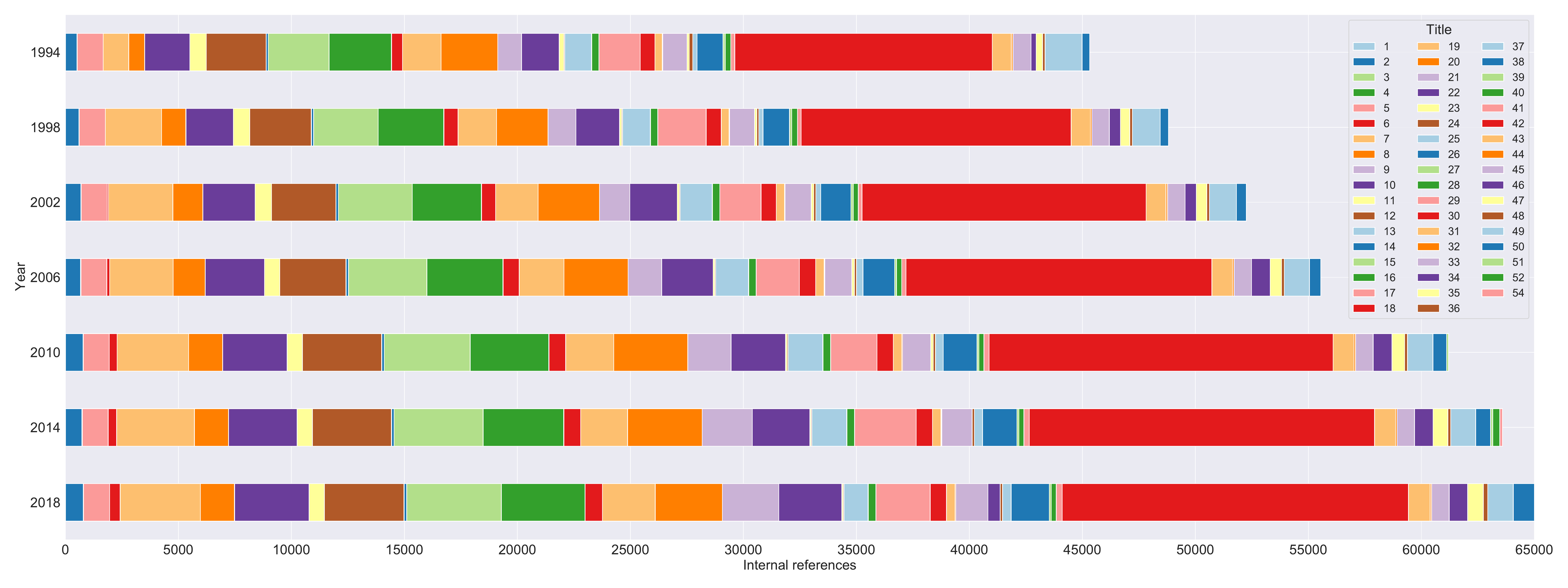}
	\end{subfigure}
	\caption{Federal legislation in the United States by Title (1994--2018), measured in structural units.}
	\label{fig:us-other-per-title} 
\end{figure}

Figure~3 from the main paper 
visualises the size of individual Titles of the US Code for every fourth year, 
starting from 1994. 
Title~53, which is missing in the legend, has been empty throughout the period under study.
In the main paper, 
we illustrate the size of Titles based on tokens.
In Figure~\ref{fig:us-other-per-title}, 
we show the size of Titles based on other measures, 
namely, structural elements, outgoing references to other Titles, ingoing references from other Titles, and internal references.

In all of the graphics, $12$ colours rotate to mark the different Titles.
Since colours are reused, the position of a bar must be taken into account when reading the legend.
The Titles are plotted in a horizontally stacked bar chart, 
following their original order from left to right, i.e., starting with Title~1 on the very left.

\subsection{Main paper Figure~4} 

Figure~4 from the main paper 
shows graphs representing the Chapters of the US Code or German laws as nodes.
If a German law contains Bücher (engl. Books) at the highest hierarchy level, 
we split the law into its Books and use the Books as nodes.
Arrows between nodes indicate that the text of one node cites the text of another.
The opacity of an arrow indicates the number of references that are summarised in this arrow.
These nodes and weighted edges visualise the major part of the data we run the \emph{Infomap} clustering algorithm on.
However, here, we hide nodes containing less than $5000$ tokens (which are included for the clustering).

In the figure, 
we colour nodes by their cluster family, and colours are comparable across years but not across countries. 
We encode the number of tokens contained in a node by its area,
which is comparable between the four graphs in Figure~4 from the main paper. 
The opacity of edges is individually scaled for each graph to cover the full range of opacity. 
The minimal and maximal values can be found in Table~\ref{tab:us-de-quotient-min-max}.

\newpage

\begin{table}[H]
	\centering
	\begin{tabular}{lllrrrr}
\toprule
\multicolumn{1}{p{20mm}}{\centering Graph\\type} & \multicolumn{1}{p{20mm}}{\centering Country\\code} & \multicolumn{1}{p{20mm}}{\centering Snapshot} &  \multicolumn{1}{p{20mm}}{\centering Edge\\weight\\max.} &  \multicolumn{1}{p{20mm}}{\centering Edge\\weight\\min.} &  \multicolumn{1}{p{20mm}}{\centering Node\\size\\max.} &  \multicolumn{1}{p{20mm}}{\centering Node\\size\\min.} \\
\midrule
                                         Chapter &                                                 US &                                          1994 &                                                182 &                                                  1 &                                             986148 &                                               5010 \\
                                         Chapter &                                                 US &                                          2018 &                                                343 &                                                  1 &                                            1155704 &                                               5002 \\
                                         Chapter &                                                 DE &                                          1994 &                                                153 &                                                  1 &                                              73548 &                                               5017 \\
                                         Chapter &                                                 DE &                                          2018 &                                                660 &                                                  1 &                                             182847 &                                               5047 \\
                                       Community &                                                 US &                                          1994 &                                                128 &                                                  1 &                                            7530609 &                                             100790 \\
                                       Community &                                                 US &                                          2018 &                                                356 &                                                  1 &                                            8850024 &                                             156513 \\
                                       Community &                                                 DE &                                          1994 &                                                179 &                                                  1 &                                            2497714 &                                             100332 \\
                                       Community &                                                 DE &                                          2018 &                                                393 &                                                  1 &                                            3943949 &                                             152300 \\
\bottomrule
\end{tabular}

	\caption{Minimal and maximal values of the raw data for each graph in Figure~4 from the main paper 
		and Figure~\ref{fig:us-de-cluster-quotient}.
		The opacity of arrows is scaled based on the edge weight extrema.
	}
	\label{tab:us-de-quotient-min-max}
\end{table}

To position the nodes, we use the Fruchterman-Reingold force-directed layout algorithm as implemented in \texttt{NetworkX} with an optimal distance between nodes of $k=2.2$ and a seed of $1234$ feeded by \texttt{numpy}.

In a similar spirit, Figure~\ref{fig:us-de-cluster-quotient} visualises the result of the clustering.
In general, the graphic is generated like Figure~4 from the main paper, 
but now, each node represents a cluster.
For the 1994 graphs, 
nodes smaller than $100000$ tokens are hidden, 
whereas for the 2018 graphs, 
we hide nodes smaller than $150000$ tokens.
Moreover, the node size scaling is $40$ times smaller than in Figure~4 from the main paper.

\subsection{Main paper Figure~5} 

Figure~5 from the main paper 
provides an overview of how the US Code developed over the last 25 years in an alluvial plot.
It is based on the family graph $F^i$ for $i=\text{US}$.

We limit the number of clusters per year to 50 and condense smaller clusters into one additional cluster per year (the \emph{miscellaneous} cluster),
to focus on large and medium size clusters.
Moreover, we combine multiple edges between clusters summarised in the miscellaneous cluster into one.

\begin{figure}[H]
	\centering
	\begin{subfigure}{0.5\linewidth}
		\includegraphics[width=\linewidth]{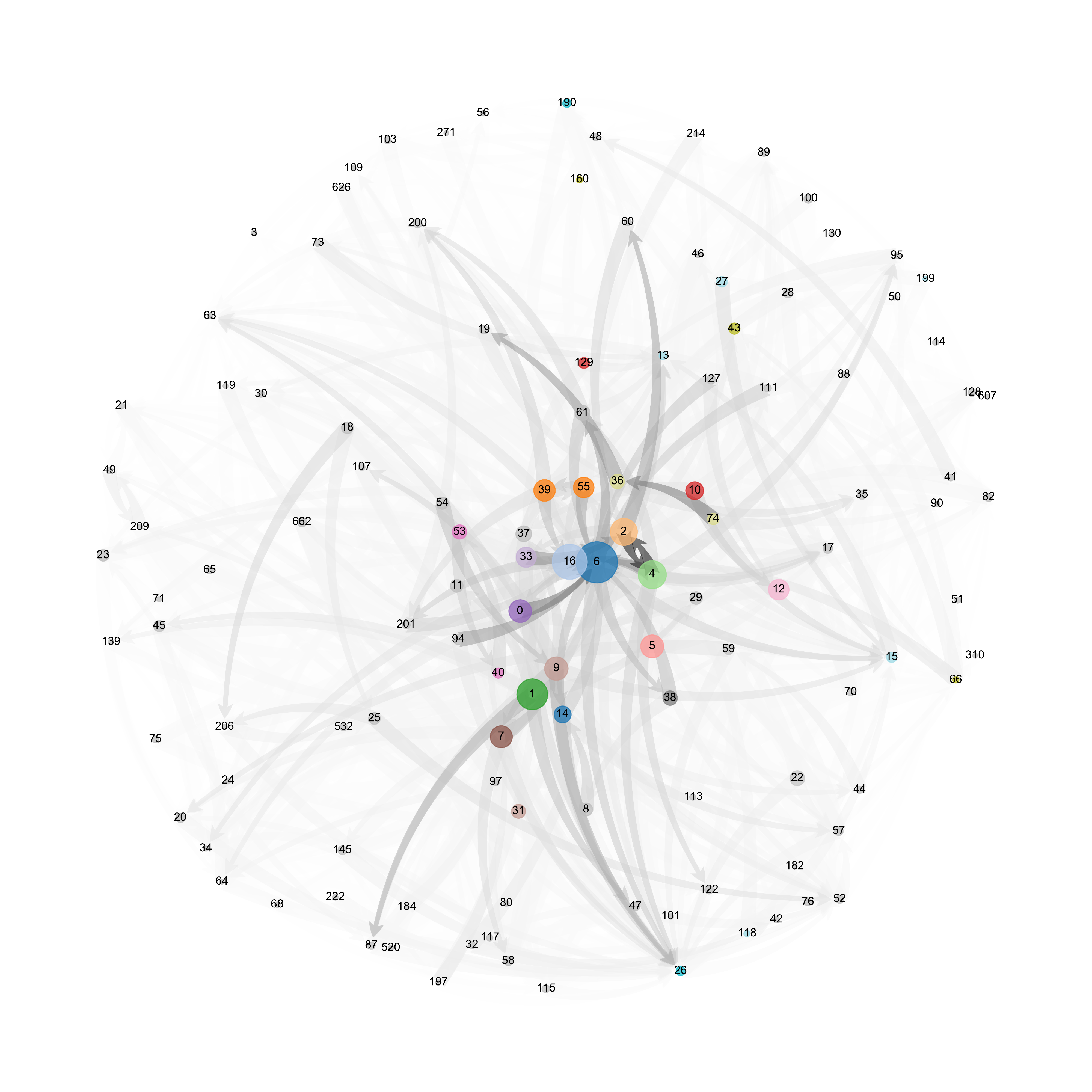}~%
		\subcaption{United States (1994)}
	\end{subfigure}~%
	\begin{subfigure}{0.5\linewidth}
		\includegraphics[width=\linewidth]{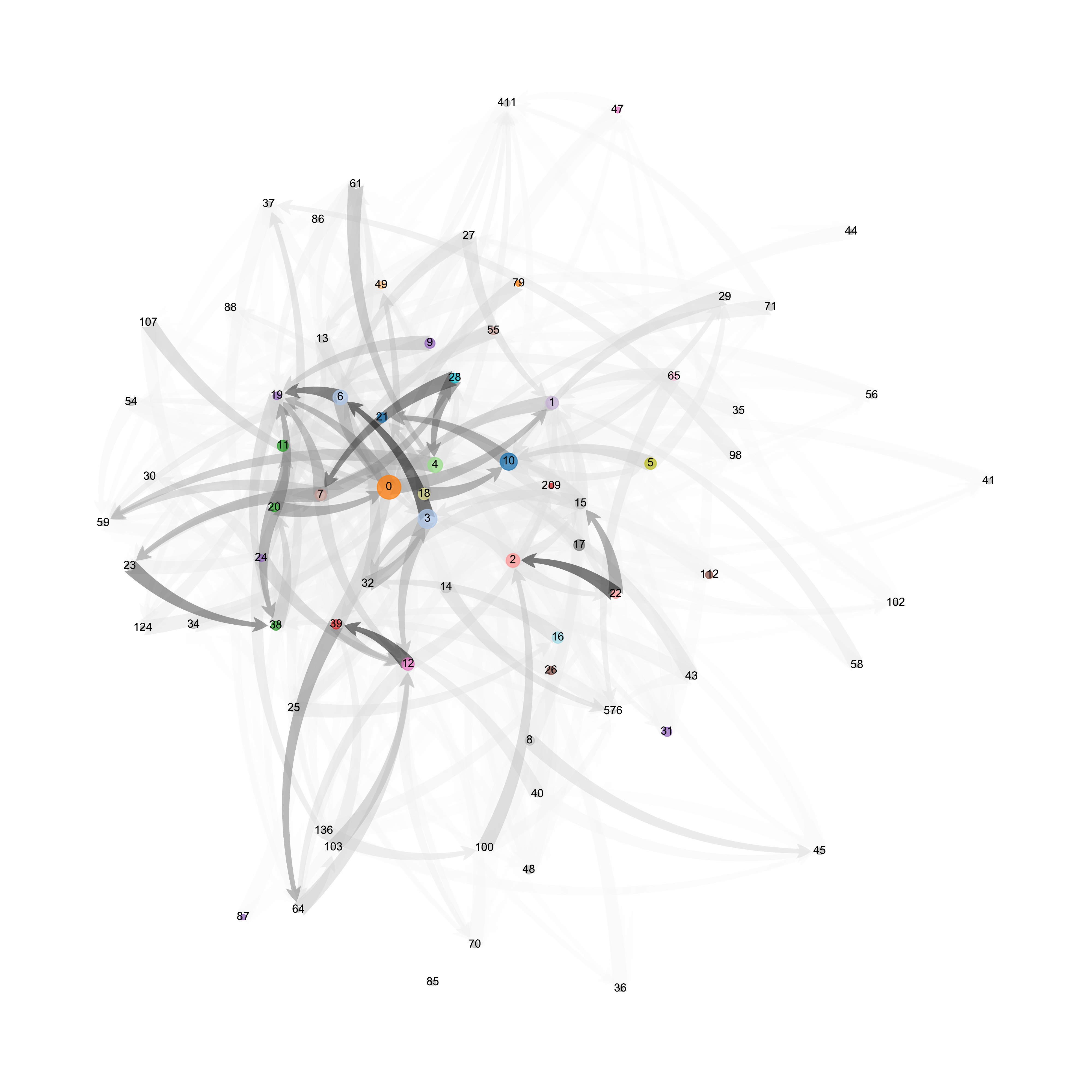}~%
		\subcaption{Germany (1994)}
	\end{subfigure}
	\begin{subfigure}{0.5\linewidth}
		\includegraphics[width=\linewidth]{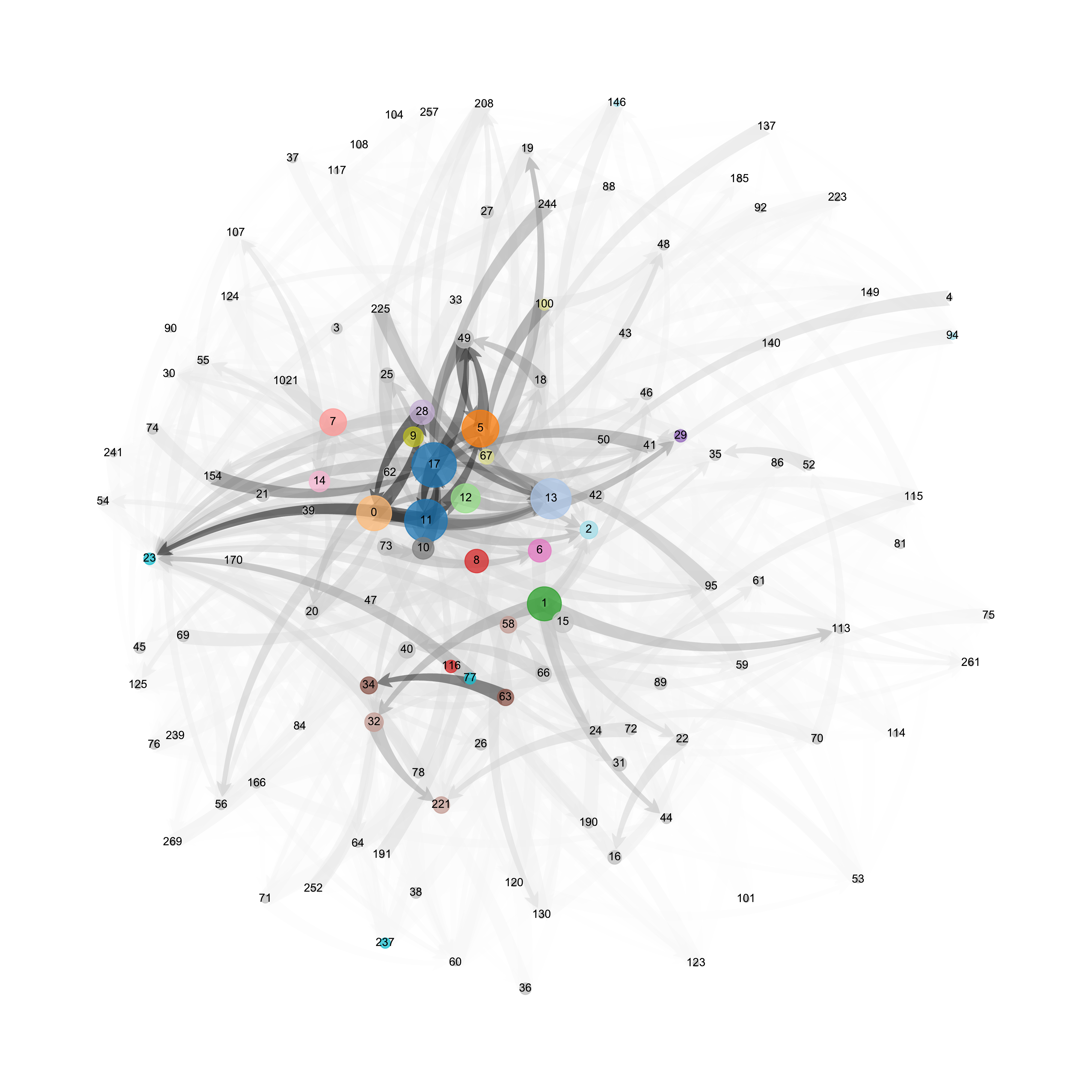}~%
		\subcaption{United States (2018)}
	\end{subfigure}~%
	\begin{subfigure}{0.5\linewidth}
		\includegraphics[width=\linewidth]{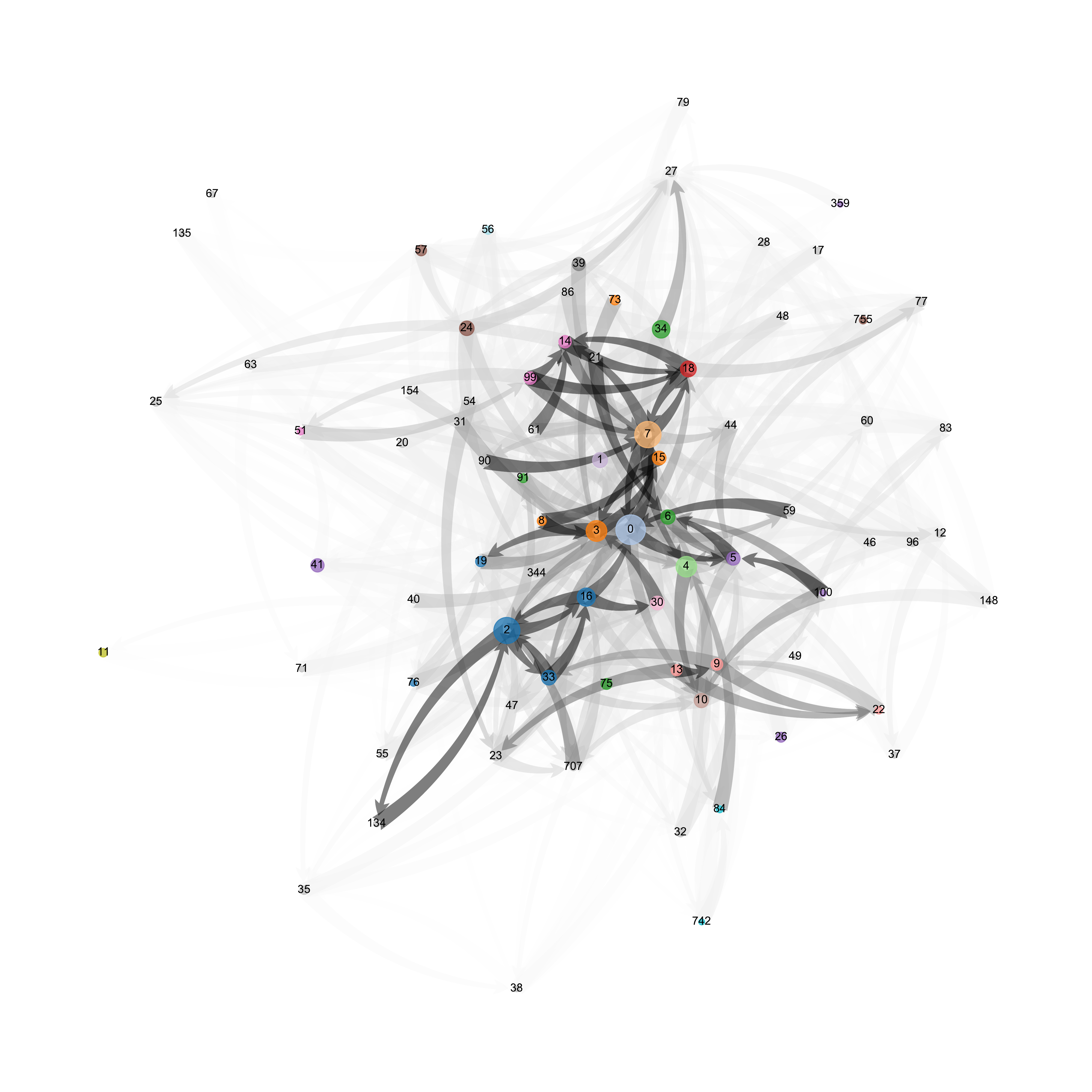}~%
		\subcaption{Germany (2018)}
	\end{subfigure}
	\captionsetup{justification=raggedright} 
	\caption{Federal legislation in the United States and Germany: quotient graphs by cluster (1994 and 2018),
		with arrows running between nodes indicating that text contained in one node cites text contained in another node.
		Node sizes indicate token counts (larger $=$ more tokens), where only nodes with at least $100000$ tokens for 1994 and $150000$ tokens for 2018
		are shown. 
		Arrow colours indicate numbers of outgoing references (darker $=$ more references). 
		For each nation separately, nodes share the same colour if they are placed in the same cluster family,
		and nodes not in one of the $20$ largest cluster families are coloured in grey.
	}
	\label{fig:us-de-cluster-quotient}
\end{figure}

The clusters are ordered vertically by year.
Horizontally, we sort them from left ot right by decreasing size, 
then force the miscellaneous cluster to the right as it condenses the smallest clusters.
The clusters for one year are visualised as horizontally stacked bars.
The height is fixed and the width indicates the size of the respective cluster in tokens.
The horizontal width of edges indicates the weight of an edge in tokens.
The scale mapping tokens to width is identical for nodes and edges in one alluvial plot but it differs across countries
because the year with the most tokens in each collection is scaled to the same width.
Clusters are labelled by numbers that represent the order in which our consensus clustering implementation reported the clustering results.
They match the numbering in Figure~\ref{fig:us-de-cluster-quotient} and should be interpreted on a nominal scale.

\begin{figure}[H]
	\centering
	\vspace*{-16pt}\includegraphics[width=0.9\textwidth]{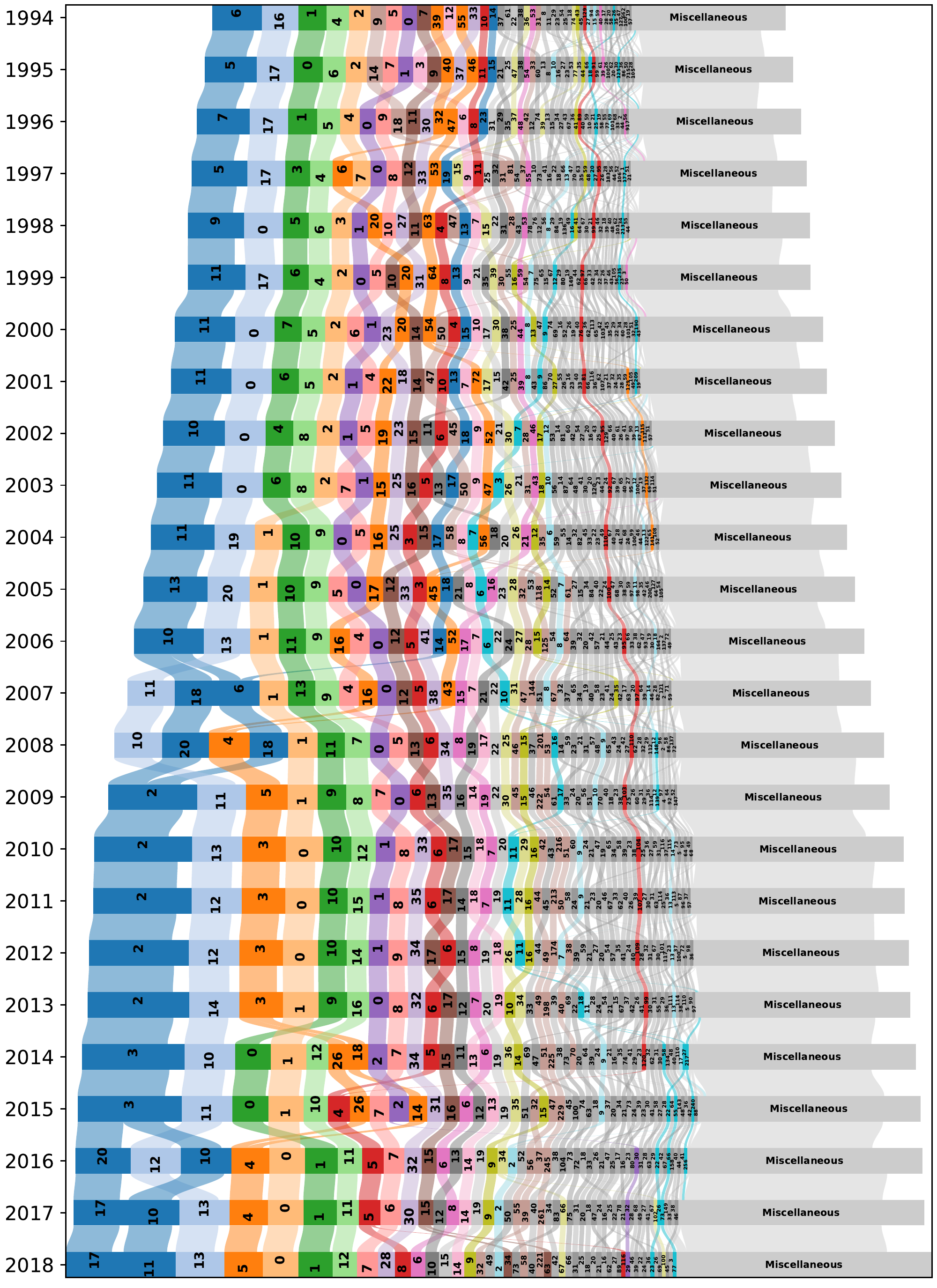}
	\caption{%
		Federal legislation in the United States by cluster (1994--2018), 
		with cluster numbers to enable content inspection (cf. Section~\ref{sec:labelling}). 
		Each block in each year represents a cluster. 
		Clusters are ordered from left to right by decreasing size (measured in tokens) and coloured by the cluster family to which they belong, 
		where clusters not in one of the $20$ largest cluster families are coloured in alternating greys. 
		Small clusters are summarised in one miscellaneous cluster, which is always the rightmost cluster and coloured in light grey.
	}
	\label{fig:sankey-us-labels}
\end{figure}

\begin{figure}[H]
	\centering
	\vspace*{-16pt}\includegraphics[width=0.9\textwidth]{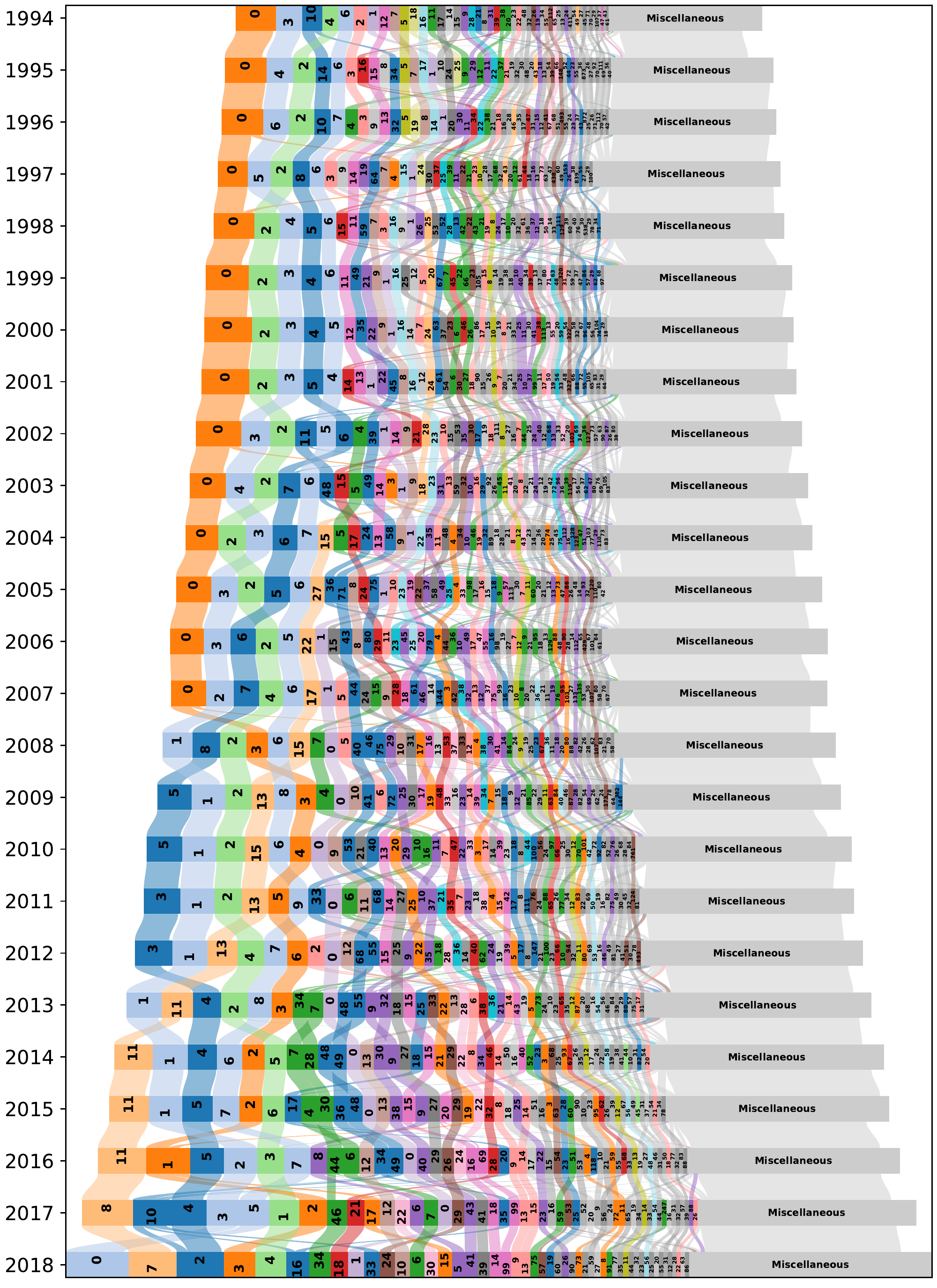}
		\caption{%
		Federal legislation in Germany by cluster (1994--2018), 
		with cluster numbers to enable content inspection (cf. Section~\ref{sec:labelling}). 
		Each block in each year represents a cluster. 
		Clusters are ordered from left to right by decreasing size (measured in tokens) and coloured by the cluster family to which they belong, 
		where clusters not in one of the $20$ largest cluster families are coloured in alternating greys. 
		Small clusters are summarised in one miscellaneous cluster, which is always the rightmost cluster and coloured in light grey.
	}
	\label{fig:sankey-de-labels}
\end{figure}

The cluster identifiers are hidden in Figure~5 from the main paper. Figure~\ref{fig:sankey-us-labels} is a copy of this figure including identifiers for detailed inspection, 
and Figure~\ref{fig:sankey-de-labels} is its analogue for Germany.
HTML files in the data repository accompanying this paper describe the content of the clusters. 
They show the absolute size in tokens of each cluster and its size relative to the whole dataset for one year and country. 
The summarised elements (Chapters or laws and Books) of each cluster are listed by their path (e.g., TITLE 42-THE PUBLIC HEALTH AND WELFARE / CHAPTER 6-THE CHILDREN'S BUREAU) and their size relative to the whole cluster.

Nodes and edges of the 20 largest cluster families are coloured by their relationship. 
Other nodes and edges are coloured in alternating greys. 
The miscellaneous node is coloured in light grey.
Edges are plotted with opacity $0.5$ and in increasing order of their weight, 
with the largest edges plotted last.

\subsection{Main paper Figure~6} 

\begin{figure}[H]
	\centering
	\begin{subfigure}{0.5\linewidth}
		\includegraphics[width=\linewidth]{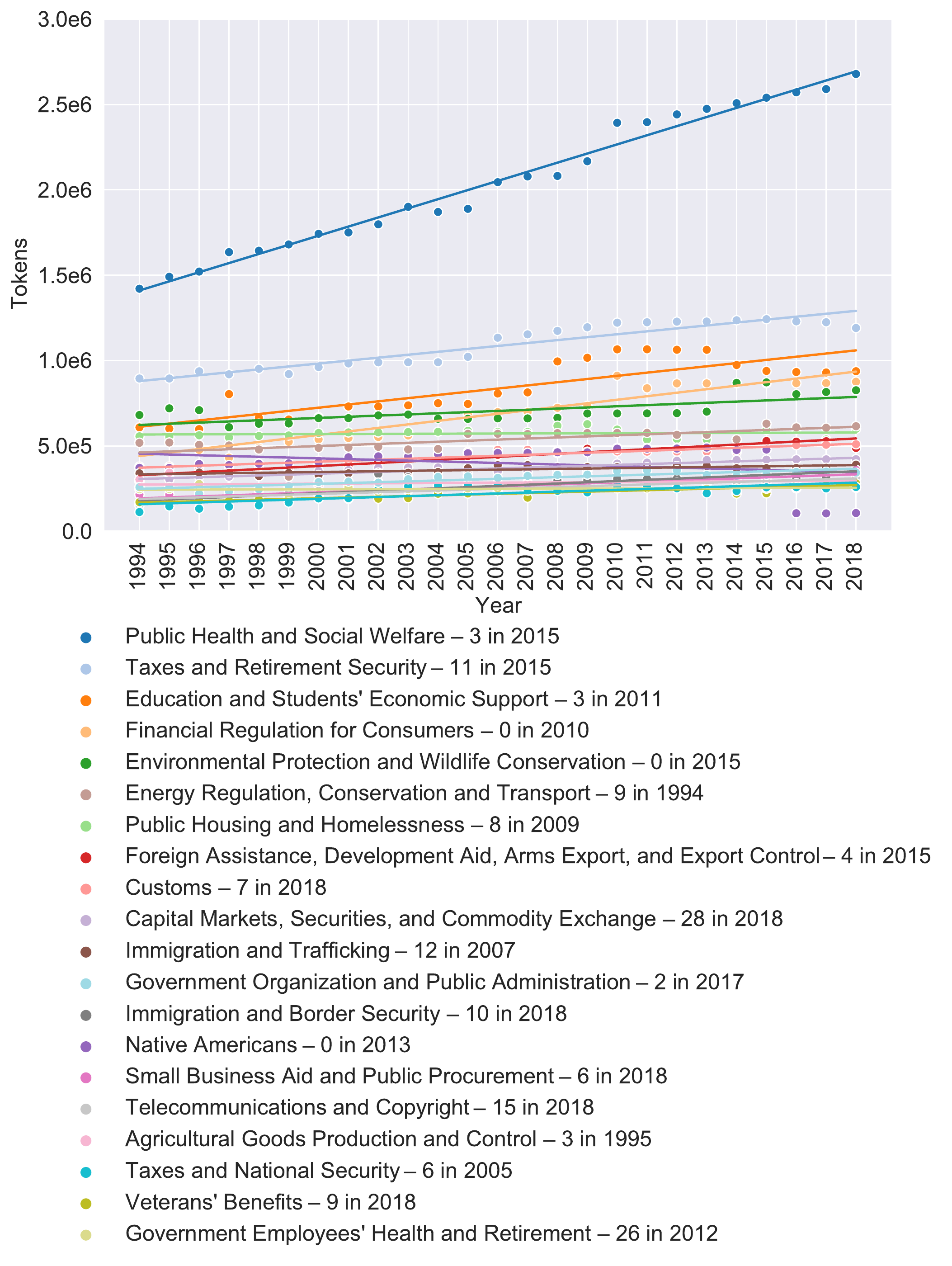}~%
		\subcaption{United States}
	\end{subfigure}~%
	\begin{subfigure}{0.5\linewidth}
		\includegraphics[width=\linewidth]{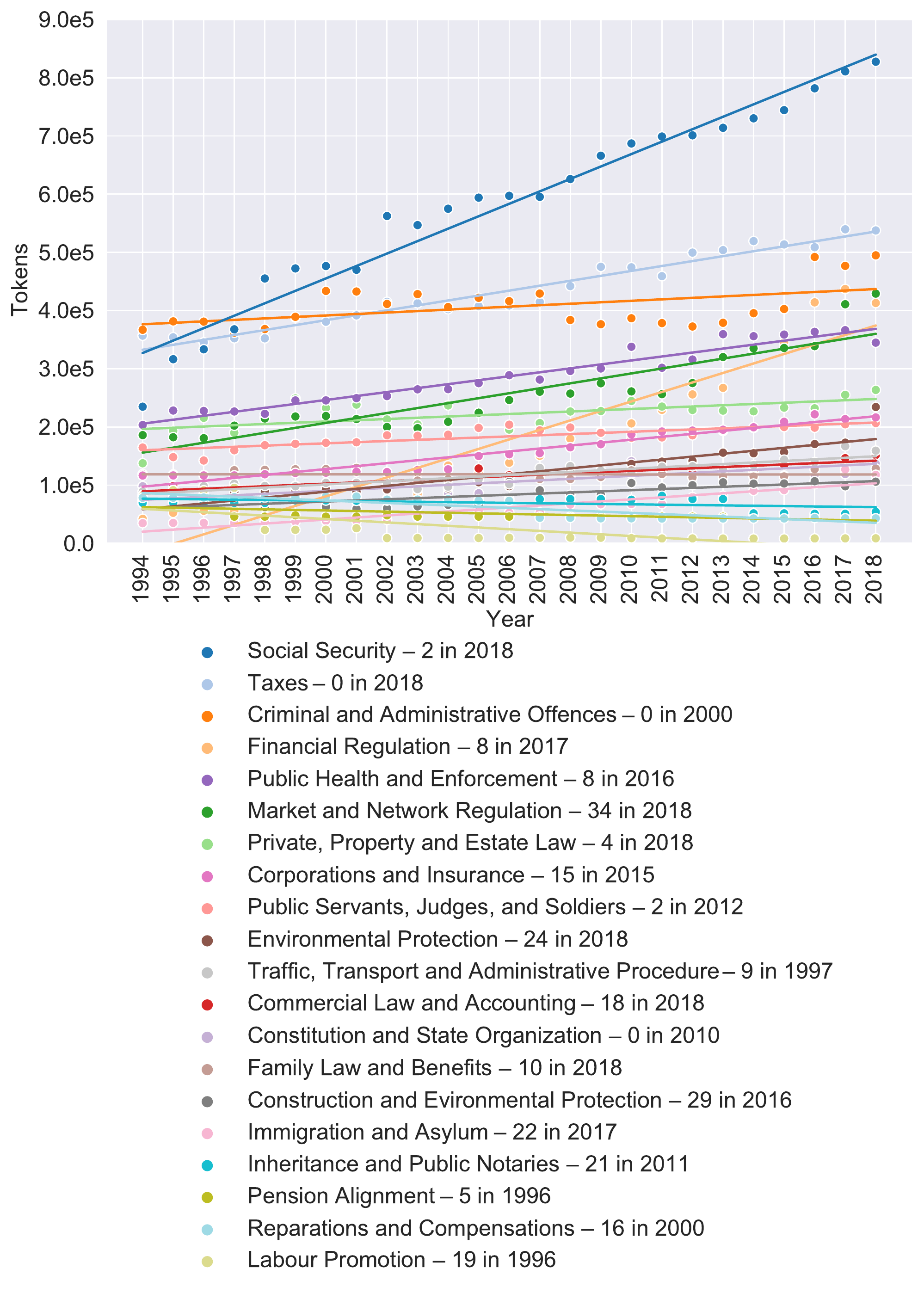}~%
		\subcaption{Germany}
	\end{subfigure}
	\caption{Federal legislation in the United States and Germany: growth statistics by cluster family (1994--2018).}
	\label{fig:legislative-growth-cluster-top20}
\end{figure}

Figure~6 from the main paper 
and Figure~\ref{fig:legislative-growth-cluster-top20} illustrate in scatter plots how the sizes of the cluster families evolve during the observation period.
Figure~\ref{fig:legislative-growth-cluster-top20} shows cluster family sizes for the 20 largest cluster families in each collection (United States and Germany),
where the size of a cluster family is the size of the largest cluster it contains. 
Figure~6 from the main paper 
visualises a selection of the most and the least growing cluster families among the 20 largest families to facilitate interpretation. 
The growth of a cluster family is determined by the slope of an OLS regression for each cluster family $V^i_{F,j}$ on the cluster family sizes $|V^i_{F,j,t}|$ at times $\{~t~\in~T~\}$, 
where $T$ is the observation period.

The resulting intercepts and regression slopes are shown as lines in Figure~6 from the main paper 
and in Figure~\ref{fig:legislative-growth-cluster-top20}.
The points and lines for one cluster family have the same colours in both graphics.
In Figure~\ref{fig:legislative-growth-cluster-top20}, 
colours are reused.
To enable a mapping of lines to the legend, 
we order the legend according to the value expected by the OLS regression in 2018.

In Figure~\ref{fig:legislative-growth-cluster-top20}, 
we label cluster families by year and number of the largest cluster they contain (which also determines the overall size of the cluster family, see Definition~6 from the main paper).
For the purposes of Figures~6 from the main paper,  
we manually assign labels to the cluster families. 
Here, we derive a topic from the names and token shares of Chapters, books, and laws comprising the cluster family.

\begin{figure}[H]
	\centering
	\begin{subfigure}{0.5\linewidth}
		\includegraphics[width=\linewidth]{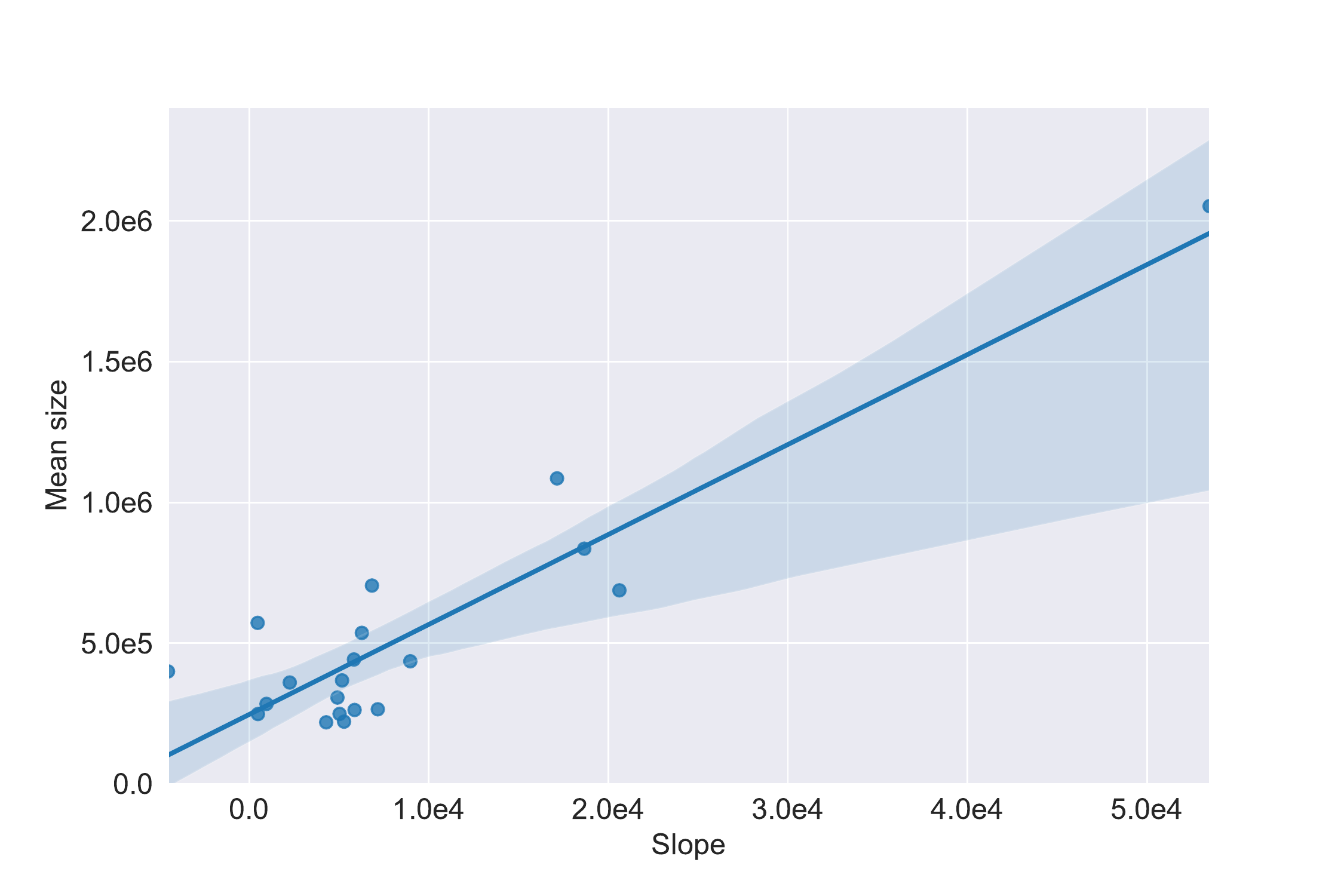}~%
		\subcaption{United States}
	\end{subfigure}~%
	\begin{subfigure}{0.5\linewidth}
		\includegraphics[width=\linewidth]{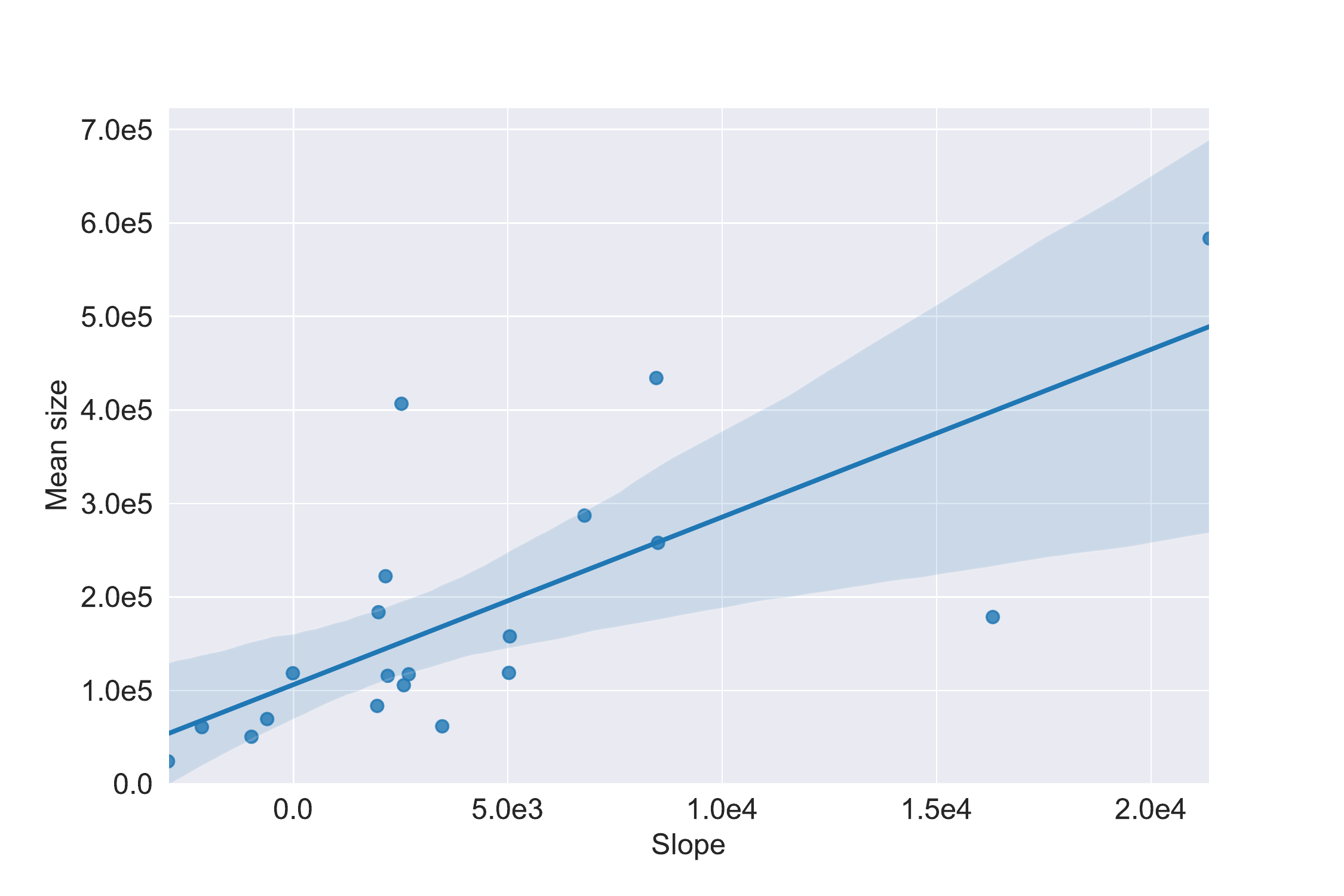}~%
		\subcaption{Germany}
	\end{subfigure}
	\caption{Federal legislation in the United States and Germany: slope-to-size correlation by cluster family (1994--2018).
	}
	\label{fig:legislative-growth-cluster-stats}
\end{figure}

Figure~\ref{fig:legislative-growth-cluster-stats} shows the mean size of the 20 largest cluster families in relation to their slope derived by the OLS regressions as a scatter plot.
We add a regression line and indicate the 95~\% confidence interval using translucent bands around that regression line.
Detailed statistics regarding the regression in Figures~6 from the main paper 
and Figure~\ref{fig:legislative-growth-cluster-top20}, 
and regarding the source data of Figure~\ref{fig:legislative-growth-cluster-stats}, 
are provided in Table~\ref{tab:cluster-family-growth-stats}.

\newpage

\newpage

\begin{table}[H]
	\centering
	\small
	\begin{tabular}{llrrrrrrr}
\toprule
   &            &  \multicolumn{1}{p{18mm}}{\centering Slope} &  \multicolumn{1}{p{18mm}}{\centering Intercept} &  \multicolumn{1}{p{10mm}}{\centering P-value} &  \multicolumn{1}{p{16mm}}{\centering Correlation\\coefficient ($r$)} &  \multicolumn{1}{p{10mm}}{\centering $r^2$} &  \multicolumn{1}{p{16mm}}{\centering Standard\\error} &  \multicolumn{1}{p{18mm}}{\centering Mean value} \\
\multicolumn{1}{p{14mm}}{\centering Country\\code} & \multicolumn{1}{p{14mm}}{\centering Cluster\\family} &                                             &                                                 &                                               &                                                                      &                                             &                                                       &                                                  \\
\midrule
US & 3 in 2015 &                                    53488.68 &                                      1410483.92 &                                          0.00 &                                               0.99 &                                        0.98 &                                            1504.06 &                                       2052348.04 \\
   & 0 in 2010 &                                    20618.14 &                                       440351.10 &                                          0.00 &                                               0.95 &                                        0.90 &                                            1423.24 &                                        687768.80 \\
   & 3 in 2011 &                                    18660.65 &                                       611768.54 &                                          0.00 &                                               0.85 &                                        0.72 &                                            2427.73 &                                        835696.28 \\
   & 11 in 2015 &                                    17144.95 &                                       879622.55 &                                          0.00 &                                               0.94 &                                        0.89 &                                            1246.52 &                                       1085362.00 \\
   & 4 in 2015 &                                     8972.70 &                                       328139.30 &                                          0.00 &                                               0.96 &                                        0.92 &                                             555.38 &                                        435811.68 \\
   & 10 in 2018 &                                     7161.79 &                                       179446.56 &                                          0.00 &                                               0.87 &                                        0.76 &                                             849.28 &                                        265388.08 \\
   & 0 in 2015 &                                     6836.02 &                                       622702.52 &                                          0.00 &                                               0.69 &                                        0.48 &                                            1497.39 &                                        704734.72 \\
   & 9 in 1994 &                                     6273.73 &                                       461446.27 &                                          0.00 &                                               0.71 &                                        0.50 &                                            1297.80 &                                        536731.04 \\
   & 6 in 2018 &                                     5868.71 &                                       192494.97 &                                          0.00 &                                               0.92 &                                        0.86 &                                             503.29 &                                        262919.44 \\
   & 7 in 2018 &                                     5833.80 &                                       372200.23 &                                          0.00 &                                               0.91 &                                        0.82 &                                             569.96 &                                        442205.88 \\
   & 6 in 2005 &                                     5282.55 &                                       158014.62 &                                          0.00 &                                               0.77 &                                        0.59 &                                             913.11 &                                        221405.20 \\
   & 28 in 2018 &                                     5171.32 &                                       305893.97 &                                          0.00 &                                               0.93 &                                        0.86 &                                             428.32 &                                        367949.84 \\
   & 15 in 2018 &                                     5032.92 &                                       188544.60 &                                          0.00 &                                               0.93 &                                        0.87 &                                             401.49 &                                        248939.64 \\
   & 2 in 2017 &                                     4913.84 &                                       248094.22 &                                          0.00 &                                               0.89 &                                        0.79 &                                             526.91 &                                        307060.24 \\
   & 9 in 2018 &                                     4287.96 &                                       167592.31 &                                          0.00 &                                               0.91 &                                        0.83 &                                             404.87 &                                        219047.84 \\
   & 12 in 2007 &                                     2259.43 &                                       333293.88 &                                          0.00 &                                               0.81 &                                        0.66 &                                             340.97 &                                        360407.00 \\
   & 3 in 1995 &                                      964.03 &                                       272922.70 &                                          0.09 &                                               0.35 &                                        0.12 &                                             544.57 &                                        284491.00 \\
   & 26 in 2012 &                                      487.62 &                                       242650.01 &                                          0.28 &                                               0.23 &                                        0.05 &                                             436.96 &                                        248501.48 \\
   & 8 in 2009 &                                      470.18 &                                       566616.93 &                                          0.52 &                                               0.13 &                                        0.02 &                                             726.27 &                                        572259.04 \\
   & 0 in 2013 &                                    -4522.89 &                                       454055.07 &                                          0.17 &                                              -0.28 &                                        0.08 &                                            3176.27 &                                        399780.44 \\
DE & 2 in 2018 &                                    21366.55 &                                       327039.44 &                                          0.00 &                                               0.98 &                                        0.96 &                                             928.87 &                                        583438.08 \\
   & 8 in 2017 &                                    16311.05 &                                       -17159.36 &                                          0.00 &                                               0.95 &                                        0.90 &                                            1128.43 &                                        178573.28 \\
   & 34 in 2018 &                                     8504.39 &                                       155890.06 &                                          0.00 &                                               0.91 &                                        0.83 &                                             803.00 &                                        257942.68 \\
   & 0 in 2018 &                                     8464.01 &                                       332669.71 &                                          0.00 &                                               0.97 &                                        0.95 &                                             407.17 &                                        434237.80 \\
   & 8 in 2016 &                                     6789.83 &                                       205624.20 &                                          0.00 &                                               0.97 &                                        0.94 &                                             343.14 &                                        287102.20 \\
   & 15 in 2015 &                                     5047.60 &                                        97312.68 &                                          0.00 &                                               0.97 &                                        0.94 &                                             275.66 &                                        157883.88 \\
   & 24 in 2018 &                                     5027.07 &                                        58539.27 &                                          0.00 &                                               0.93 &                                        0.87 &                                             407.10 &                                        118864.16 \\
   & 22 in 2017 &                                     3471.46 &                                        20050.75 &                                          0.00 &                                               0.94 &                                        0.88 &                                             265.79 &                                         61708.28 \\
   & 9 in 1997 &                                     2690.08 &                                        85071.52 &                                          0.00 &                                               0.92 &                                        0.84 &                                             246.24 &                                        117352.48 \\
   & 0 in 2010 &                                     2576.20 &                                        74740.47 &                                          0.00 &                                               0.83 &                                        0.69 &                                             361.66 &                                        105654.88 \\
   & 0 in 2000 &                                     2520.76 &                                       376433.06 &                                          0.01 &                                               0.49 &                                        0.24 &                                             928.82 &                                        406682.20 \\
   & 18 in 2018 &                                     2200.20 &                                        89329.56 &                                          0.00 &                                               0.89 &                                        0.80 &                                             232.43 &                                        115731.96 \\
   & 4 in 2018 &                                     2150.07 &                                       196437.59 &                                          0.00 &                                               0.62 &                                        0.39 &                                             566.59 &                                        222238.48 \\
   & 2 in 2012 &                                     1985.76 &                                       159928.37 &                                          0.00 &                                               0.84 &                                        0.70 &                                             269.83 &                                        183757.48 \\
   & 29 in 2016 &                                     1954.78 &                                        60064.11 &                                          0.00 &                                               0.82 &                                        0.67 &                                             289.18 &                                         83521.44 \\
   & 10 in 2018 &                                      -12.21 &                                       118595.63 &                                          0.95 &                                              -0.01 &                                        0.00 &                                             209.39 &                                        118449.08 \\
   & 21 in 2011 &                                     -611.70 &                                        76818.35 &                                          0.02 &                                              -0.47 &                                        0.22 &                                             237.07 &                                         69478.00 \\
   & 5 in 1996 &                                     -976.76 &                                        62273.86 &                                          0.00 &                                              -0.58 &                                        0.33 &                                             287.79 &                                         50552.76 \\
   & 16 in 2000 &                                    -2133.60 &                                        86460.10 &                                          0.00 &                                              -0.89 &                                        0.80 &                                             225.41 &                                         60856.92 \\
   & 19 in 1996 &                                    -2923.61 &                                        59316.50 &                                          0.00 &                                              -0.72 &                                        0.52 &                                             585.81 &                                         24233.20 \\
\bottomrule
\end{tabular}

	\caption{Statistics regarding the OLS regressions in Figure~6 from the main paper 
		and Figure~\ref{fig:legislative-growth-cluster-top20}.
		The p-value column shows a two-sided p-value for a hypothesis test whose null hypothesis is that the slope is zero, using the Wald Test with t-distribution of the test statistic.} 
	\label{tab:cluster-family-growth-stats}
\end{table}


\newpage

\section{Clustering algorithm parametrisation}

In the following, we analyse the performance of our clustering algorithm under different parameter choices to ensure that our results are not artefacts of our parametrisation. 
The statistics we report are based on the Normalized Mutual Information (NMI) and Adjusted Rand Index (ARI)%
---two metrics that are commonly used for pairwise comparisons of clustering results.

NMI is an information-theoretic measure expressing how much information is shared between two clusterings.
It is scaled to range between $0$ (not similar at all) and 1 (identical), and defined as
\begin{align*}
\text{NMI}(X;Y) = \frac{I(X;Y)}{\sqrt{H(X)H(Y)}}~,
\end{align*}
where $I(X;Y) = H(X;Y)-H(X\mid Y)-H(Y\mid X)$ is the mutual information between $X$ and $Y$, $H(X;Y)$ is the joint entropy of $X$ and $Y$, 
$H(X)$ and $H(Y)$ are the individual entropies, 
and $H(X\mid Y)$ and $H(Y\mid X)$ are the conditional entropies.
For more information on this measure, see \cite{strehl2002}.

The ARI is variant of the Rand Index (RI) adjusted for chance. 
The Rand Index is based on counting how many pairs of nodes are in the same clusters or in different clusters in both clusterings. 
It is defined as
\begin{align*}
\text{RI}(X;Y) = \frac{a+b}{a+b+c+d}~,
\end{align*}
where $a$ is the number of node pairs that appear in the same cluster in both clusterings, 
$b$ is the number of node pairs that appear in different clusters in both clusterings, 
$c$ is the number of node pairs that appear in the same cluster in $X$ but in different clusters in $Y$, and
$d$ is the number of node pairs that appear in different clusters in $X$ but in the same cluster in $Y$. 
The ARI is defined as
\begin{align*}
\text{ARI}(X;Y) = \frac{\text{RI}(X;Y) - \mathbb{E}[\text{RI}(X;Y)]}{1 - \mathbb{E}[\text{RI}(X;Y)]}~,
\end{align*}
where $\mathbb{E}[\text{RI}(X;Y)]$ is the expected RI when assuming that the $X$ and $Y$ partitions are constructed randomly, 
subject to having the original number of clusters and the original number of nodes in each cluster.
While the RI ranges between $0$ and $1$, the ARI is bounded from above by $1$ but may take negative values when the agreement between the clusterings is less than expected. 
Unrelated clusterings have an ARI close to $0$ and identical clusterings have an ARI of $1$. 
More information on this measure can be found in \cite{hubert1985}.

\subsection{Sensitivity analysis}

\begin{figure}[H]
	\centering
	\begin{subfigure}{0.5\linewidth}
		\includegraphics[width=\linewidth]{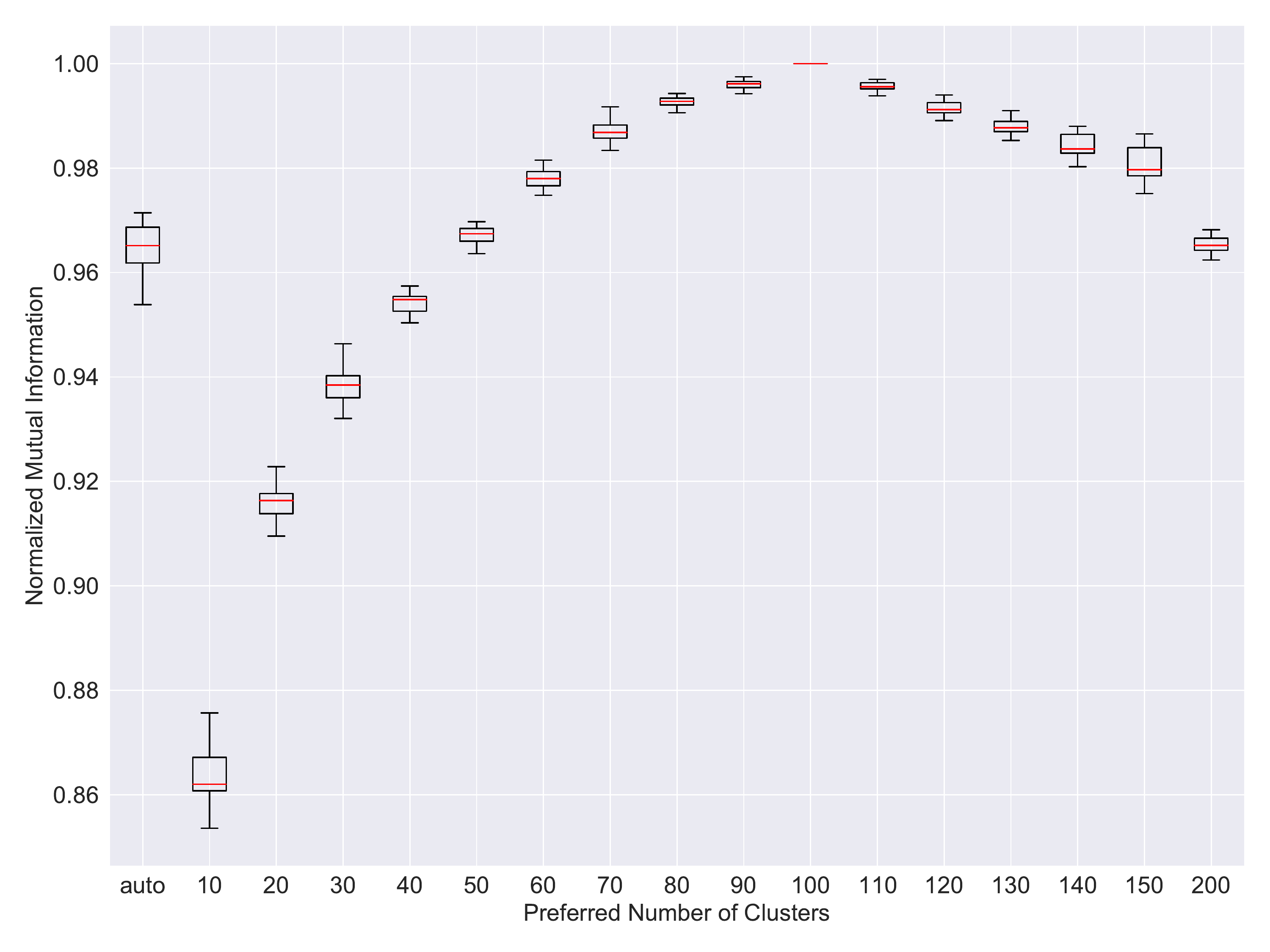}
		\subcaption{United States (Normalised Mutual Information)}
	\end{subfigure}~%
	\begin{subfigure}{0.5\linewidth}
		\includegraphics[width=\linewidth]{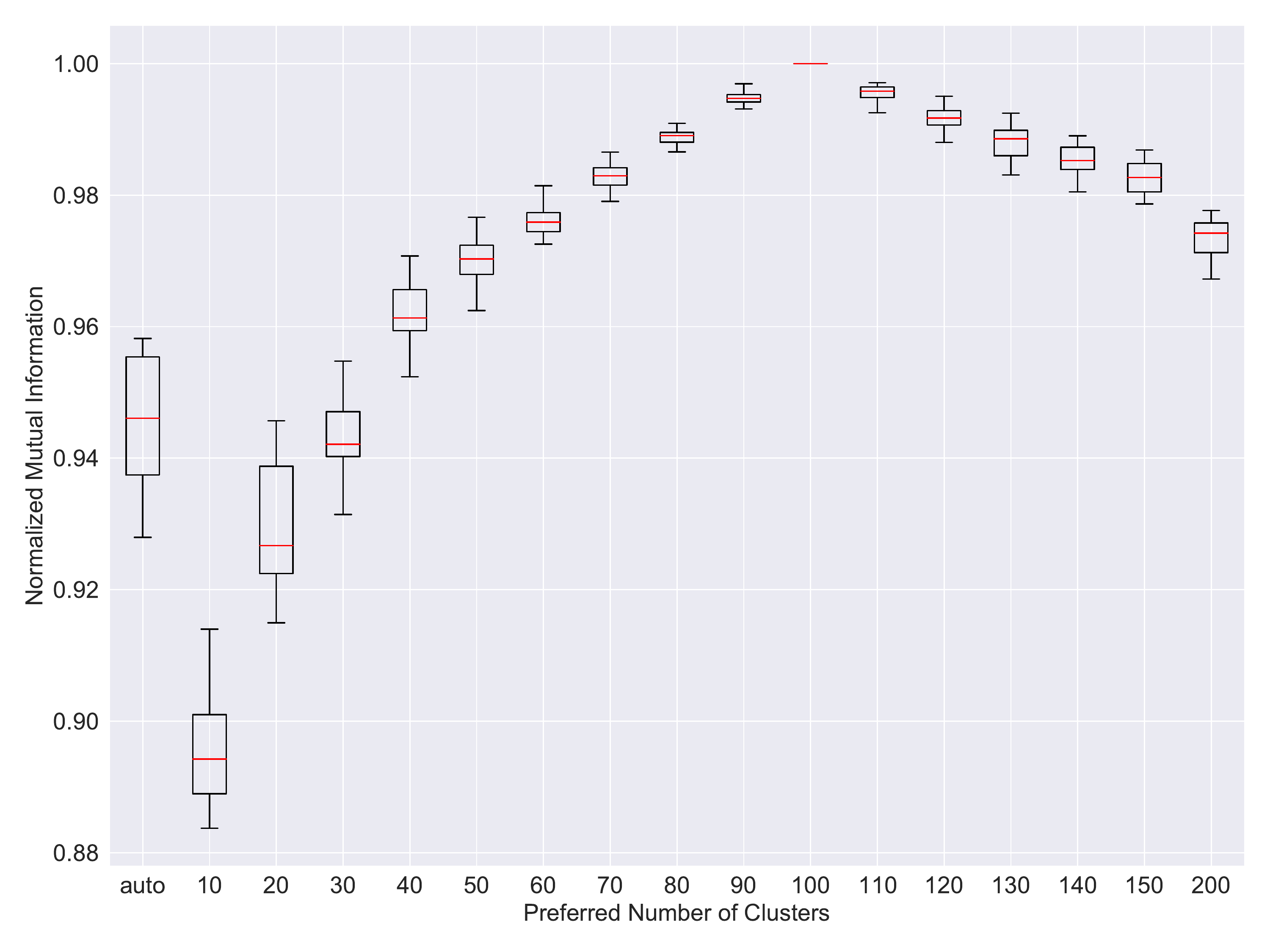}
		\subcaption{Germany (Normalised Mutual Information)}
	\end{subfigure}
	\begin{subfigure}{0.5\linewidth}
		\includegraphics[width=\linewidth]{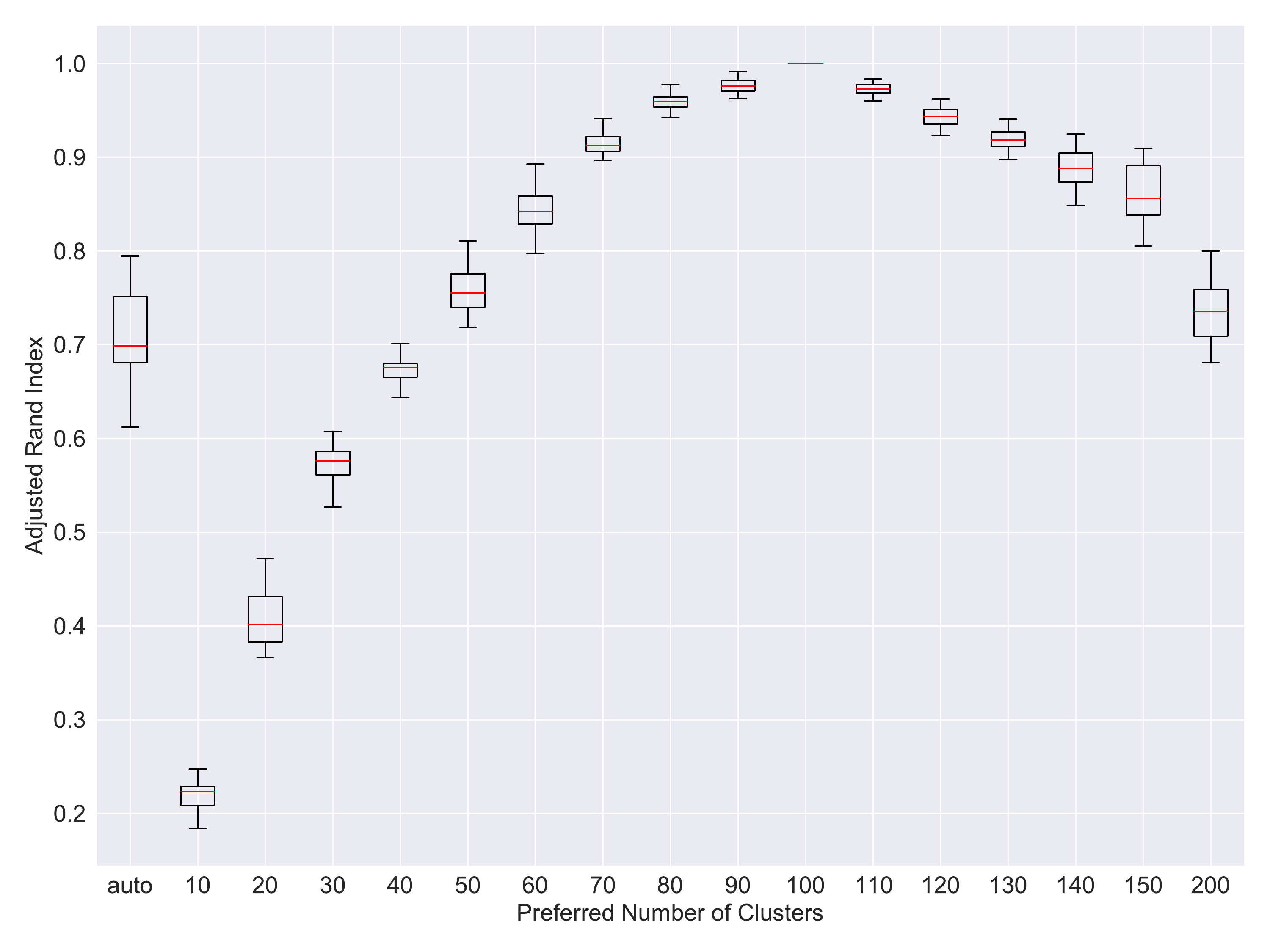}
		\subcaption{United States (Adjusted Rand Index)}
	\end{subfigure}~%
	\begin{subfigure}{0.5\linewidth}
		\includegraphics[width=\linewidth]{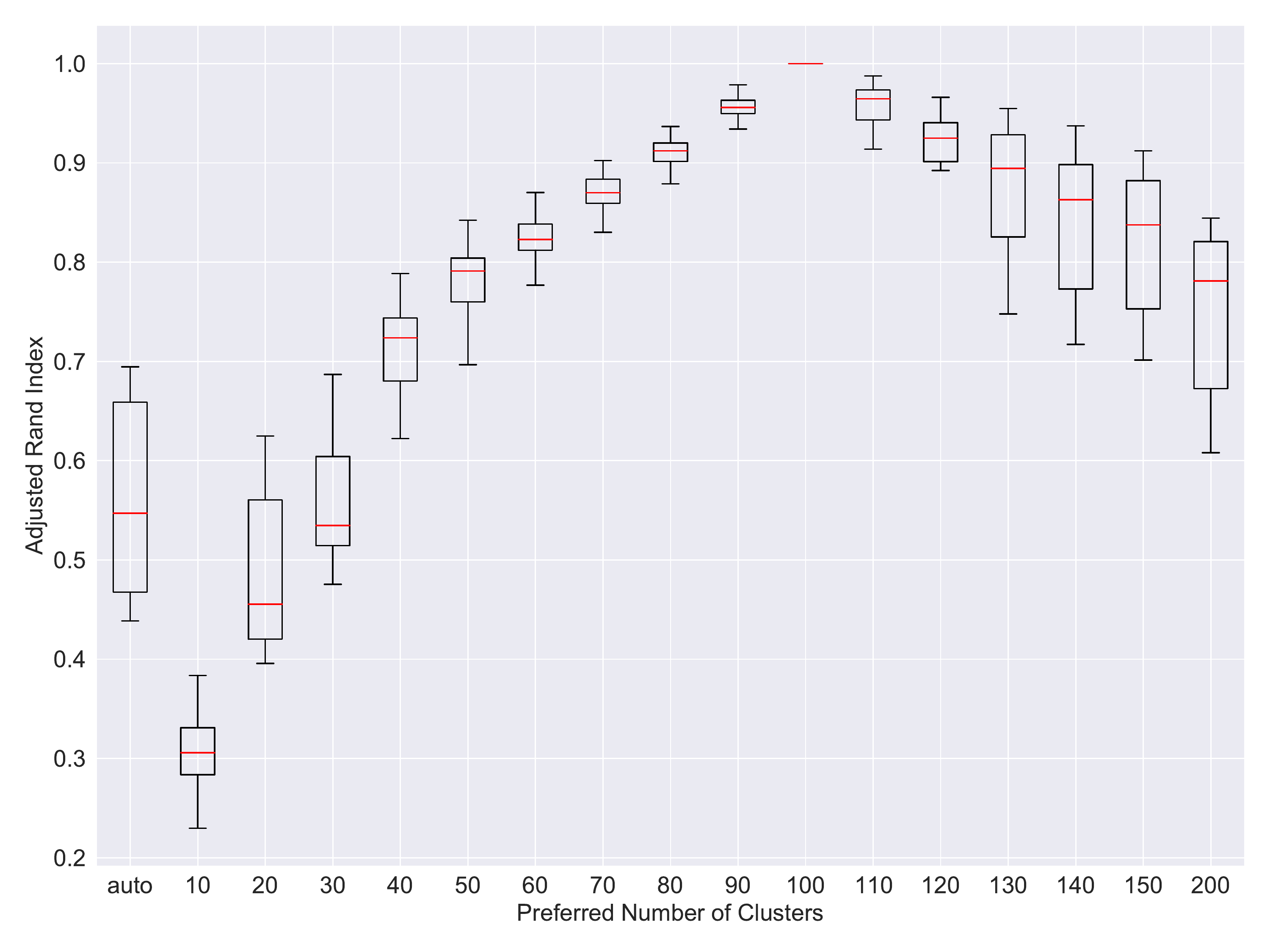}
		\subcaption{Germany (Adjusted Rand Index)}
	\end{subfigure}
	\caption{Distribution of pairwise similarities between clusterings with different preferred cluster sizes in the same year over the $25$ years from $1994$ to $2018$. 
		\emph{Auto} indicates that the \emph{Infomap} algorithm chooses the preferred number of clusters.
		Note that only the box plots labelled $10$ through $150$ are equidistant to each other on the real line. 
		The $y$-coordinates of the box boundaries indicate the second and fourth quartile, while the red line indicates the median. 
		Upper whiskers extend to the last data point less than $1.5$ times the box height above the fourth quartile, 
		while lower whiskers extend to the first data point less than $1.5$ times the box height below the first quartile.
	}\label{fig:sensitivity-preferred-clusters}
\end{figure}

\notereviewone{5}\notereviewtwo{8}
Figure~\ref{fig:sensitivity-preferred-clusters} shows how the clustering results change when we alter the preferred number of clusters, 
with our chosen number of clusters as the baseline. 
As preferred numbers of clusters, we test all numbers divisible by $10$ from $10$ to $150$ as well as the number $200$. 
In one experiment, labelled \emph{auto}, we let the \emph{Infomap} algorithm choose the preferred number of clusters.
Unsurprisingly, the box plots show that clusterings become more similar to our baseline clustering with $100$ preferred clusters as we approach this number. 
At the same time, clusterings with $50$ or $200$ preferred clusters are already relatively similar to the baseline, with NMI values over $0.96$ and ARI values over $0.7$. 

Note that the spread in clustering similarities is largest for comparisons of the baseline with \emph{auto}, i.e., the clusterings in which \emph{Infomap} chooses the preferred number of clusters.
This is likely due to the jumps in clustering granularity that sometimes occur in \emph{Infomap} due to small differences in the minimum description length of competing models with different resolutions.
Avoiding these jumps is our primary motivation for specifying a preferred number of clusters.

\begin{figure}[H]
	\vspace*{-16pt}
	\centering
	\begin{subfigure}{0.45\linewidth}
		\includegraphics[width=\linewidth]{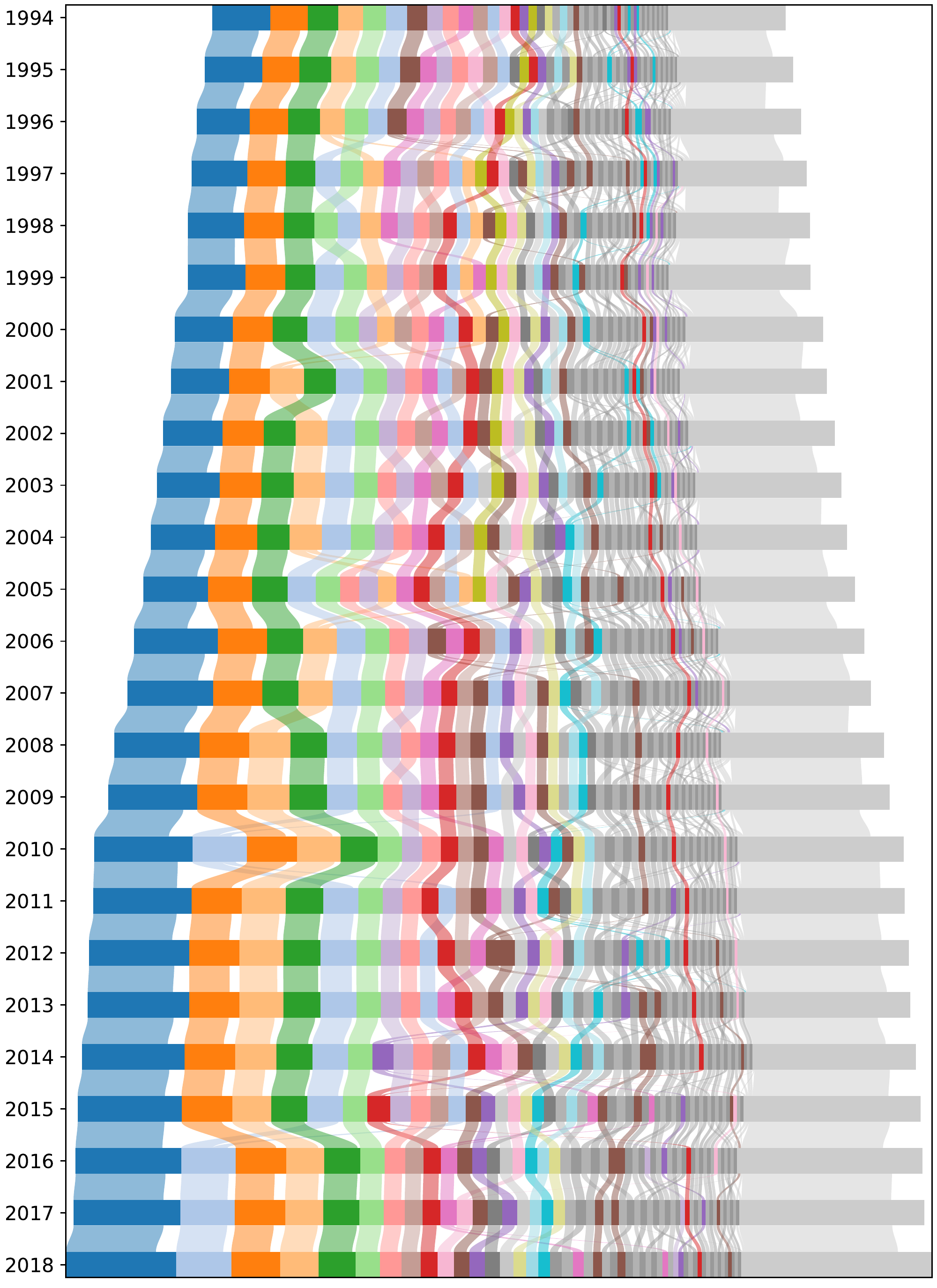}
		\subcaption{50 preferred clusters}
	\end{subfigure}~%
	\begin{subfigure}{0.45\linewidth}
		\includegraphics[width=\linewidth]{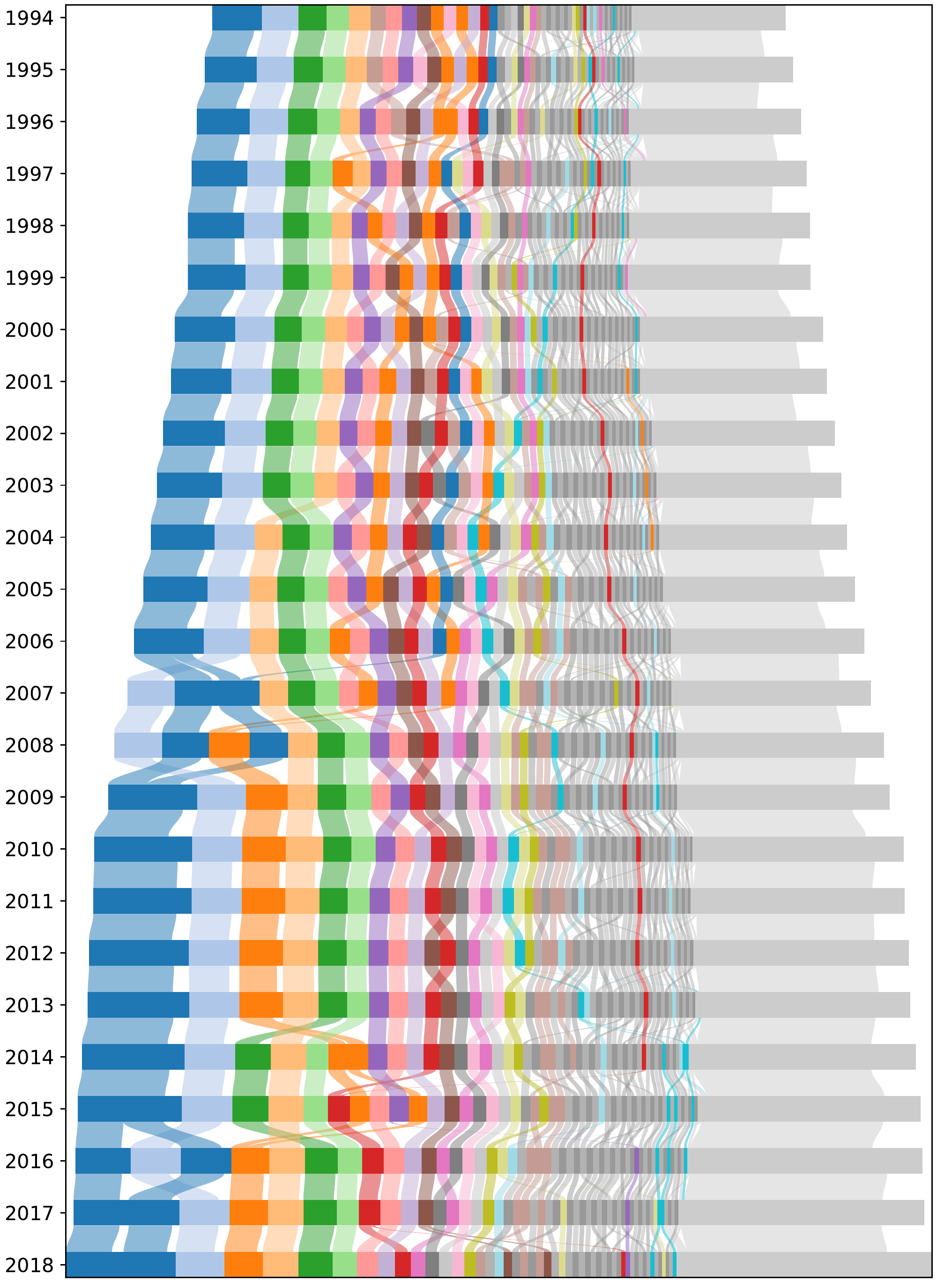}
		\subcaption{100 preferred clusters}
	\end{subfigure}

	\begin{subfigure}{0.45\linewidth}
		\includegraphics[width=\linewidth]{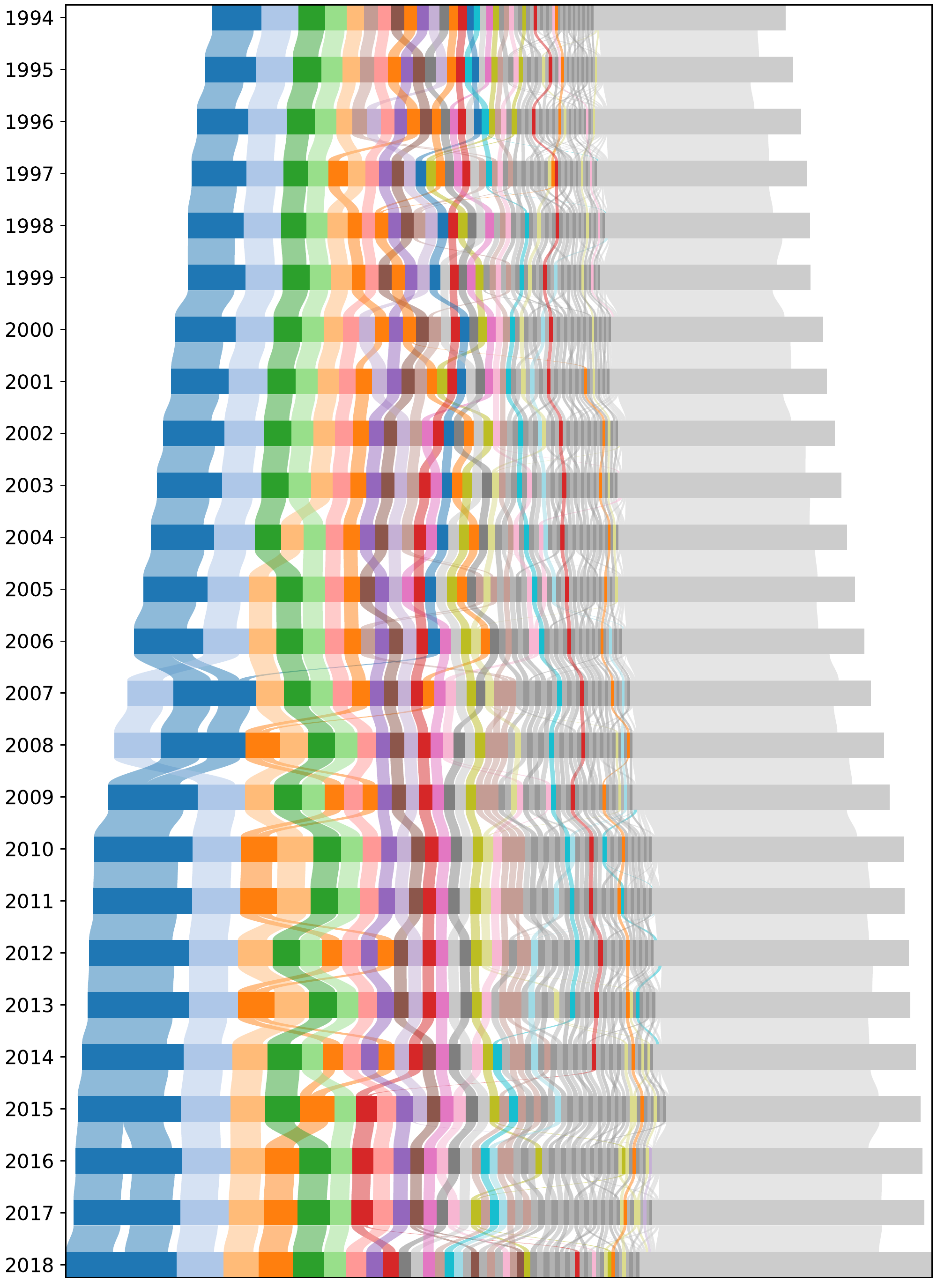}
		\subcaption{200 preferred clusters}
	\end{subfigure}~%
	\begin{subfigure}{0.45\linewidth}
		\includegraphics[width=\linewidth]{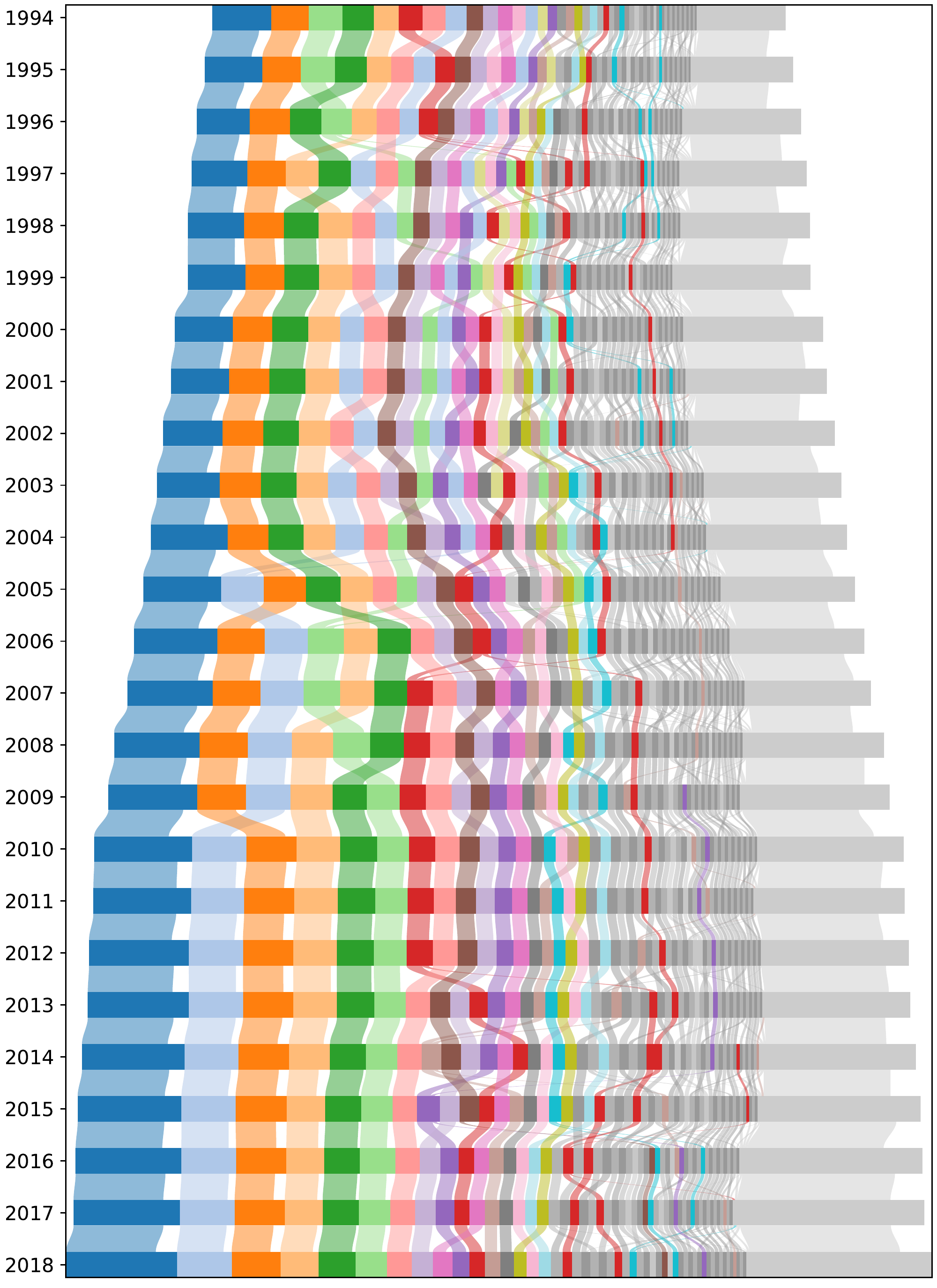}
		\subcaption{Algorithm determines number of clusters}
	\end{subfigure}
\caption{Federal legislation in the United States by cluster (1994--2018), depicted as in Figure~5 from the main paper, for different preferred numbers of clusters.}
\label{fig:sankey-preferred-variants}
\end{figure}

Figure~\ref{fig:sankey-preferred-variants} shows how the compositions of cluster families change if we choose $50$ or $200$, rather than $100$ preferred clusters, or let the algorithm determine the preferred number of clusters.
As should become clear by visual inspection, the overall picture remains the same.

\newpage

\newpage

\subsection{Robustness checks}

\begin{figure}[H]
	\centering
	\begin{subfigure}{0.5\linewidth}
		\includegraphics[width=\linewidth]{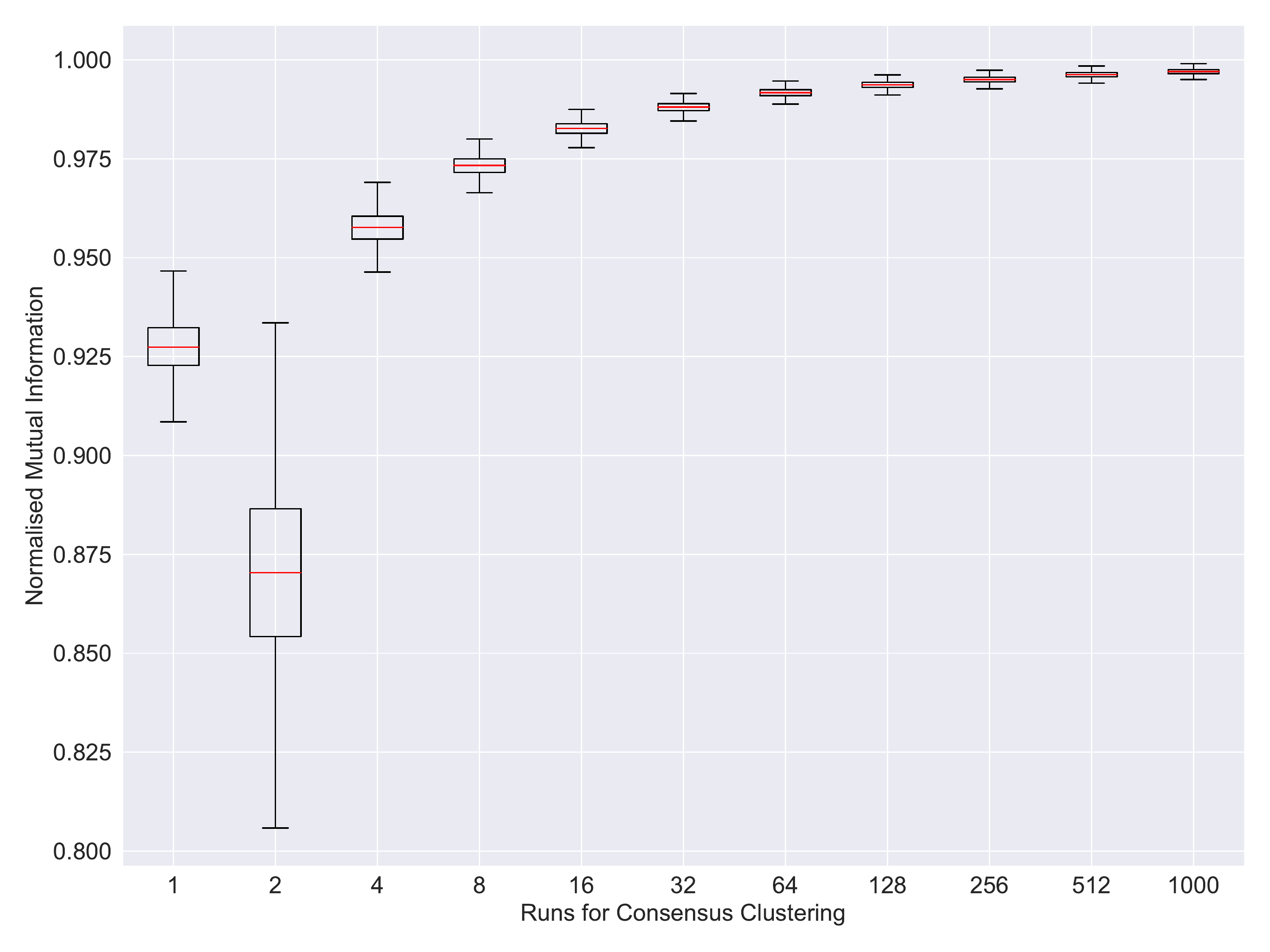}
		\subcaption{United States (Normalised Mutual Information)}
	\end{subfigure}~%
	\begin{subfigure}{0.5\linewidth}
		\includegraphics[width=\linewidth]{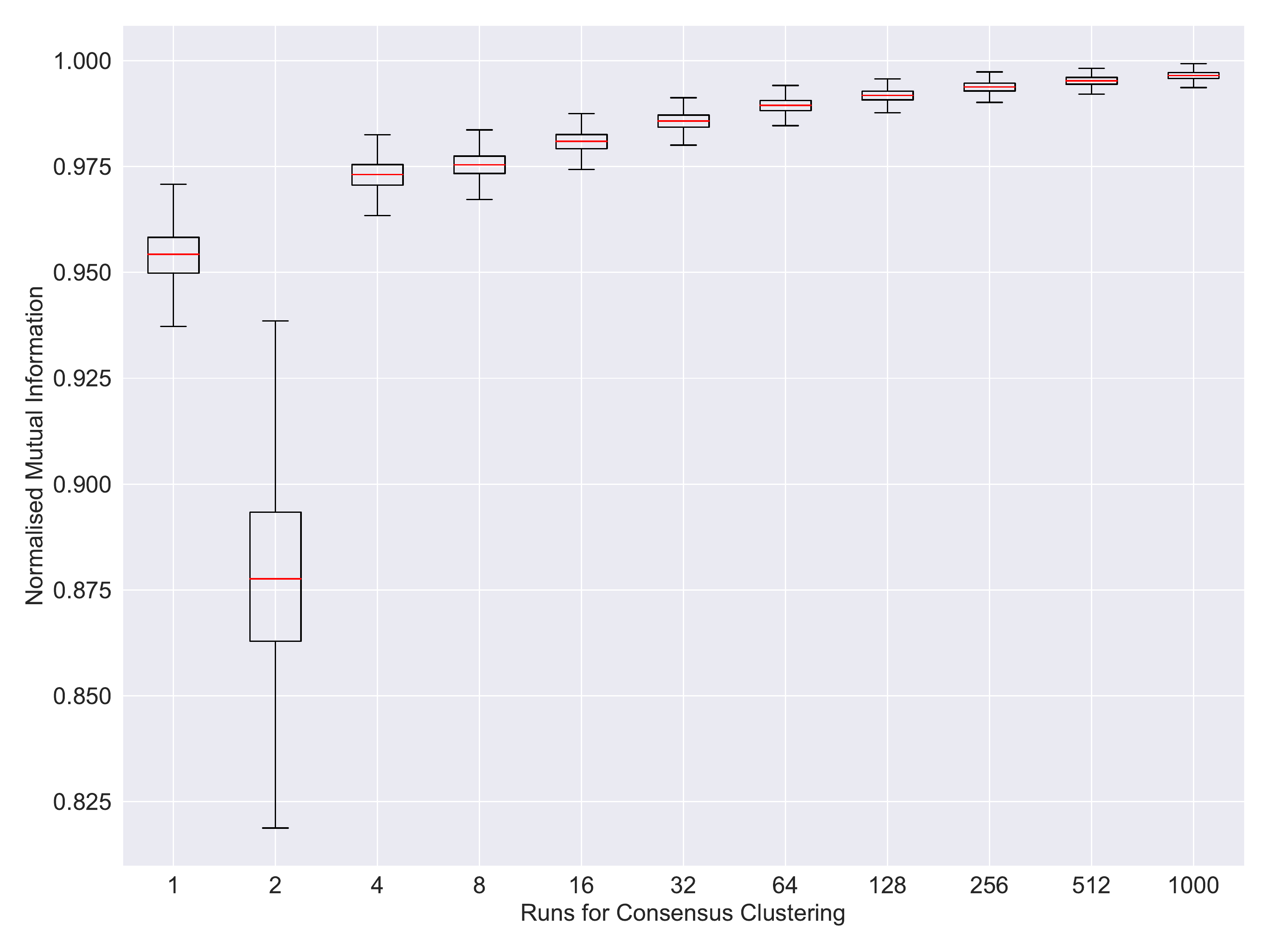}
		\subcaption{Germany (Normalised Mutual Information)}
	\end{subfigure}
	\begin{subfigure}{0.5\linewidth}
		\includegraphics[width=\linewidth]{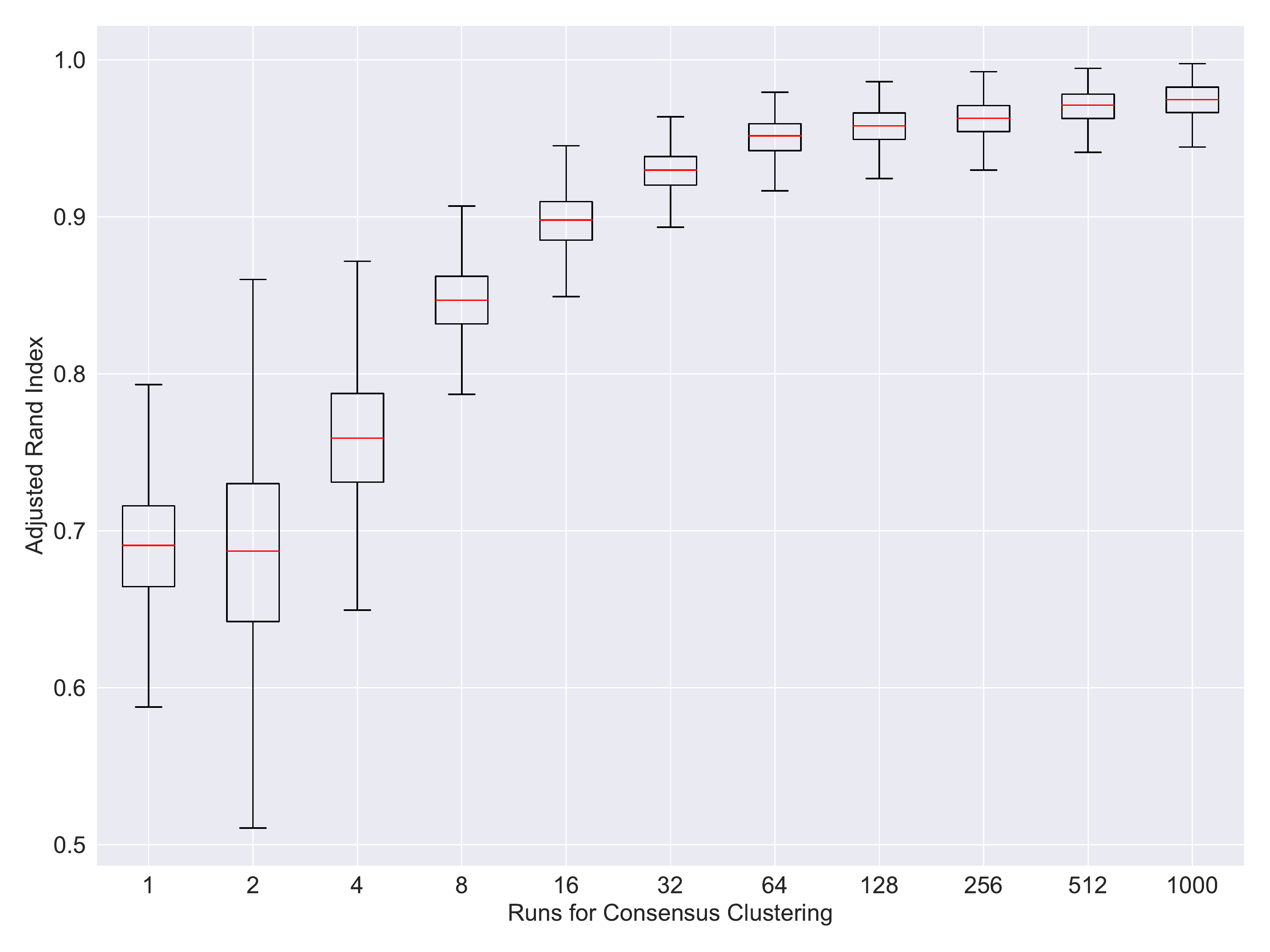}
		\subcaption{United States (Adjusted Rand Index)}
	\end{subfigure}~%
	\begin{subfigure}{0.5\linewidth}
		\includegraphics[width=\linewidth]{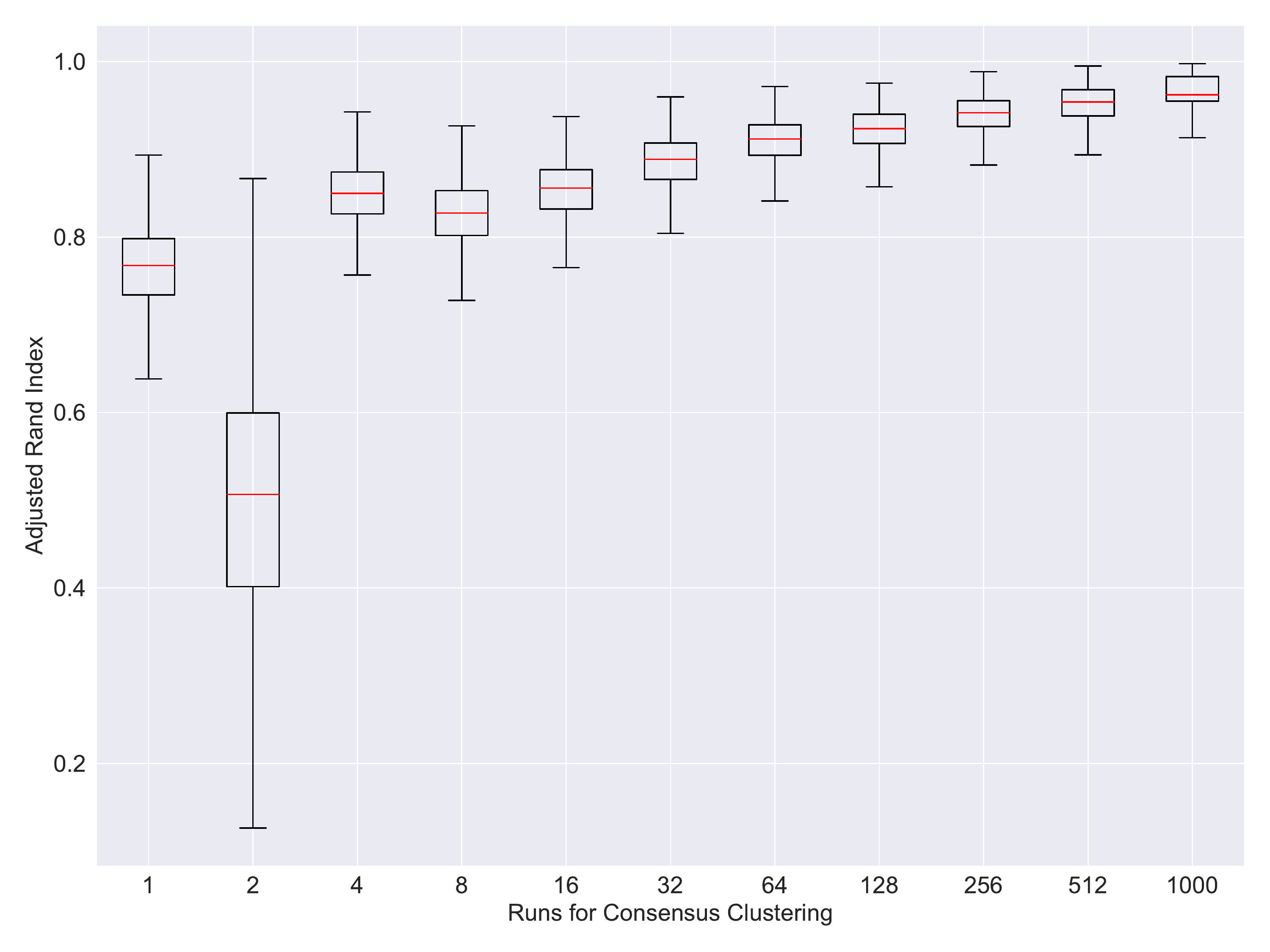}
		\subcaption{Germany (Adjusted Rand Index)}
	\end{subfigure}
	\caption{Pairwise similarity between $100$ consensus clustering results by number of clusterings used for finding the consensus (box plot interpretation as described in the caption of  Figure~\ref{fig:sensitivity-preferred-clusters}).
	}
\label{fig:consensus-effect}
\end{figure}

\notereviewtwo{9}
Figure~\ref{fig:consensus-effect} shows the distribution of pairwise similarities between $100$ consensus clustering results for different numbers of clusterings used in the consensus. 
The plots show that using a higher number of clusterings to form the consensus increases the overall similarity level and reduces the spread between the observed similarities. 
When choosing $1000$ clusterings to form the consensus (as we do in the main paper), 
the consensus clusterings we obtain in different runs are almost identical.

\begin{figure}[H]
	\centering
	\begin{subfigure}{0.5\linewidth}
		\includegraphics[width=\linewidth]{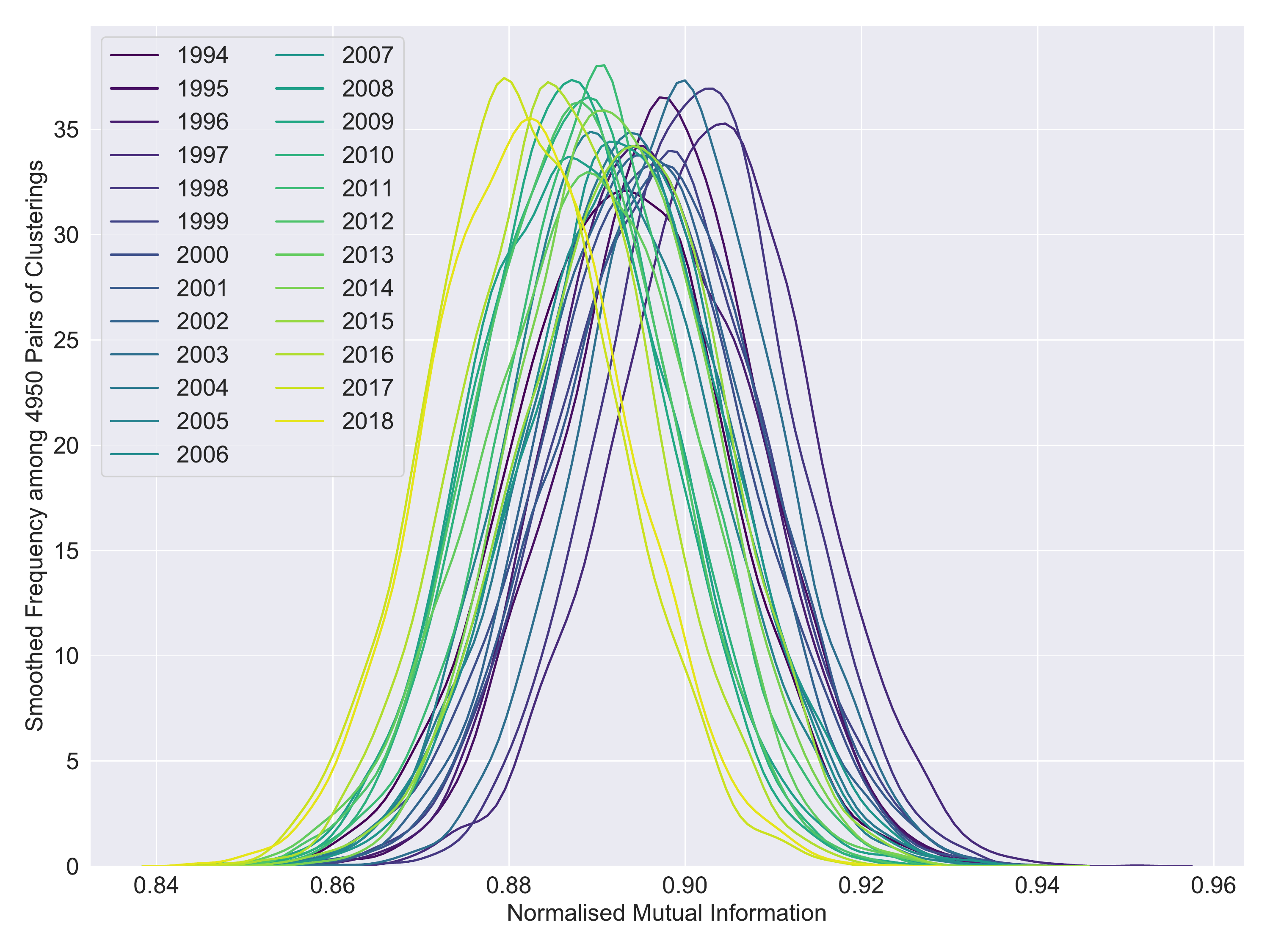}
		\subcaption{United States (Normalised Mutual Information)}
	\end{subfigure}~%
	\begin{subfigure}{0.5\linewidth}
		\includegraphics[width=\linewidth]{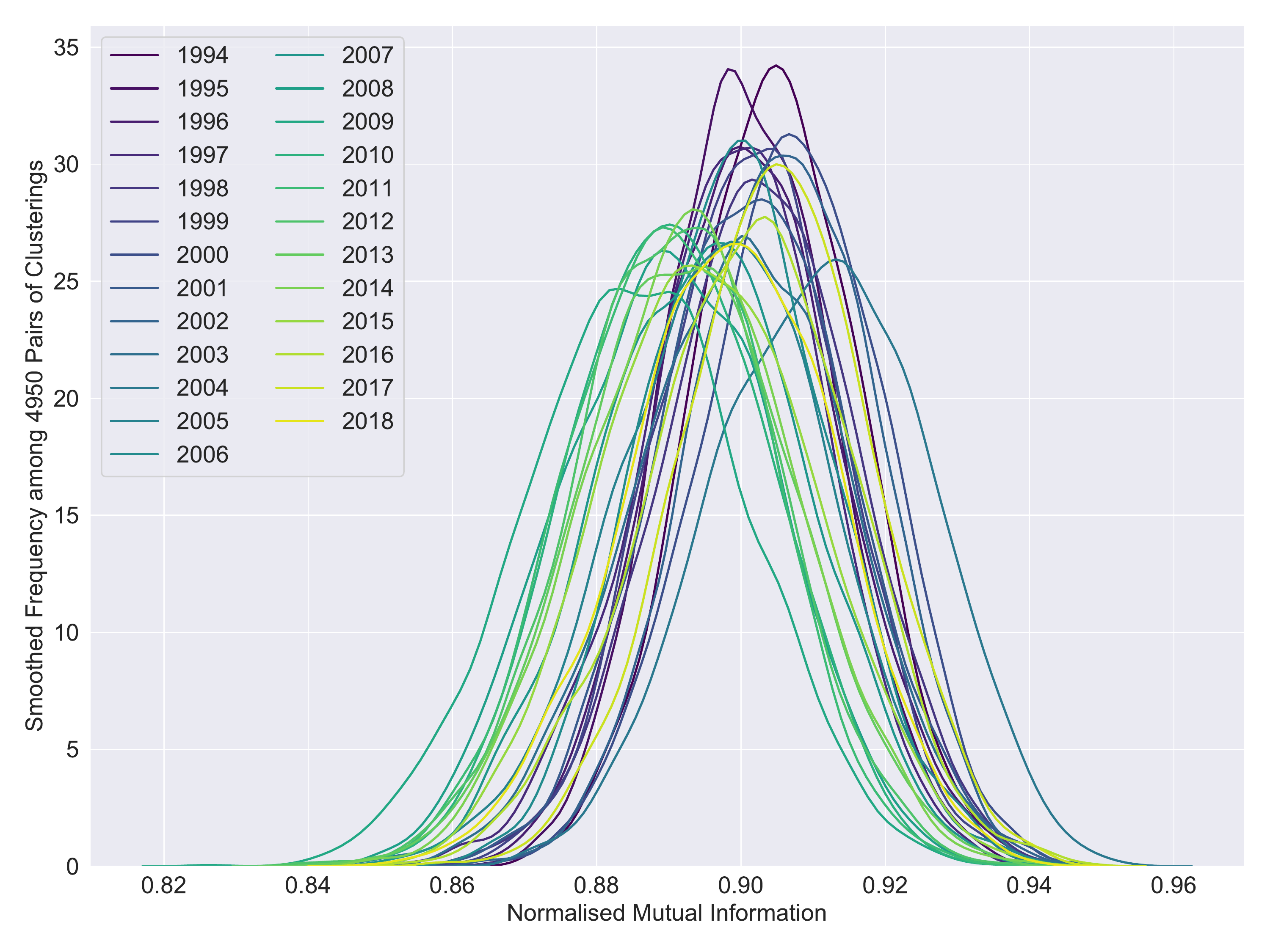}
		\subcaption{Germany (Normalised Mutual Information)}
	\end{subfigure}
	\begin{subfigure}{0.5\linewidth}
		\includegraphics[width=\linewidth]{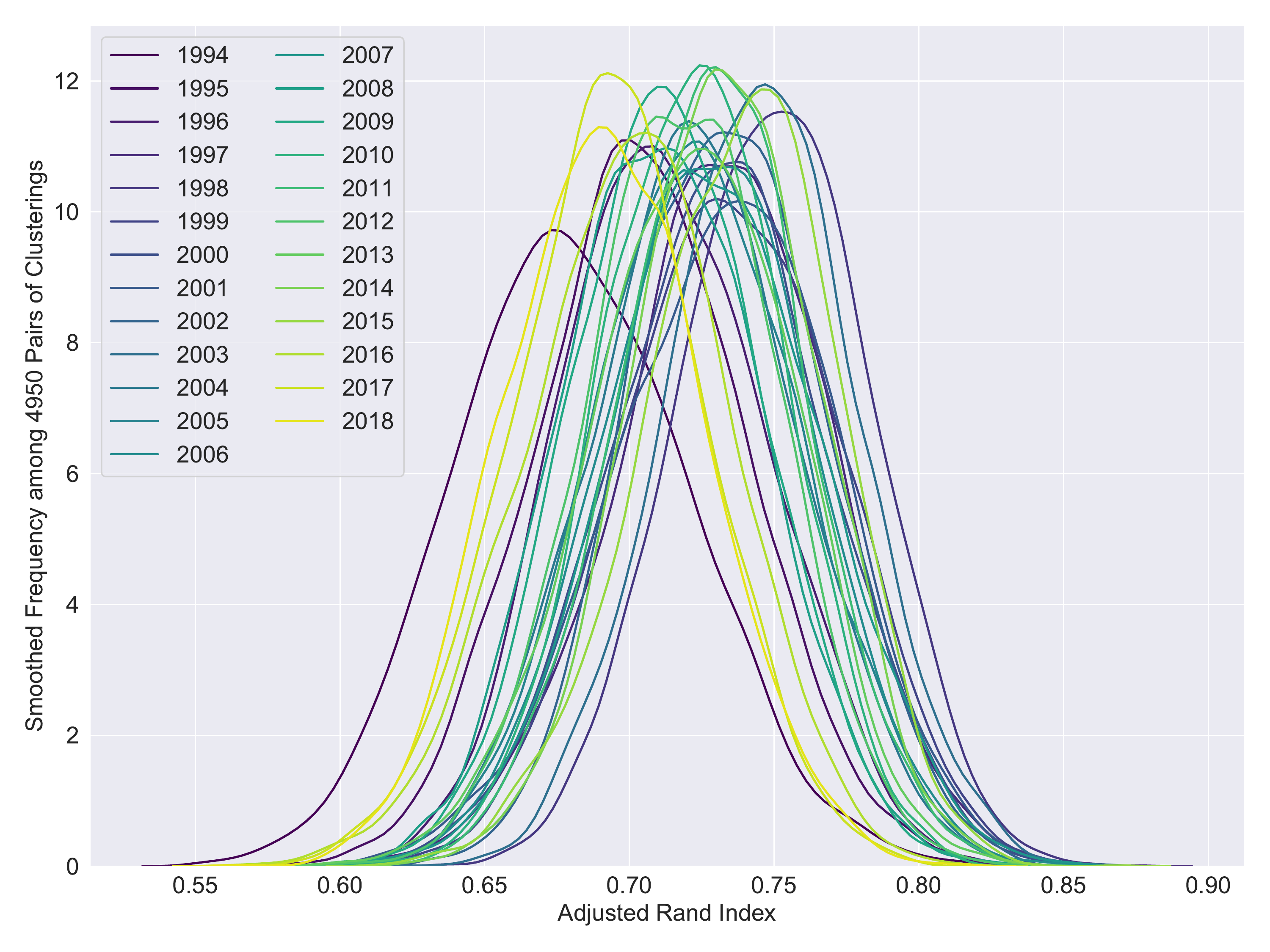}
		\subcaption{United States (Adjusted Rand Index)}
	\end{subfigure}~%
	\begin{subfigure}{0.5\linewidth}
		\includegraphics[width=\linewidth]{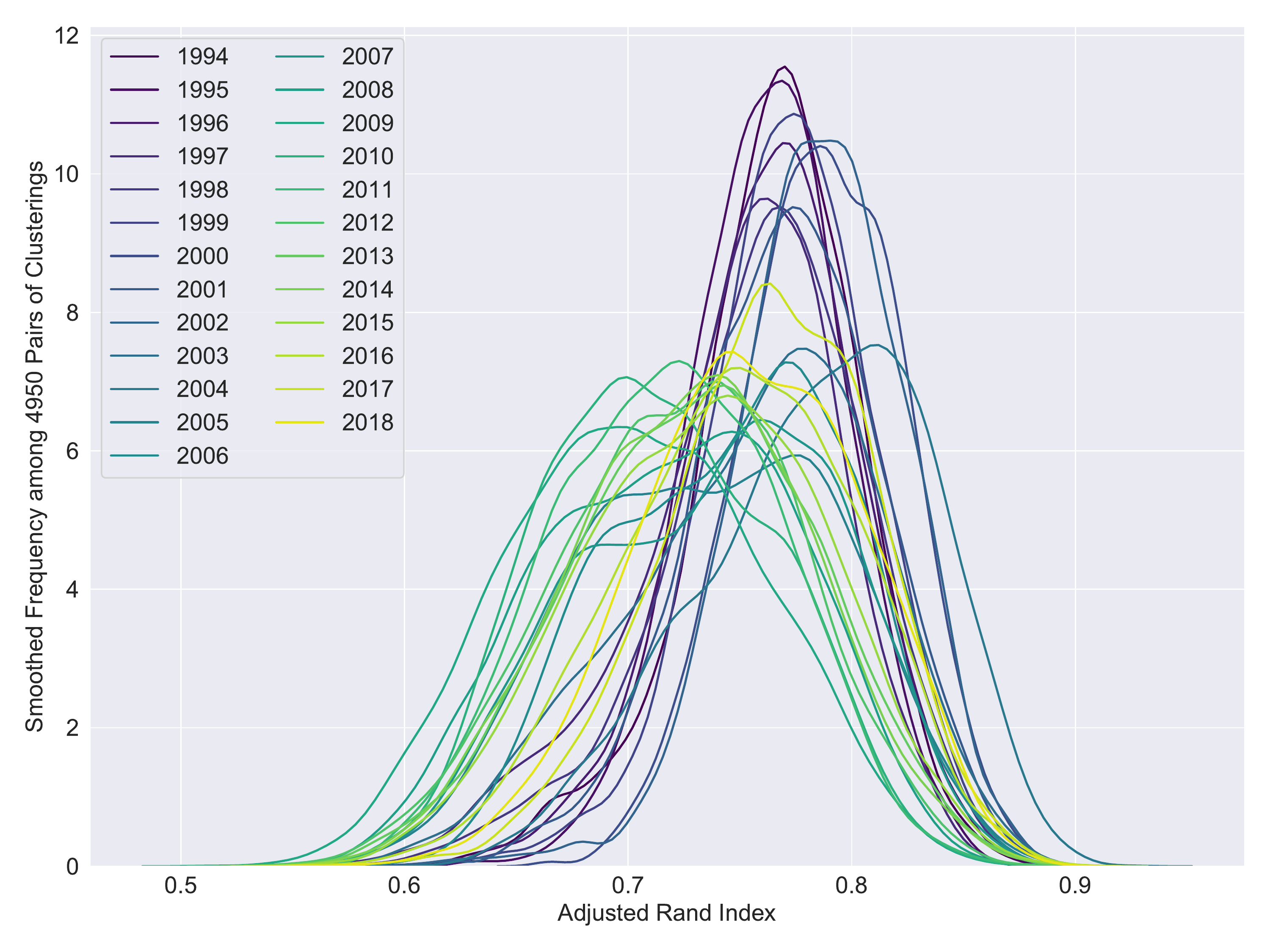}
		\subcaption{Germany (Adjusted Rand Index)}
	\end{subfigure}
	\caption{Pairwise similarity between $100$ clusterings with $100$ as the preferred number of clusters, depicted as kernel-density estimates rather than frequency histograms to reduce visual clutter.}
	\label{fig:consensus-within}
\end{figure}

Figure~\ref{fig:consensus-within} shows the distribution of pairwise similarities between $100$ pairs of clusterings (i.e., a total of $4950$ similarities) with $100$ as the preferred number of clusters. 
The NMI values for the United States clusterings mostly range between $0.86$ and $0.94$, while the NMI values for Germany mostly range between $0.84$ and $0.96$.
The ARI values for the United States clusterings mostly range between $0.55$ and $0.85$ (with the majority lying between $0.65$ and $0.80$), 
while the ARI values for Germany mostly range between $0.60$ and $0.90$.
All similarity distributions seem to shift towards the left over time, 
i.e., clusterings in earlier years tend to be more similar to each other than clusterings in later years. 
This is likely due to the growth in complexity reported in the main paper.

\newpage

\section{Cluster families}\label{sec:labelling}

\subsection{Labelling the twenty largest cluster families}

\notereviewone{9}
To arrive at the labels of the $20$ largest cluster families, 
we leverage our subject matter expertise, 
inspecting the content of the cluster families based on automatically generated summaries that show what percentage of the cluster is made up by which particular Chapter, Book, or law (measured in tokens). 
These summaries contain the full paths to each node, including the names of all structural elements in which it is contained in the hierarchy graph. 
As such, they provide just enough dimensionality reduction for humans to be able to assign the final label with confidence, and they are part of the data provided with the paper. 
For comparison, we also provide basic TF-IDF (term frequency-inverse document frequency; for more information, see \cite{schutze2008}) statistics for the $20$ largest cluster families as CSV files in the data repository accompanying this paper (with very little preprocessing and without stopwords removed). 
 
Table~\ref{tab:cluster-labeling-alt} contains the Top $10$ nouns according to TF-IDF for the United States cluster families depicted in Figure~6~(a) from the main paper. 
Table~\ref{tab:cluster-labeling} shows a (reformatted) excerpt from the summary of the largest cluster family in the United States as used in our manual labelling process.
The full lists of our labels for the $20$ largest clusters are reproduced in Tables~\ref{tab:labels-all-us} and~\ref{tab:labels-all-de}.

\begin{table}[H]
	\centering\footnotesize
	\bgroup
	\def\arraystretch{1.5}
	\begin{tabular}{r|p{0.115\textwidth}p{0.115\textwidth}p{0.12\textwidth}p{0.115\textwidth}p{0.095\textwidth}p{0.12\textwidth}p{0.095\textwidth}}
		&\textbf{2015-3}&\textbf{2015-11}&\textbf{2011-3}&\textbf{2010-0}&\textbf{2015-4}&\textbf{2009-8}&\textbf{1995-3}\\\hline
		1&pesticide&multiemployer&student&depository&cuba&mortgagee&acreage\\
		2&secretary&year&teachers&thrift&hiv&mortgagor&cotton\\
		3&state&taxpayer&youth&conservator&states&homeownership&wheat\\
		4&medicare&plan&secretary&sipc (*)&pakistan&secretary&crop\\
		5&drug&purposes&school&bank&mtcr (**)&housing&quota\\
		6&services&amount&state&depositor&afghanistan&mortgages&peanuts\\
		7&physician&dividend&teacher&banking&nato&mortgage&sugar\\
		8&pediatric&corporation&childhood&institution&hungary&dwelling&upland\\
		9&health&income&agency&board&democracy&homebuyers&tobacco\\
		10&vaccine&distributee&education&corporation&israel&paint&milk\\
	\end{tabular}
	\egroup
	\caption{Top $10$ nouns for the $7$ cluster families depicted in Figure~6~(a) from the main paper, labelled by leading cluster (Year-Cluster Identifier).
	Nouns referring to structural elements of legal texts (e.g., title, section, subsection, paragraph) are excluded. 
	(*)~Securities Investor Protection Corporation. (**)~Missile Technology Control Regime.
	}
	\label{tab:cluster-labeling-alt}
\end{table}

\subsection{Inspecting the \emph{miscellaneous} clusters}
\notereviewtwo{6}

Recall that although we use $100$ as the preferred number of clusters, we end up with more than $100$ clusters due to the presence of nodes without any incoming or outgoing references (\emph{singletons}) and our use of consensus clustering. 
To reduce visual clutter, in Figure~5 from the main paper, we limit the number of clusters drawn per year to $50$, 
summarising the remaining clusters in one additional miscellaneous cluster. 

In both the United States and Germany, in all years,
the miscellaneous cluster contains mostly singletons or near-singletons corresponding to small Chapters or laws that are largely self-contained. 
Its composition remains fairly stable over time (i.e., nodes in the miscellaneous cluster seldom get pulled into a different cluster), 
and its growth is primarily driven by the addition of new, relatively independent Chapters or laws.
Since its contents are very diverse, 
the growth of the miscellaneous cluster could be interpreted as an indicator that our legal corpora grow not only in volume but also in diversity.

\begin{figure}[H]
	\vspace*{-16pt}
	\centering
	\begin{subfigure}{0.45\linewidth}
		\includegraphics[width=\linewidth]{sankey_us_0-0_1-0_-1_a-infomap_n100_m1-0_s0_c1000}
		\subcaption{United States (50 + 1 clusters drawn per year)}
	\end{subfigure}~%
	\begin{subfigure}{0.45\linewidth}
		\includegraphics[width=\linewidth]{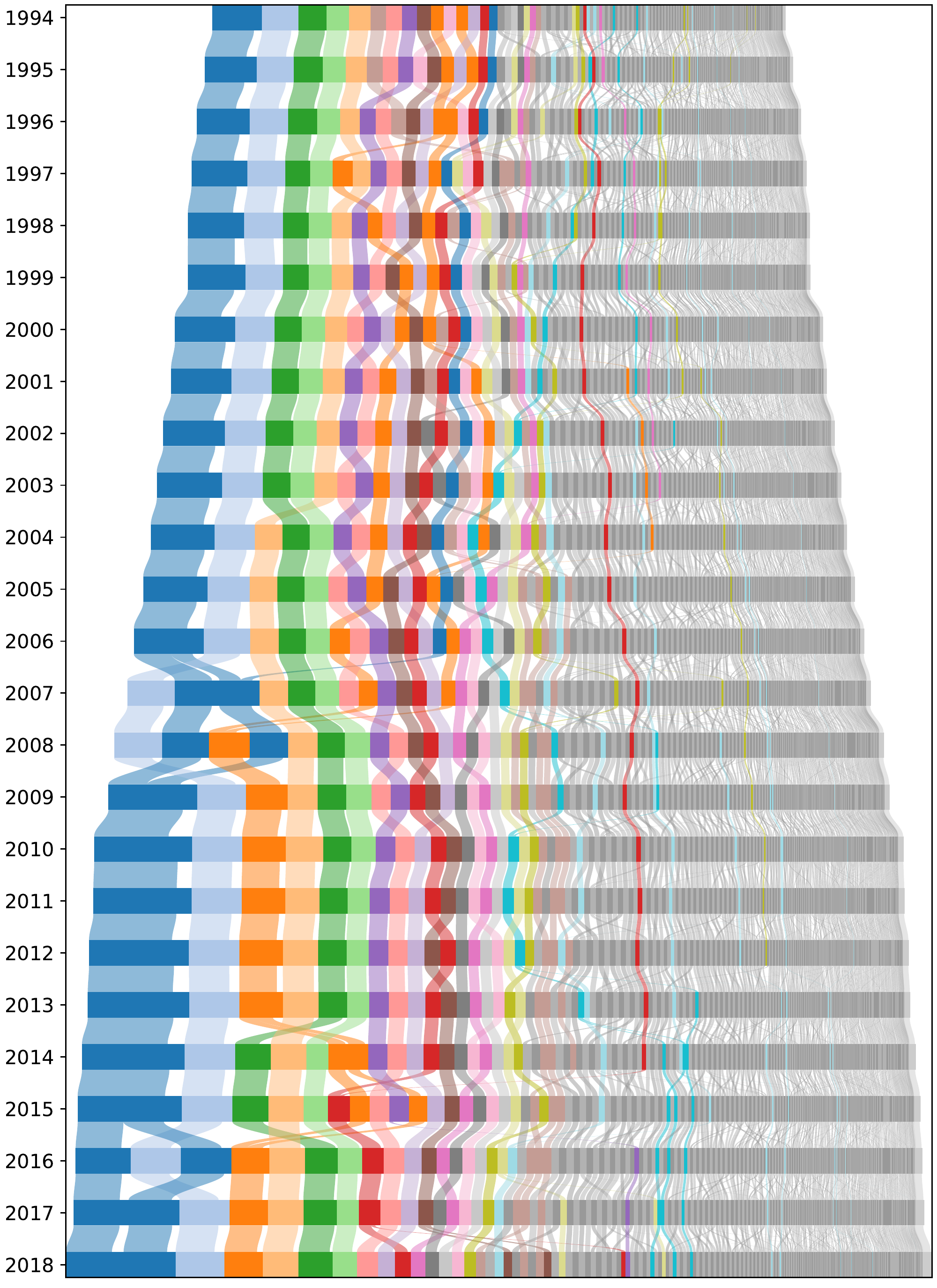}
		\subcaption{United States (500 + 1 clusters drawn  per year)}
	\end{subfigure}
	
	\begin{subfigure}{0.45\linewidth}
		\includegraphics[width=\linewidth]{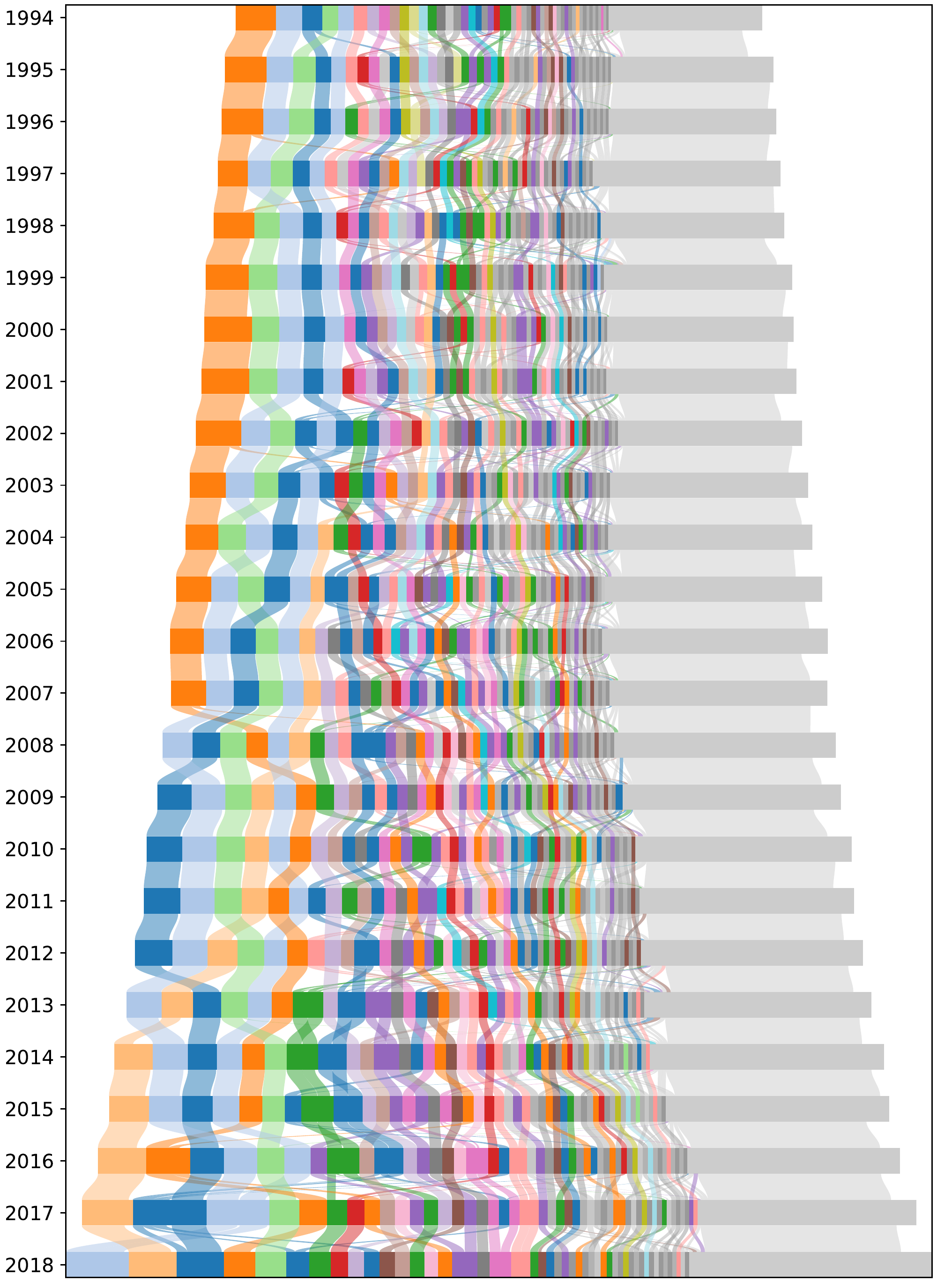}
		\subcaption{Germany (50 + 1 clusters drawn  per year)}
	\end{subfigure}~%
	\begin{subfigure}{0.45\linewidth}
		\includegraphics[width=\linewidth]{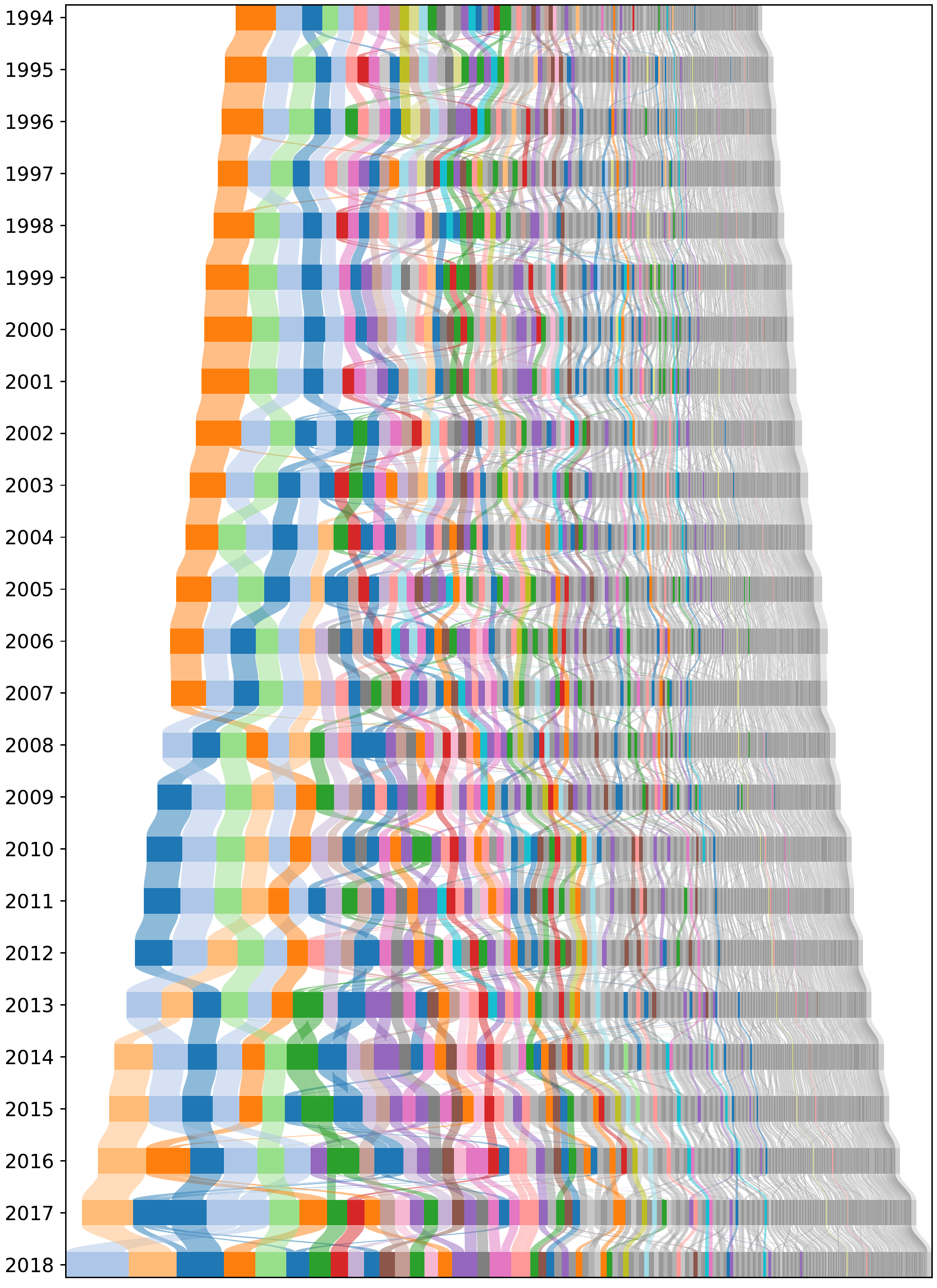}
		\subcaption{Germany (500 + 1 clusters drawn per year)}
	\end{subfigure}
	\caption{Federal legislation in the United States and Germany by cluster (1994--2018), depicted as in Figure~5 from the main paper, with different thresholds for summarising small clusters in one miscellaneous cluster.}
	\label{fig:sankey-500}
\end{figure}

To illustrate that the clusters we summarise in the miscellaneous clusters have little impact on our results, 
Figure~\ref{fig:sankey-500} juxtaposes analogues of Figure~5 from the main paper 
that summarise only clusters behind the $500$\textsuperscript{th} largest cluster in a miscellaneous cluster with their original counterparts that summarise all clusters behind the $50$\textsuperscript{th} cluster, where clusters are sorted in decreasing order of their size. 


\begin{table}[H]
	\centering\footnotesize
\bgroup
\def\arraystretch{1.08}
\hspace*{-4em}\begin{tabular}{p{0.08\textwidth}R{0.08\textwidth}p{0.9\textwidth}}
\textbf{Leading Cluster}&\textbf{Percentage}&\textbf{Chapter Path}\\[6pt]
1994-6&81.35 &	TITLE 42-THE PUBLIC HEALTH AND WELFARE / CHAPTER 6-THE CHILDREN'S BUREAU\\
&6.19 &	TITLE 42-THE PUBLIC HEALTH AND WELFARE / CHAPTER 34-ECONOMIC OPPORTUNITY PROGRAM\\
&3.30 &	TITLE 45-RAILROADS / CHAPTER 9-RETIREMENT OF RAILROAD EMPLOYEES\\
1994-14&53.21 &	TITLE 21-FOOD AND DRUGS / CHAPTER 8-NARCOTIC FARMS\\
&14.77 &	TITLE 7-AGRICULTURE / CHAPTER 6-INSECTICIDES AND ENVIRONMENTAL PESTICIDE CONTROL\\
&10.91 &	TITLE 15-COMMERCE AND TRADE / CHAPTER 47-CONSUMER PRODUCT SAFETY\\[6pt]
1998-9&49.30 &	TITLE 42-THE PUBLIC HEALTH AND WELFARE / CHAPTER 7-SOCIAL SECURITY\\
&32.88 &	TITLE 42-THE PUBLIC HEALTH AND WELFARE / CHAPTER 6A-PUBLIC HEALTH SERVICE\\
&4.21 &	TITLE 42-THE PUBLIC HEALTH AND WELFARE / CHAPTER 35-PROGRAMS FOR OLDER AMERICANS\\
1998-13&50.87 &	TITLE 21-FOOD AND DRUGS / CHAPTER 9-FEDERAL FOOD, DRUG, AND COSMETIC ACT\\
&14.81 &	TITLE 7-AGRICULTURE / CHAPTER 6-INSECTICIDES AND ENVIRONMENTAL PESTICIDE CONTROL\\
&8.33 &	TITLE 15-COMMERCE AND TRADE / CHAPTER 47-CONSUMER PRODUCT SAFETY\\[6pt]
2002-10&48.65 &	TITLE 42-THE PUBLIC HEALTH AND WELFARE / CHAPTER 7-SOCIAL SECURITY\\
&35.10 &	TITLE 42-THE PUBLIC HEALTH AND WELFARE / CHAPTER 6A-PUBLIC HEALTH SERVICE\\
&3.69 &	TITLE 7-AGRICULTURE / CHAPTER 51-FOOD STAMP PROGRAM\\
2002-18&54.65 &	TITLE 21-FOOD AND DRUGS / CHAPTER 9-FEDERAL FOOD, DRUG, AND COSMETIC ACT\\
&13.77 &	TITLE 7-AGRICULTURE / CHAPTER 6-INSECTICIDES AND ENVIRONMENTAL PESTICIDE CONTROL\\
&7.82 &	TITLE 15-COMMERCE AND TRADE / CHAPTER 47-CONSUMER PRODUCT SAFETY\\[6pt]
2006-10&50.39 &	TITLE 42-THE PUBLIC HEALTH AND WELFARE / CHAPTER 7-SOCIAL SECURITY\\
&34.42 &	TITLE 42-THE PUBLIC HEALTH AND WELFARE / CHAPTER 6A-PUBLIC HEALTH SERVICE\\
&3.67 &	TITLE 42-THE PUBLIC HEALTH AND WELFARE / CHAPTER 35-PROGRAMS FOR OLDER AMERICANS\\
2006-14&53.84 &	TITLE 21-FOOD AND DRUGS / CHAPTER 9-FEDERAL FOOD, DRUG, AND COSMETIC ACT\\
&13.28 &	TITLE 7-AGRICULTURE / CHAPTER 6-INSECTICIDES AND ENVIRONMENTAL PESTICIDE CONTROL\\
&6.79 &	TITLE 15-COMMERCE AND TRADE / CHAPTER 47-CONSUMER PRODUCT SAFETY\\[6pt]
2010-2&43.74 &	TITLE 42-THE PUBLIC HEALTH AND WELFARE / CHAPTER 7-SOCIAL SECURITY\\
&29.89 &	TITLE 42-THE PUBLIC HEALTH AND WELFARE / CHAPTER 6A-PUBLIC HEALTH SERVICE\\
&10.85 &	TITLE 21-FOOD AND DRUGS / CHAPTER 9-FEDERAL FOOD, DRUG, AND COSMETIC ACT\\[6pt]
2014-3&42.90 &	TITLE 42-THE PUBLIC HEALTH AND WELFARE / CHAPTER 7-SOCIAL SECURITY\\
&28.56 &	TITLE 42-THE PUBLIC HEALTH AND WELFARE / CHAPTER 6A-PUBLIC HEALTH SERVICE\\
&12.75 &	TITLE 21-FOOD AND DRUGS / CHAPTER 9-FEDERAL FOOD, DRUG, AND COSMETIC ACT\\[6pt]
2018-17&82.07 &	TITLE 42-THE PUBLIC HEALTH AND WELFARE / CHAPTER 7-SOCIAL SECURITY\\
&5.05 &	TITLE 7-AGRICULTURE / CHAPTER 51-SUPPLEMENTAL NUTRITION ASSISTANCE PROGRAM\\
&4.47 &	TITLE 42-THE PUBLIC HEALTH AND WELFARE / CHAPTER 35-PROGRAMS FOR OLDER AMERICANS\\
2018-11&59.96 &	TITLE 42-THE PUBLIC HEALTH AND WELFARE / CHAPTER 6A-PUBLIC HEALTH SERVICE\\
&27.72 &	TITLE 21-FOOD AND DRUGS / CHAPTER 9-FEDERAL FOOD, DRUG, AND COSMETIC ACT\\
&4.26 &	TITLE 7-AGRICULTURE / CHAPTER 6-INSECTICIDES AND ENVIRONMENTAL PESTICIDE CONTROL\\
\end{tabular}
\egroup
	\caption{Top $3$ contents of clusters in family $0$ (leading cluster: 2015-3), labelled ``Public Health and Social Welfare'', in four-year intervals from $1994$ to $2018$.
}
	\label{tab:cluster-labeling}
\end{table}

\begin{table}[H]
	\centering
\bgroup
\def\arraystretch{1.5}
\begin{tabular}{r|p{0.1\textwidth}cp{0.7\textwidth}}
	&\textbf{Leading Cluster}&\textbf{Color}&\textbf{Label}\\\hline
	1&2015-3&\colorbox{tab1}{\makebox[2em]{\strut}}&Public Health and Social Welfare\\
	2&2010-0&\colorbox{tab4}{\makebox[2em]{\strut}}&Financial Regulation for Consumers\\
	3&2011-3&\colorbox{tab3}{\makebox[2em]{\strut}}&Education and Students' Economic Support\\
	4&2015-11&\colorbox{tab2}{\makebox[2em]{\strut}}&Taxes and Retirement Security\\
	5&2015-4&\colorbox{tab7}{\makebox[2em]{\strut}}&Foreign Assistance, Development Aid, Arms Export, and Export Control\\
	6&2018-10&\colorbox{tab15}{\makebox[2em]{\strut}}&Immigration and Border Security\\
	7&2015-0&\colorbox{tab5}{\makebox[2em]{\strut}}&Environmental Protection and Wildlife Conservation\\
	8&1994-9&\colorbox{tab12}{\makebox[2em]{\strut}}&Energy Regulation, Conservation, and Transport\\
	9&2018-6&\colorbox{tab13}{\makebox[2em]{\strut}}&Small Business Aid and Public Procurement\\
	10&2018-7&\colorbox{tab8}{\makebox[2em]{\strut}}&Customs\\
	11&2005-6&\colorbox{tab19}{\makebox[2em]{\strut}}&Taxes and National Security\\
	12&2018-28&\colorbox{tab10}{\makebox[2em]{\strut}}&Capital Markets, Securities, and Commodity Exchange\\
	13&2018-15&\colorbox{tab16}{\makebox[2em]{\strut}}&Telecommunications and Copyright\\
	14&2017-2&\colorbox{tab20}{\makebox[2em]{\strut}}&Government Organization and Public Administration\\
	15&2018-9&\colorbox{tab17}{\makebox[2em]{\strut}}&Veterans' Benefits\\
	16&2007-12&\colorbox{tab11}{\makebox[2em]{\strut}}&Immigration and Trafficking\\
	17&1995-3&\colorbox{tab14}{\makebox[2em]{\strut}}&Agricultural Goods Production and Control\\
	18&2012-26&\colorbox{tab18}{\makebox[2em]{\strut}}&Government Employees' Health and Retirement\\
	19&2009-8&\colorbox{tab6}{\makebox[2em]{\strut}}&Public Housing and Homelessness\\
	20&2013-0&\colorbox{tab9}{\makebox[2em]{\strut}}&Native Americans\\
\end{tabular}
\egroup
	\caption{Labels assigned to the $20$ largest cluster families in the United States, ordered by regression slope (cf. Table~\ref{tab:cluster-family-growth-stats}).
	}
	\label{tab:labels-all-us}
\end{table}

\begin{table}[H]
	\centering
\bgroup
\def\arraystretch{1.5}
\begin{tabular}{r|p{0.1\textwidth}cp{0.7\textwidth}}
	&\textbf{Leading Cluster}&\textbf{Color}&\textbf{Label}\\\hline
	1&2018-2&\colorbox{tab1}{\makebox[2em]{\strut}}&Social Security\\
	2&2017-8&\colorbox{tab3}{\makebox[2em]{\strut}}&Financial Regulation\\
	3&2018-34&\colorbox{tab5}{\makebox[2em]{\strut}}&Market and Network Regulation\\
	4&2018-0&\colorbox{tab2}{\makebox[2em]{\strut}}&Taxes\\
	5&2016-8&\colorbox{tab9}{\makebox[2em]{\strut}}&Public Health and Enforcement\\
	6&2015-15&\colorbox{tab13}{\makebox[2em]{\strut}}&Corporations and Insurance\\
	7&2018-24&\colorbox{tab11}{\makebox[2em]{\strut}}&Environmental Protection\\
	8&2017-22&\colorbox{tab14}{\makebox[2em]{\strut}}&Immigration and Asylum\\
	9&1997-9&\colorbox{tab16}{\makebox[2em]{\strut}}&Traffic, Transport and Administrative Procedure\\
	10&2010-0&\colorbox{tab10}{\makebox[2em]{\strut}}&Constitution and State Organization\\
	11&2000-0&\colorbox{tab3}{\makebox[2em]{\strut}}&Criminal and Administrative Offences\\
	12&2018-18&\colorbox{tab7}{\makebox[2em]{\strut}}&Commercial Law and Accounting\\
	13&2018-4&\colorbox{tab6}{\makebox[2em]{\strut}}&Private Law, Property Law, and Estate Law\\
	14&2012-2&\colorbox{tab8}{\makebox[2em]{\strut}}&Public Servants, Judges, and Soldiers\\
	15&2016-29&\colorbox{tab15}{\makebox[2em]{\strut}}&Construction and Environmental Protection\\
	16&2018-10&\colorbox{tab12}{\makebox[2em]{\strut}}&Family Law and Benefits\\
	17&2011-21&\colorbox{tab19}{\makebox[2em]{\strut}}&Inheritance and Public Notaries\\
	18&1996-5&\colorbox{tab17}{\makebox[2em]{\strut}}&Pension Alignment\\
	19&2000-16&\colorbox{tab20}{\makebox[2em]{\strut}}&Reparations and Compensations\\
	20&1996-19&\colorbox{tab18}{\makebox[2em]{\strut}}&Labour Promotion\\
\end{tabular}
\egroup
	\caption{Labels assigned to the $20$ largest cluster families in Germany, ordered by regression slope (cf. Table~\ref{tab:cluster-family-growth-stats}).
	}
	\label{tab:labels-all-de}
\end{table}

\newpage

\bibliography{bibliography}






